%% file: superfgd.tex
\pdfoutput=1
\documentclass[preprint,12pt]{elsarticle}

\usepackage{amssymb}
\usepackage{amsmath}
\usepackage{float}
\usepackage{xcolor}

\usepackage{siunitx}
\usepackage{tikz}

\usepackage{subcaption}
\usepackage{pdfpages}

\usepackage{hyperref}
\usepackage{comment}

\graphicspath{{figures/}{figures/dEdx/}}

\journal{Nuclear Instruments and Methods}

\newcommand{\sonoarrivatoqui}[1]{\textcolor{red}{\\ \%\%\% sono arrivato qui \%\%\% \\\\}}

\begin{document}

\begin{frontmatter}




\title{The Super Fine-Grained Detector for the T2K neutrino oscillation experiment}
\input{author.tex}

\begin{abstract}
The magnetised near detector ND280 of the long-baseline neutrino experiment T2K has been upgraded to improve its detection performance and, consequently, enhance our understanding of neutrino-nucleus interactions, reducing the systematic uncertainties in measurements of the neutrino oscillation parameters. A key component of the upgrade is a novel segmented plastic scintillator detector, called the Super Fine-Grained Detector (SuperFGD), made of approximately 2 million optically isolated 1~cm$^3$ cubes read out by three orthogonal wavelength-shifting~(WLS) fibres. Scintillation photons are detected by 55,888 Hamamatsu Multi-Pixel Photon Counters (MPPCs). The SuperFGD provides 3D images of neutrino interactions by tracking the final-state charged particles produced isotropically, including protons down to a threshold of around 330 MeV/$c$. The high light yield of SuperFGD greatly improves particle identification and the sub-nanosecond time resolution provides an excellent identification of Michel electrons. The SuperFGD is also able to detect neutrons from neutrino interactions and, for the first time in a neutrino experiment, to reconstruct their kinetic energy using a fine detector segmentation and  by measuring the time-of-flight with sub-nanosecond precision. In this article the details of the detector design, construction and performance are described. The detector was installed in ND280 and successfully commissioned with cosmic data in 2023 and, later, with the T2K neutrino beam. The detector response has been characterised with the 2023 and 2024 data and the results are reported in this article. \\
\end{abstract}

\begin{keyword}
 Neutrino oscillations \sep T2K experiment \sep scintillating segmented detector \sep wavelength shifting fibres \sep  micropixel photon counters  \sep LED calibration system \sep electronics Front-End Boards
\end{keyword}




\end{frontmatter}

\setcounter{footnote}{0}

\tableofcontents
\newpage
\section{Introduction }
\input{intro.tex}
\label{sec:introduction}
\section{General description of the SuperFGD }
\label{sec:general}
\input{general_description.tex}
\section{Scintillator cubes }
\label{sec:cubes}
\input{cubes.tex}
\section{SuperFGD mechanics }
\label{sec:mechanics}
\input{mechanics.tex}
\section{Assembly procedure  }
\label{sec:assembly}
\input{assembly.tex}
\section{Photosensors and WLS fibres }
\label{sec:mppc}
\input{photosensors.tex}
\section{LED calibration system }
\label{sec:led}
\input{ledsystem.tex}
\section{SuperFGD electronics }
\label{sec:electronics}
\input{electronics.tex}
\section{DAQ and triggers }
\label{sec:daq}
\input{daq.tex}
\section{Slow control system}
\label{sec:slowcontrol}
\input{slowcontrol.tex}
\section{Detector performance }
\label{sec:detector_performance}
\input{detector_performance.tex}
\section{Conclusions }
\label{sec:conclusion}
\input{conclusions.tex}
\section*{Acknowledgements}

This work was supported in the framework of the State project ``Science'' by the Ministry of Science and Higher Education of the Russian Federation under contract No. 075-15-2024-541. We acknowledge the support of JSPS KAKENHI Grant Numbers 16H06288 and 20H00149 and the SNSF grant PCEFP2\_203261, Switzerland. We gratefully acknowledge the support of CNRS/IN2P3, France; the Ministry of Science and Higher Education (2023/WK/04) and the National Science Centre (UMO-2018/30/E/ST2/00441), Poland; DOE, USA; the STFC and UKRI, UK. The research leading to these results has received funding from the Spanish Ministry of Science and Innovation PID2022-136297NB-I00 /AEI/10.13039/501100011033/ FEDER, UE. IFAE is partially funded by the CERCA program of the Generalitat de Catalunya.

In addition, participation of individual researchers has been supported by the European Union’s Horizon 2020 Research and Innovation Programme under the grant numbers RISE-GA872549-SK2HK and RISE-GA822070-JENNIFER2 and the Horizon Europe Marie Sklodowska-Curie Staff Exchange project JENNIFER3 Grant Agreement no.101183137.

For the purposes of open access, the authors have applied a Creative Commons Attribution licence to any Author Accepted Manuscript version arising. Representations of the data relevant to the conclusions drawn here are provided within this paper.





\bibliographystyle{elsarticle-num}
\bibliography{bibliography}

\end{document}

%% file: author.tex
\author[icrr]{S.~Abe}
\author[lancs]{H.~Alarakia-Charles}
\author[lpi]{I.~Alekseev}
\author[utokyo]{T.~Arai}
\author[tmu]{T.~Arihara}
\author[kyoto]{S.~Arimoto}
\author[jinr]{A.M.~Artikov}
\author[tmu]{Y.~Awataguchi}
\author[lsu]{N.~Babu}
\author[jinr]{V.~Baranov}
% ...
\address[icrr]{University of Tokyo, Institute for Cosmic Ray Research, Kamioka Observatory, Kamioka, Japan}
\address[lancs]{Lancaster University, Physics Department, Lancaster, United Kingdom}
\address[lpi]{Lebedev Physical Institute of the Russian Academy of Sciences, Moscow, Russia}
\address[utokyo]{University of Tokyo, Tokyo, Japan}
\address[tmu]{Tokyo Metropolitan University, Department of Physics, Tokyo, Japan}
\address[kyoto]{Kyoto University, Department of Physics, Kyoto, Japan}
\address[jinr]{Joint Institute for Nuclear Research, Dubna, Moscow Region, Russia}
\address[lsu]{Louisiana State University, Baton Rouge, USA}
% ...
\cortext[cor]{Corresponding author}
\fntext[fnref1]{tristan.doyle@physics.ox.ac.uk}
\fntext[fnref2]{kudenko@inr.ru}
\fntext[fnref3]{tsuna@post.kek.jp}
\fntext[fnref4]{davide.sgalaberna@cern.ch}

%% file: intro.tex
The main goals of current and future long-baseline (LBL) neutrino oscillation experiments are to search for leptonic CP violation (CPV), determine the neutrino mass ordering, and perform precise measurements of the neutrino oscillation parameters. The LBL accelerator neutrino experiment Tokai-to-\-Kami\-oka~(T2K)~\cite{Abe:2011ks} is studying neutrino oscillations using a neutrino beam produced at the Japan Proton Accelerator Research Complex (J-PARC) in Tokai, Japan, and directed towards the 50 kt water Cherenkov detector Super-Kamiokande~\cite{Super-Kamiokande:2002weg}, located 295 km away in Kamioka. A magnetised near detector ND280 is located 280 m downstream of the hadron production target at J-PARC and measures the properties of the neutrino beam before oscillations occur. ND280 precisely measures the unoscillated neutrino spectra to constrain systematic uncertainties related to the neutrino flux and interaction cross section, which is required to predict the neutrino event rate and energy  spectrum at Super-Kamiokande.

\begin{figure}[htb] 
\centering\includegraphics[width=13cm,angle=0]{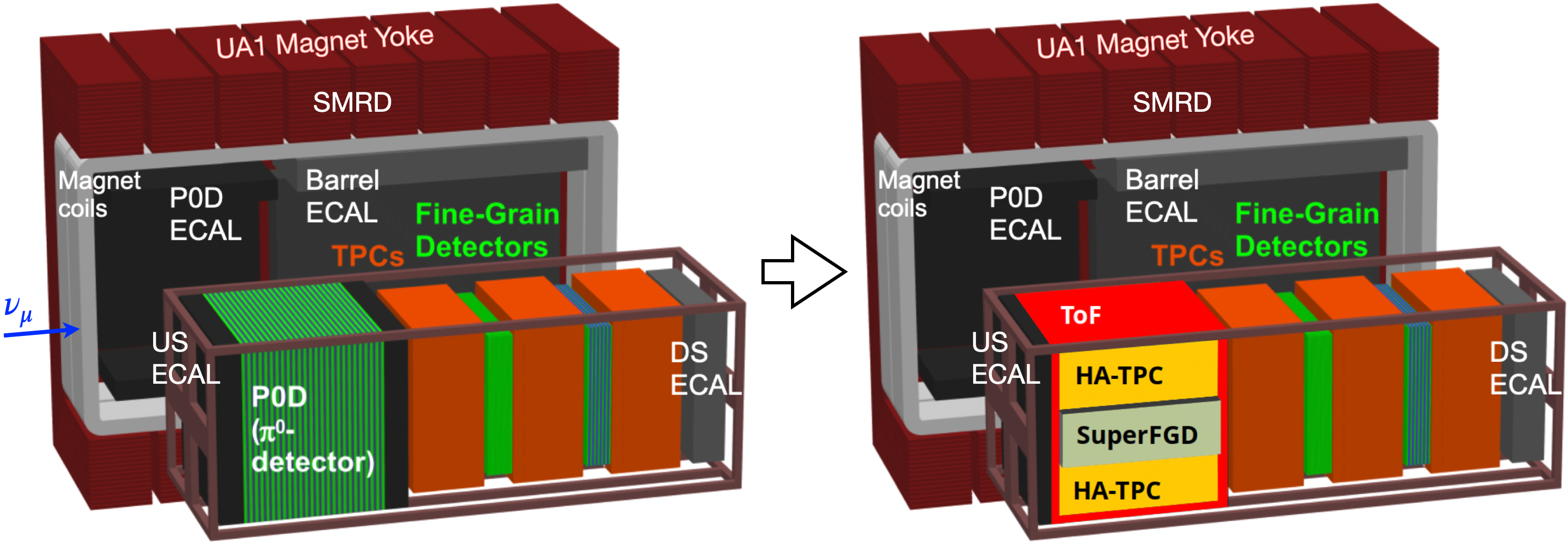} 
\caption{The T2K near detector ND280 before the upgrade (left). The P\O D  is replaced by new detectors HA-TPCs, SuperFGD, and ToF    after the upgrade (right).}
\label{fig:nd280}
\end{figure}

The original ND280 detector comprises a Pi-Zero Detector (P\O D), three low-density gaseous Time Projection Chambers (TPCs), two Fine-Grained Detectors (FGDs) composed of scintillator bars which serve as targets for neutrino interactions, an Electromagnetic Calorimeter (ECal), and a Side Muon Range Detector (SMRD). All of these detectors are enclosed inside the magnet recycled from the UA1 experiment~\cite{UA1:1983crd} at CERN operating with a magnetic field of 0.2 T, with the exception of the SMRD which sits in the air gaps between the iron plates of the magnet yoke. An illustration of the original ND280 is shown in Fig.~\ref{fig:nd280} and is described in detail in~\cite{Kudenko:2008ia,Amaudruz:2012agx,Assylbekov:2011sh, Allan:2013ofa,Aoki:2012mf}. ND280 has been used in T2K oscillation measurements to reduce the uncertainty on the predicted event rates at the far detector from $\sim 13$\% to $\sim 3\%$~\cite{T2K:2023smv}.

The main limitation with the original ND280 design is that in order to precisely determine the properties of the leptons produced in neutrino interactions, they have to be reconstructed in at least one of the TPCs. As a result, the reconstruction efficiency in the forward region is excellent, but it drops considerably for scattering angles greater than $\sim 50^{\circ}$ with respect to the beam direction. However at Super-Kamiokande the reconstruction efficiency is flat as a function of lepton angle with respect to the beam direction. Moreover, ND280 has limited tracking efficiency and a relatively high momentum threshold, especially for protons at approximately 450 MeV/$c$. The reconstruction of low-momentum pions and protons is fundamental for studies of nuclear effects in neutrino interactions. Like all other accelerator neutrino detectors, ND280 is also not able to reconstruct the kinetic energy of neutrons which is of high importance for LBL experiments.

To overcome these limitations an upgrade of ND280 was performed. The goals of the upgrade were to increase the efficiency of reconstructing high-angle and low-momentum particles, accumulate a larger sample of $\nu_e$ interactions, and detect and measure the kinetic energy of neutrons. Therefore the requirements for the upgraded near detector are:
\begin{itemize}
    \item Full polar angle acceptance for particles produced in neutrino interactions with similar performance in terms of momentum resolution, $\text{d}E/\text{d}x$, and charge measurement as the original ND280.
    \item High tracking efficiency for low-energy pions and protons, with proton-pion separation to allow accurate determination of the neutrino event topology.
    \item High light yield and sub-nanosecond time resolution for identification of Michel electrons.
    \item Sufficient spatial and time resolution to efficiently detect neutrons and measure their kinetic energy by time-of-flight. 
    \item Reduction of the number background interactions originating in the areas surrounding SuperFGD.
\end{itemize}

To meet these requirements and improve the performance of ND280 new detectors were designed and constructed. The P\O D was replaced with a novel fully-active fine-grained plastic scintillator detector, the Super Fine-Grained Detector (SuperFGD)~\cite{Blondel_2018,Abe:2019whr}, and two TPCs~\cite{Attie:2021yeh}, surrounded by six Time Of Flight (TOF) planes~\cite{Korzenev:2021mny}, as shown in Fig.~\ref{fig:nd280}. The SuperFGD will operate as the target for neutrino interactions as well as the detector in which tracks around the interaction vertex will be reconstructed.

An intensive R\&D was carried out to optimize the configuration of SuperFGD and examine the expected performance. Parameters of individual scintillator cubes using WLS fibre/MPPC readout were  measured using cosmic muons~\cite{Blondel_2018}, a UV LED~\cite{Artikov2022}, and a UV laser~\cite{Alekseev:2022jki}.  Several SuperFGD prototypes comprised of $1\times 1\times 1~\text{cm}^3$ optically-isolated scintillator cubes with a three directional WLS fibre readout were tested with charged particles at CERN~\cite{Mineev:2019,Blondel_2020} and neutrons at LANL~\cite{Agarwal:2022kiv,Riccio:2023hwu}. Obtained parameters of cubic scintillators and prototypes met the above-mentioned physics requirements  and and opened up the possibility of construction a large 3D segmented scintillator neutrino detector.

%% file: general_description.tex
The SuperFGD, a two-tonne polystyrene-based plastic scintillator detector, will serve as an active neutrino target. It is segmented into $1\times1\times1~\text{cm}^3$ optically-isolated cubes. When a charged particle, such as a muon or an electron, passes through a cube, it produces an amount of scintillation light that depends on its energy and the distance crossed in the scintillator. In order to minimise the light leakage between adjacent cubes, each scintillator cube is covered by a white diffuser produced with chemical etching. Wavelength-shifting (WLS) optical fibres along the three orthogonal axes are used to capture and convey the light inside a cube to silicon Multi-Pixel Photon Counters (MPPCs) to count the number of photons produced by the particle. The SuperFGD contains 56 planes, each an array composed of $192\times182$ cubes. A total of 1,956,864 sensitive elements are read out by 55,888 electronics channels.

The 3D cubic geometry is innovative and provides isotropic particle tracking. The key benefits are improved particle tracking, improved angular acceptance, and a lower momentum threshold for detection. Furthermore, the cubic geometry increases the scintillation light output by almost a factor of four over the original ND280 scintillator detectors.

The detector is homogeneous in construction, i.e. the material inside the fiducial volume is exclusively made of plastic (primarily polystyrene), an important factor in precision measurements of neutrino cross sections. This new concept~\cite{Blondel_2018} is an improvement over traditional detectors using long scintillator bars~\cite{MINOS:2008hdf,Nitta:2004nt, MINERvA:2004gta}.

%% file: cubes.tex
\subsection{Design and production of cubes}

\begin{figure}[h]
\centering
\includegraphics[width=0.44\textwidth]{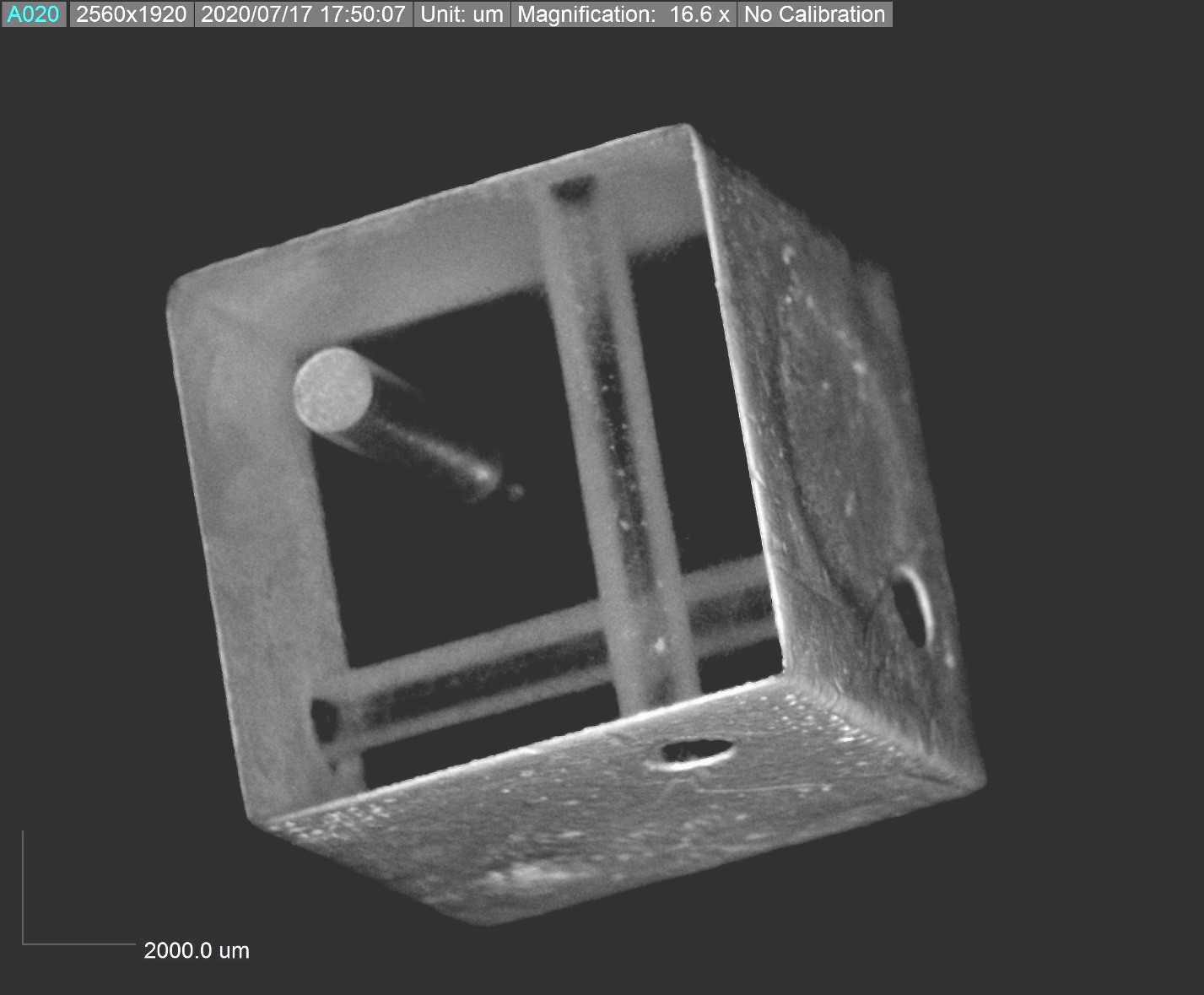}
\includegraphics[width=0.49\textwidth]{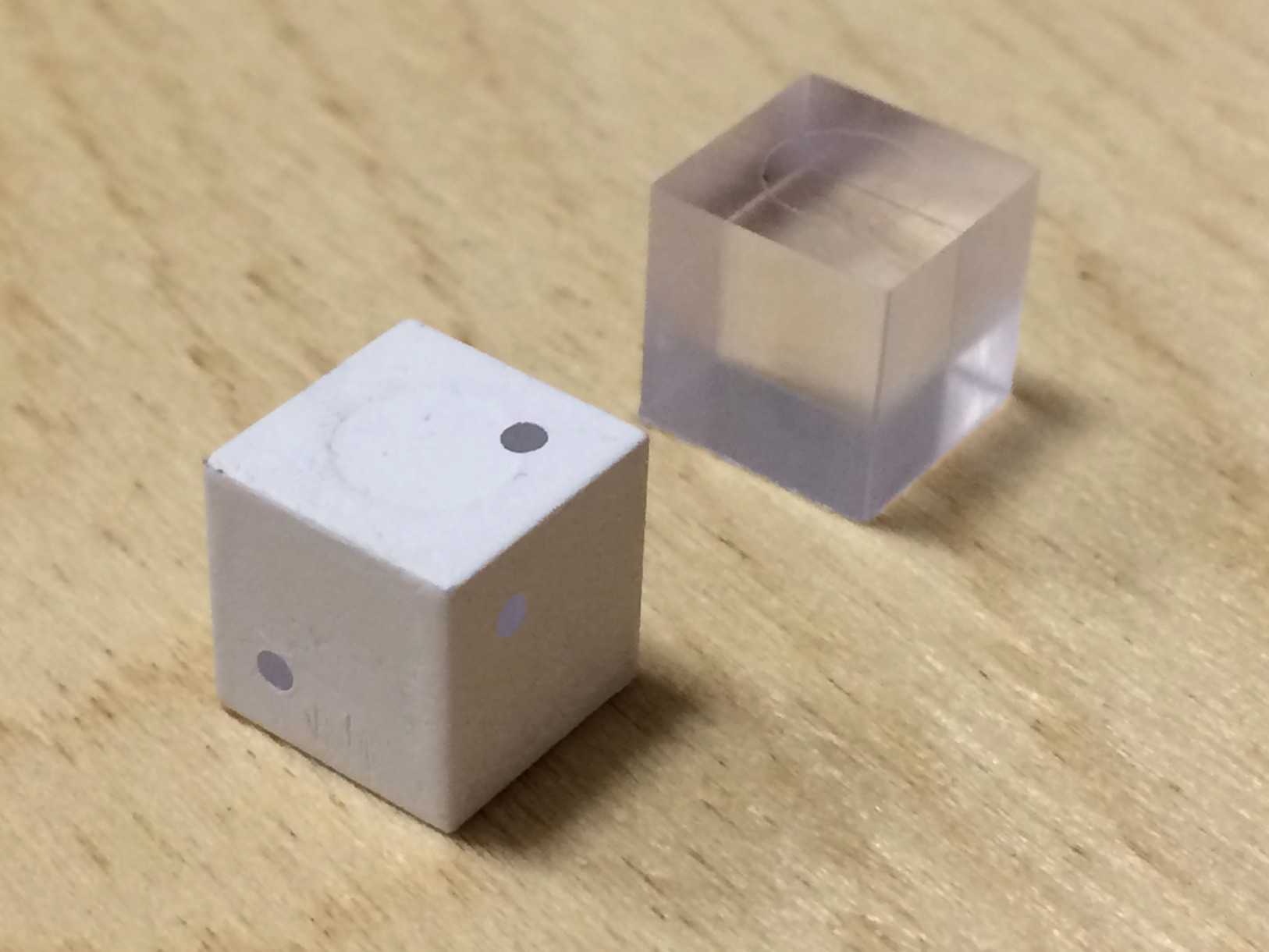}
\caption{Left: A zoom view a scintillator cube with three orthogonal holes with a diameter of 1.5~mm. Two sides of the cube polished to see the holes, four sides a covered by a reflector. Right: Scintillator cubes before and after chemical etching. A trace from the injection mould pusher is seen at the top of both cubes.}
\label{fig:cube}
\end{figure}

The active element of SuperFGD is a $10\times 10\times 10~{\rm mm}^3$ plastic scintillator cube as shown in Fig.~\ref{fig:cube}. This size provides the required fine granularity of the SuperFGD. Each cube has three orthogonal cylindrical holes of 1.5~mm diameter drilled along X, Y and Z axes to accommodate 1~mm diameter WLS fibres.
The scintillator cubes were produced at the UNIPLAST Co. (Vladimir, Russia)~\cite{Uniplast_Website} by injection moulding. The scintillator is composed of optical quality polystyrene doped with 1.5\% of paraterphenyl (PTP) and 0.01\% of POPOP. After fabrication the cubes were covered by a white diffusing layer by etching the scintillator surface with a chemical agent, see Fig.~\ref{fig:cube}. The etching results in the formation of a white polystyrene micropore deposit over the scintillator with good diffuse reflective performance.  

\subsection{Geometry of the cubes}

In order to assemble the SuperFGD, which contains about $2\times 10^6$ cubes, it was estimated  that the  cube three-dimensional size and positioning of the holes should vary within 0.1~mm.  Fluctuations in cube size and hole positions are expected during the manufacturing process. Table~\ref{tab:table_cubesize} 
\begin{table}[h!]
  \caption{Cube size after moulding with and without a chemical reflector. Size is presented as the average width between sides with rms error.}
\vspace{0.1 cm}
  \centering
  \begin{tabular}{llll} \hline \hline
   Type& width 1 [{\textmu}m]& width 2 [{\textmu}m]&  width 3 [{\textmu}m]\\ \hline
tr   No reflector & 10026$\pm$5 &  10026$\pm$6& 10066$\pm$14\\
   With reflector & 10166$\pm$31 &  10167$\pm$28& 10193$\pm$28\\ 
       \hline \hline 
  \end{tabular}
    \label{tab:table_cubesize}
\end{table}
summarises the size measurements of the cubes using a digital micrometer. The specified accuracy is 2~{\textmu}m. Three measurements were made per cube: two side-to-side widths (width 1 and width 2) in random order and one selected side-to-side width (width 3) where a trace from a pusher of a mould is visible at the cube surface as shown in Fig.~\ref{fig:cube}. The thickness of the reflector is determined to be 60--70~{\textmu}m.
We attribute the variation of the average cube size to reflector thickness, which is not well controlled during the manufacturing process. On the other hand, no systematic trend in the cube size was observed.  

While we can consider the cube size as relatively stable for the purpose of detector assembly, the hole positions relative to the cube perimeter show larger fluctuations that create a challenge to insert the WLS fibres. Drilled holes  to accommodate WLS fibres in a cube are shown in Fig.~\ref{fig:cube}.

\begin{figure}[h!]
\centering
\includegraphics[scale=0.20]{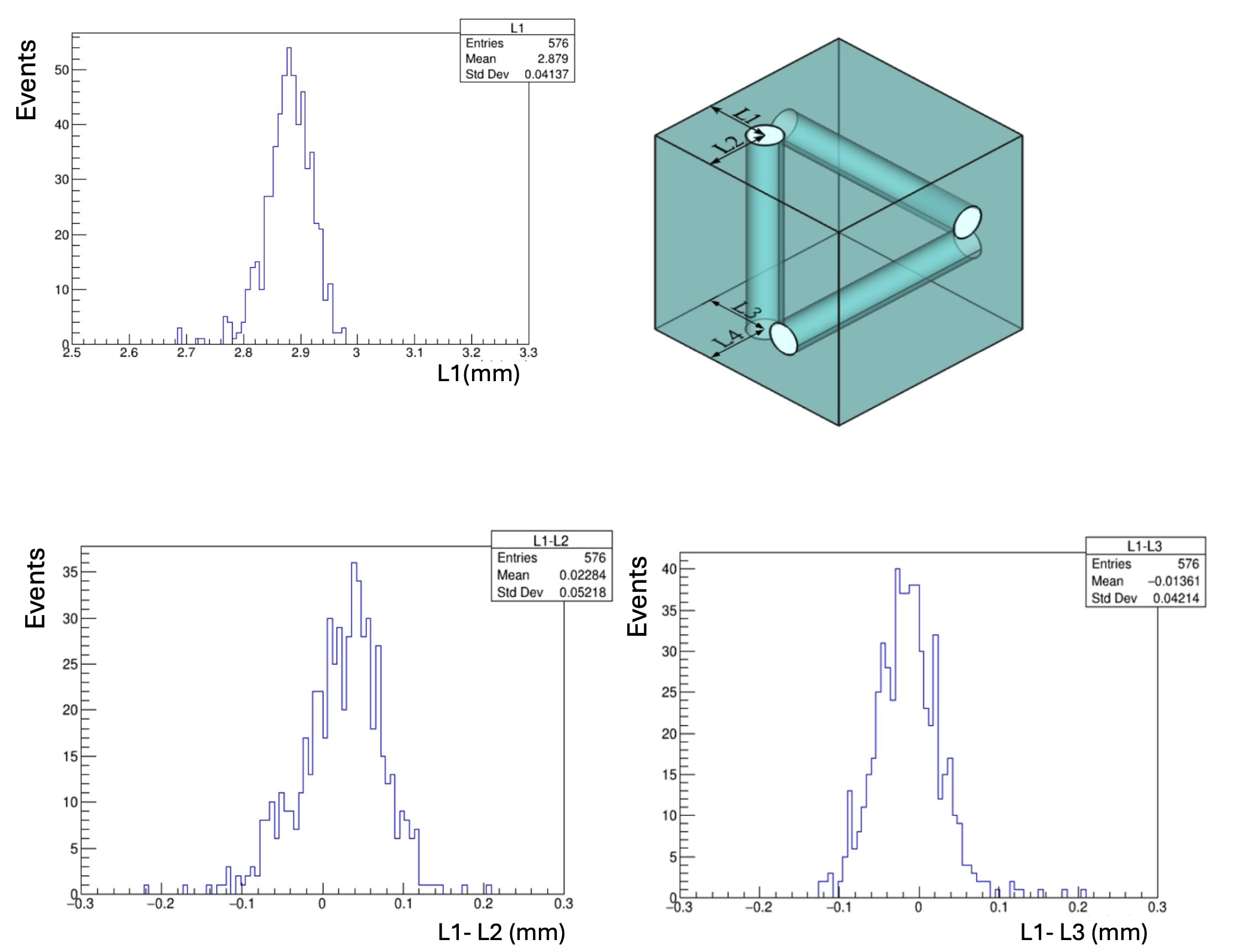}
\caption{Variation of the hole position at the drill entrance point. Top left: the variation of length L1. Bottom left: displacement of holes L1--L2. Bottom right: tilt of channels L1--L3. Top right: cube sketch with the definitions of each length.}
\label{fline_L}
\end{figure}

The geometry of the holes was measured using a digital microscope. Fig.~\ref{fline_L} shows the measurements of hole positions. The measured sample consisted of 192 cubes, with all six surface holes per cube (from three fibre channels) being measured. All 192 cubes had passed the quality check procedure described below  so large hole offsets were already excluded. Each channel in a cube gives four parameters: two distances between the drill entrance hole and adjacent cube edges (L1 and L2), and two distances for the exit hole at the opposite end of the channel (L3 and L4). The entrance point can be visually identified by a small halo around the hole left by a drill. Then we can calculate the displacement of the hole center (L1--L2) from the specified position and channel tilt angle (L1--L3 and L2--L4).

Table~\ref{table_holes} summarises the obtained channel geometry data. Lengths L1, L2 and L3 are close to the specified value of 3.0~mm, deviating by $\sim$0.1~mm which is well below the tolerance level. The length L4 is shifted by 0.1~mm from the other linear parameters. This systematic displacement is caused by a tilted drill direction, while the drill entrance point appears to be accurate.  

\begin{table}[h!]
   \caption{Parameters of the cube channels for a sample of 192 cubes with the holes drilled at the same machine.}
\vspace{0.5 cm}
  \centering
  \begin{tabular}{lllllll} \hline \hline
   & L1 &  L2 & L3 & L4 & L1-L3 & L2-L4 \\ \hline
   Average length [mm]& 2.88 & 2.86 & 2.89 & 2.76 & 0.01 &  0.09\\
   $\sigma$ [{\textmu}m]& 41 & 45& 60 & 77 & 42 & 47\\ 
       \hline \hline 
  \end{tabular}
    \label{table_holes}
\end{table}

From this table in can be concluded that  spread of cube dimensions and hole positions of $\sim 0.1$~mm allows us to draw 1~mm diameter WLS fibres through 192 cubes in the 3D array configuration of the SuperFGD. The main task was then to select cubes with parameters which meet the tolerance criteria during mass production.

\subsection{Selection of cubes using rods and fishing lines}

The selection of cubes was carried out at both the production site (UNIPLAST) and the pre-assembly site (INR RAS). It included a visual inspection to reject obvious issues: cubes with damaged surfaces, cubes with no holes, or cubes with holes in the wrong place. Due to time constraints in the production workflow, this visual inspection was performed during the SuperFGD pre-assembly procedure.

Since the final SuperFGD assembly with optical fibres was to be done at J-PARC, 56 layers of 192$\times$182 cubes were made using fishing lines during pre-assembly and transported to Japan. This pre-assembly procedure consisted of several stages: first, making cube strings on fishing lines (192 cubes per fishing line), second, formation of a layer of 182 strings on fishing lines (192 $\times$ 182 cubes), and third, pre-assembly of a 3D array out of 56 cube layers. Uncalibrated fishing lines of $\sim$1.3 mm diameter (actual average diameter was measured to be 1.35~mm) were used at every stage of the pre-assembly procedure.

\subsubsection{Making strings with 192 cubes}

Before making a string, each cube candidate was visually checked, then each hole of the cube was cleaned using knitting needles (stainless steel, 1.4 mm diameter, 19 cm length) to remove dust and chips after drilling.

\begin{figure}[h!] 
\centering\includegraphics[width=0.7\textwidth]
{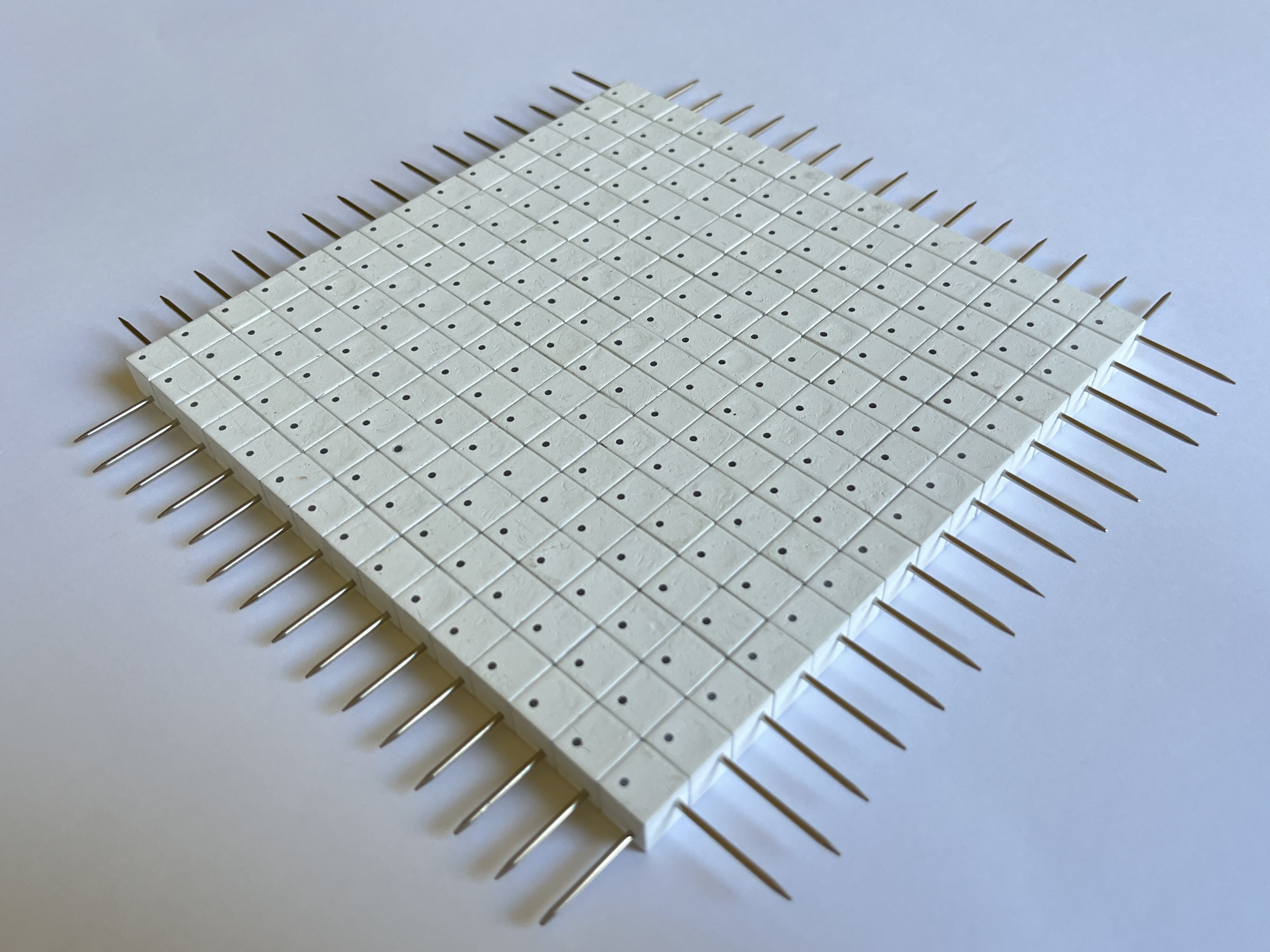} 
\caption{A 15$\times$15 array of cubes on knitting needles used to check the orthogonality of cube holes.}
\label{fig:15x15}
\end{figure}

Then a row of 15 preselected cubes were put on one knitting needle and the flatness was checked by looking for tilted cubes. This simple test allows cubes with non-orthogonal holes to be identified. The tilted cubes were replaced with good ones, then the procedure was repeated to make 15 knitting needles with cubes. After this a group of 15 knitting needles were placed into orthogonal holes of the checked cubes on a needle to form a 15$\times$15 array of cubes, as shown in Fig.~\ref{fig:15x15}. 
These additional needles should go smoothly through all the holes if the cubes and their holes are aligned. Otherwise, the cube(s) responsible for misalignment were removed and replaced. In total, about 20\% of produced cubes did not pass the test with knitting needles and were rejected.

In order to check different holes we dismounted this 15$\times$15 array of cubes, turned 15 needles with cubes by 90 degrees, and repeated the procedure. This completed a 3D test of 225 cubes. Then we inserted a prepared 2.5~m fishing line into the holes of 192 cubes from this 15$\times$15 array to make a string of cubes. We repeated this procedure to produce 182 strings of cubes per layer.

\subsubsection{Making a layer of strings with 192$\times$182 cubes}

\begin{figure}[h!] 
\centering\includegraphics[width=0.9\textwidth]
{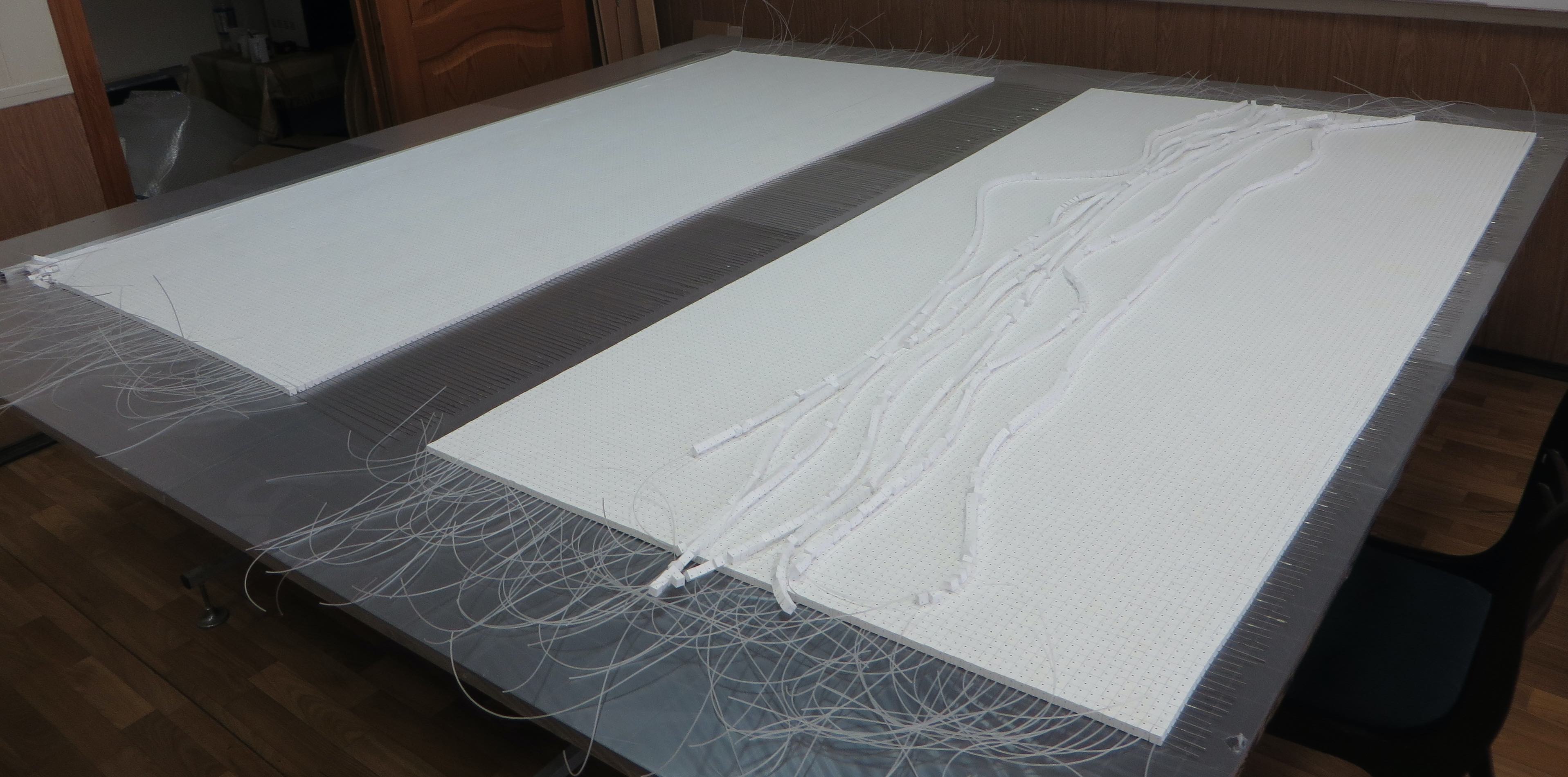} 
\caption{Two halves of a cube layer (192$\times$91 cubes each) with welding rods just before assembly of a full layer. One can see strings of 192 cubes on fishing lines on the top of the right half.}
\label{fig:layer}
\end{figure}

The prepared 182 strings were divided into two equal parts which were placed on a clean, flat surface. The strings were aligned in such a way that the horizontal holes of the corresponding cubes in adjacent strings were aligned and a welding rod (1.2 mm diameter, $\sim$1~m length) was inserted. One can insert up to 192 welding rods 
and make one half of a layer with 192$\times$91 cubes well aligned. The rods are rigid enough that alignment of the structure can be achieved even with a much smaller number of rods. After repeating this procedure with the second half of the layer, we moved these two halves as close to each other as possible, as shown in Fig.~\ref{fig:layer}. 
Then we sequentially removed the rods and inserted fishing lines in their places. After this we pushed the cubes from all sides to make the array of $192\times182$ cubes as compact as possible. Then we visually checked the flatness and alignment of the assembled cube array and replaced any tilted or damaged cubes that had passed the previous selection.

\subsubsection{Pre-assembly of a 3D array out of 56 cube layers}

\begin{figure}[h!] 
\centering\includegraphics[width=0.7\textwidth]{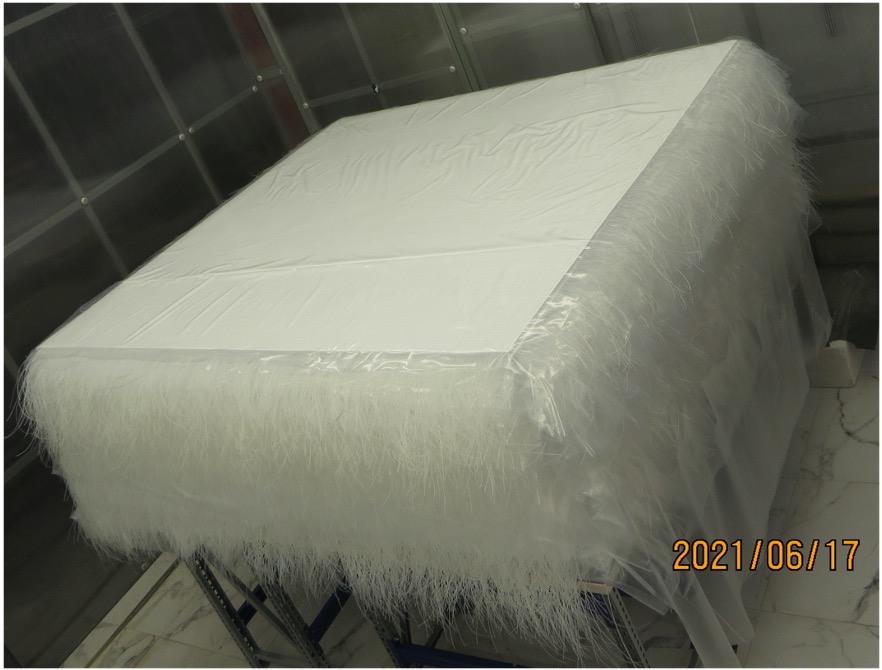} 
\caption{56 cube layers of 192$\times$182 cubes on fishing lines.}
\label{fig:56layers}
\end{figure}

After the assembly of each $192\times182$ cube layer it was placed on top of the previous one. Teflon tape strips were put on the surface of the previous layer which played the role of a solid lubricant. These strips were then removed and two layers were aligned using knitting needles (and welding rods at later stages). This allowed cubes with non-orthogonal holes which passed the previous selections to be identified, removed, and replaced. In total a 3D array of $56\times192\times182$ selected cubes was pre-assembled, as shown in Fig.~\ref{fig:56layers}.

\subsection{Evaluation of assembly procedure  }
\subsubsection{3D cube assemblies}
To smoothly insert the 1~mm diameter WLS fibres, all 56 vertical holes in all 34944 cube columns ($182\times192$) must be aligned in the vertical direction. To identify possible problems, several 3D cube assemblies of different size and shape were constructed. The prototypes also allowed us to accurately measure the achievable pitch between cube holes in the horizontal and vertical directions, which was a crucial parameter in the design of the mechanical box that holds the complete $56\times192\times182$ cube array.

In the first prototype (A), five cube planes were stacked and aligned within a right angle aluminium corner bar, where a matrix of 3~mm diameter holes was drilled with a pitch of 10.30~mm so fishing lines could be inserted. We checked the vertical alignment of cube holes with the 1.4~mm diameter needles used in the string assembly quality check.  We found no channels where the needle could not be inserted. It was possible to remove and reinsert the horizontal fishing lines. After one year the assembly procedure was repeated to check if the total plane size changed. We found that the total size of the cube assembly was the same as in the initial test.

\begin{figure}[h!]
\centering
\includegraphics[width=0.7\textwidth]{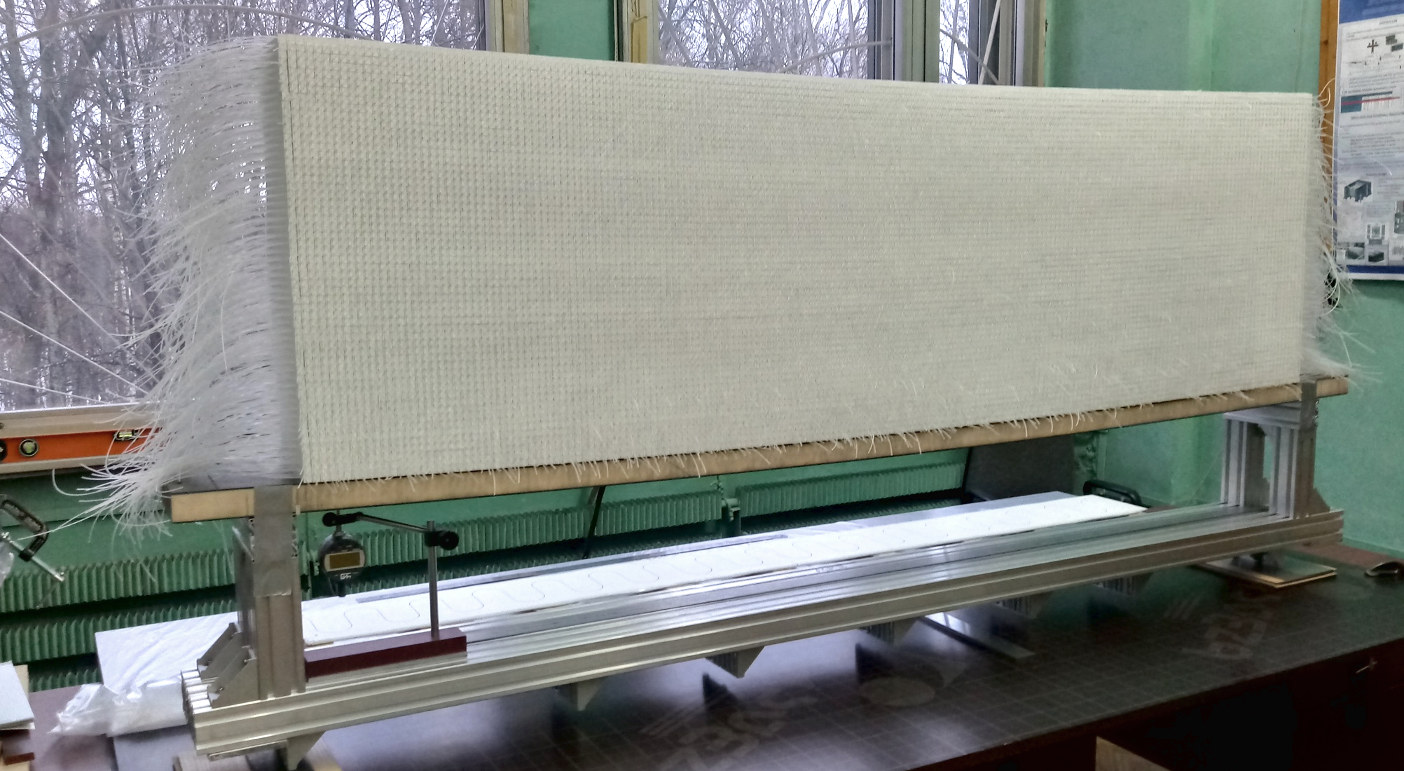}
\caption{The wall prototype during a sag test of a sandwich panel. The sandwich panel was tested to measure possible sag of the detector box bottom. }
\label{fig:wall}
\end{figure}

Another prototype (B) was built in the shape of a wall to have the 56 cubes height of the real detector, as shown in Fig.~\ref{fig:wall}.
56 narrow planes of $15\times192$ cubes on fishing lines were stacked as follows.  A batch of 5-6 narrow planes was placed at the lower layers, then we vertically aligned all the layers with 1.4~mm diameter needles at the maximum needle depth of 19~cm. The needles were then removed and the next batch was stacked on top. In the end all 56 layers were pierced with 1.2~mm diameter, 1~m length spokes. We found no vertical channel where the spoke could not be inserted. Then 1~mm fishing lines were inserted smoothly into all vertical channels. Some of the 192-cube-long horizontal channels were also checked. A 1.3~mm diameter fishing line was removed and a 1~mm diameter WLS fibre was inserted. The fishing line was then reinserted. Reinsertion of the fishing line was difficult in some cases, however the fibre was easily inserted inside the 192 cube-long channel.

\begin{figure}[h!]
\centering
\includegraphics[scale=0.8]{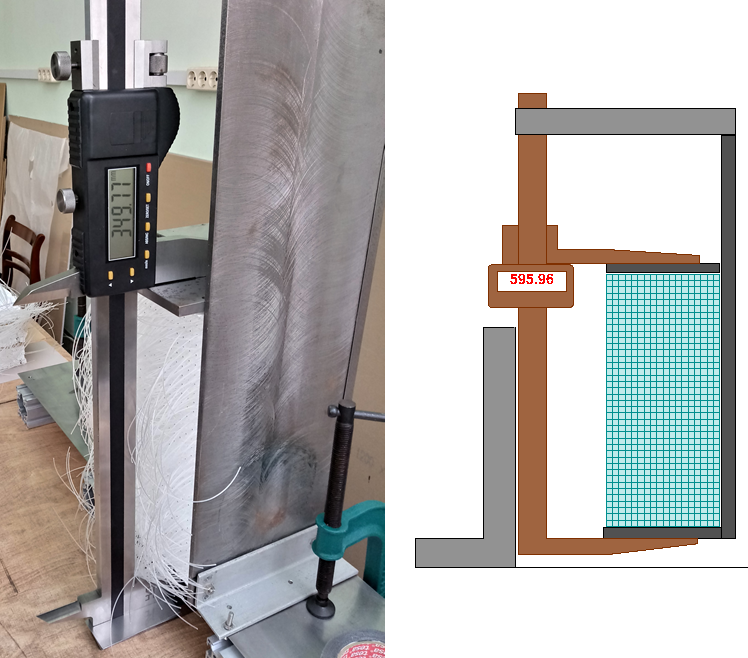}
\caption{Photo of the tower setup (left) and sketch of the height measurement of a 56 cube column (right). }
\label{fig:tower}
\end{figure}

A third prototype (C) for measuring the vertical pitch between the cube holes is shown in Fig.~\ref{fig:tower}. 
The tower of small planes ($15\times18$ cubes) was built within a precise rigid frame made of milled steel plates to provide a stable reproducible geometry of the array with the real 56 cube height. Fishing lines of 1~mm diameter were used to imitate the horizontal fibres. Vertical fishing lines were not used in this setup. The height is measured by an attached digital caliper with accuracy of around 0.01~mm.

The tower setup allowed us to measure both the full height and the height of an array with a smaller number of planes to see misalignment between the calculated position of a particular plane and its real position. In this way we could estimate the vertical pitch gradient. We measured the height of three different samples of cubes in the tower setup to see the range of fluctuations in geometry.

\subsubsection{Total size of the SuperFGD cube array} 

An array of many cubes includes variation of the following factors: cube size, gaps between the cubes, and cube elasticity caused by the diffuse reflector. Our measurements of a line of 192 cubes showed that the line length varies within 6~mm depending on the different cube combinations and pressure applied to compress the cubes. While we could control to some extent the outer size of the cube array because of intrinsic elasticity, we needed to know the achievable size of the SuperFGD cube assembly to design the mechanical box. A few prototypes were assembled on fishing lines and their sizes were measured with a 2~m long metallic ruler with least count of 0.5~mm and a digital caliper with least count of 0.01~mm.

Measurement of the size  of the  horizontal plane was done with five plane assemblies (prototype A). The total size corresponds to a pitch between cubes of 10.297~mm during the first assembly and 10.301~mm for the second assembly after one year. We also measured the deviation of the measured cube position from the calculated one for every 10th cube assuming a 10.30~mm pitch in a 192 cube line. The deviations were within 0.5~mm which was the precision defined by the least count of the 2~m long ruler. These deviations are smaller than the specified tolerance between cube position, WLS fibre and position of the corresponding hole in the mechanical box.

Measurement of 192 cube lines was performed using prototype B with a 2~m long digital caliper at 11 different heights. Without outer walls the cube array size was not well controlled. The length of the 192 cube lines varied from 1975.02~mm to 1978.45~mm with an average value of 1976.96~mm which corresponds to a pitch of 10.297~mm. The horizontal pitch for the mechanical box was set to be 10.301~mm.

We supposed that the vertical pitch can differ from the horizontal pitch because of gravity and different methods of cube alignment (fishing lines in horizontal planes vs steel needles in vertical).  We had to build dedicated setups to measure the vertical pitch with high accuracy. The average height of 56 cubes was measured in the wall prototype (B) to produce the pitch of 10.30~mm when the wall rests at the flat horizontal support.  After the prototype rested for one month at the tested sandwich panel, the wall prototype had a  central sag of 18~mm. A new measurement gave an average vertical pitch of 10.27~mm. We set the vertical pitch for the mechanical box to be 10.28~mm based on the data obtained in the measurements of the tower (prototype C).

\subsubsection{Tower tests} 

The tower setup of $15 \times 18 \times 56$ cubes is shown in Fig.~\ref{fig:tower}. The height of 56 cubes was measured between steel plates at the top and bottom of the cube array. A 14~kg load was placed on top of the tower to compensate the springiness of short horizontal fishing lines and accelerate settling of the cube columns due to gravity.

\begin{figure}[h!]
\centering
\includegraphics[scale=0.28]{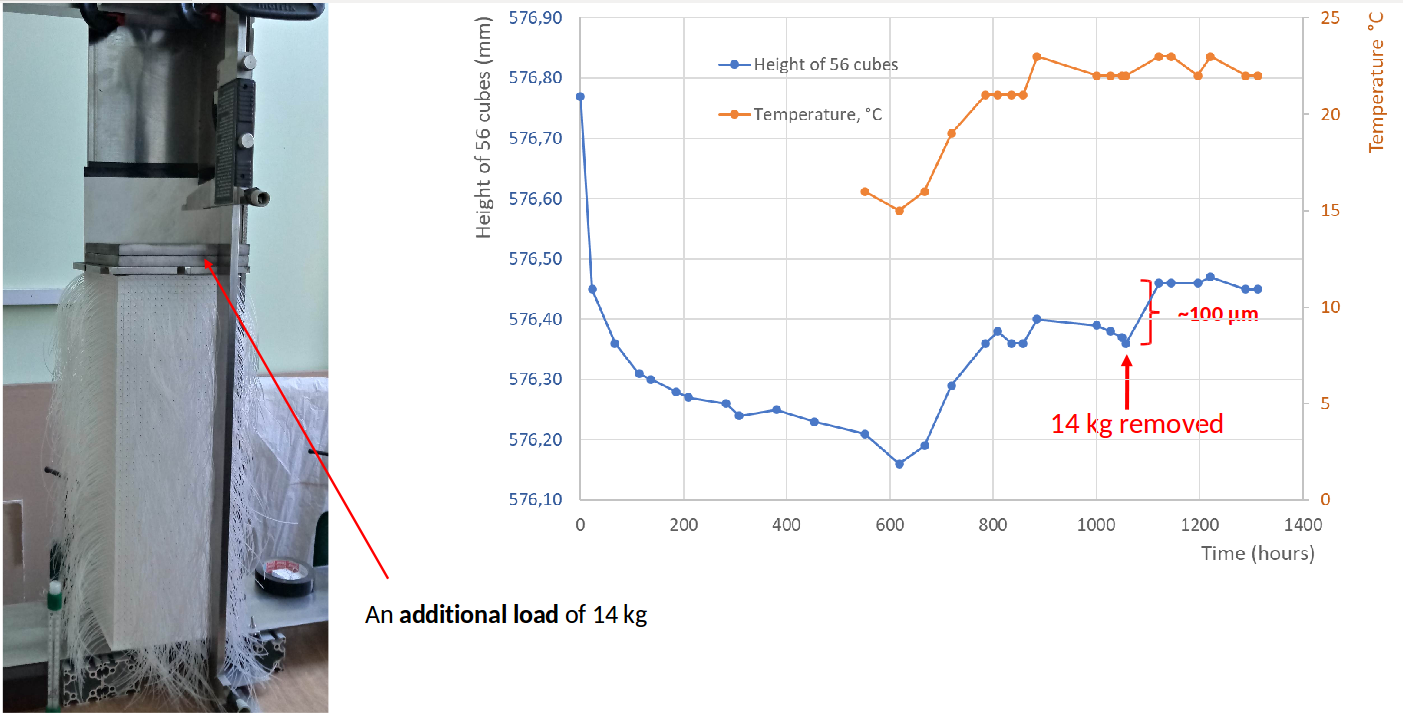}
\caption{Height of 56 cubes in the tower setup over two months.}
\label{56stability}
\end{figure}

The average pitch was found to be  10.30, 10.28 and 10.29~mm for three different samples of the cubes. The third sample was observed over two months to measure drift of the full height. As shown in Fig.~\ref{56stability}, the height stabilised during the first few days and then the height changed with the ambient temperature.  We calculated the linear thermal expansion as 52$\times 10^{-6}/^\circ$C while the specifications for polystyrene gives a value in the range  60-80$\times 10^{-6}/^\circ$C.

After the 14~kg load had been removed, the height of the tower increased by 0.1~mm, less than 0.02\%. To check the limits of compression of the cubes we performed the measurement of the full height with different loads on the tower. A maximum load of 96~kg compressed the tower by only 0.6~mm. The weight of the cubes themselves is 15~kg. After loading the tower we waited 15 minutes and repeated the measurements by unloading the tower. The tower recovered the full height with residual compression of about 120~{\textmu}m that attests some elasticity of the tower.

We suppose that compression happens mainly because of space rearrangement between the cubes. The effect of the elasticity of the reflector must only be visible at a much higher pressure than was applied in this test. According to the manufacturer the reflector begins to shrink under pressure of 1~kg/cm$^2$. We suppose that the whole cube array after assembling in a  tight structure acquires high resistance to outside pressure. It means the cube positions in the vertical direction could not be adjusted by applying pressure. The right pitch value had to be specified with an accuracy of 0.01~mm to ensure the maximum systematic misalignment between WLS fibres and holes in the mechanical box was smaller than 0.5~mm. The cube positions could be aligned to some extent in the horizontal directions because of relatively loose coupling in the cube planes.

Due to the protruding fishing lines, it was not possible to measure the positions of the cubes directly in the full height tower accurately. Therefore the vertical pitch gradient was measured by removing some planes and measuring the height of the remaining planes. The measurements were also performed by increasing the number of the planes. The gradient is characterised by a maximum deviation of the measured height from the calculated height assuming constant average pitch. We made these measurements five times with two different samples of the cubes. The maximum misalignment was found to be 0.35~mm, which is below the tolerance for the WLS fibres.

%% file: mechanics.tex
The mechanical box has to support the load of the 1,956,864 plastic scintillator cubes, allow the 55,888 WLS fibres to exit the box inner volume from both sides, be precisely coupled with both the MPPCs (surface mounted on PCBs, see Sec.~\ref{sec:mppc}) and the light guide plates of the LED calibration system (see Sec.~\ref{sec:led}). Moreover, it has to be safely installed into the ND280 basket without interfering with the other detectors and resist against possible stresses and vibrations during the installation or due to an earthquake. The total supported weight, including the box itself and the front-end electronics, is about 2.5 tonnes.

\subsection{Box plates and structure}
\label{sec:box}

The scintillator cubes inside the box are not glued together to avoid damaging the optical diffuser, therefore air gaps may exist between them. The cubes are crossed by three WLS fibres that should not be stressed by a potential misplacement of adjacent cubes that may cause cracks. Another potential cause of WLS fibre breakage is misalignment between the hole of the outermost scintillator cubes and the hole on the mechanical box. 

In order to overcome these possible issues, the mechanical box should precisely constrain the cubes in the three orthogonal directions and, thus have an internal volume of the same size as that occupied by the cubes. At the same time, the tolerances related to the cube three-dimensional size and the diameter and position of the WLS fibre hole has to be small enough to ensure that stresses are not applied to the WLS fibres when the mechanical box is closed. Moreover, the box must maintain a rectangular shape throughout the detector assembly, its move to the experimental hall, its installation in the ND280 basket, and for the duration of the experiment. The sagging of the bottom face must also be minimised.
Finally, the cubes need to be compressed enough by the box to prevent air gaps between them and prevent them from freely moving, for example if exposed to the vibrations produced by an earthquake. At the same time, the compression should not compress the cubes too much to avoid damaging the crossing WLS fibres.

The box is composed of six sub-assemblies, one for each side of the box. They consist of a structure of different layers glued with epoxy: two carbon fibre plates, a Divinycell H250 rigid foam core inside an aluminium frame, 
a fibre glass datum (bottom,left and upstream panels) or an EPDM soft surface (top, right and downstream panels) inside the box, and
fibre glass mounting plates for MPPC-PCBs and LED calibration modules on the outside surfaces.  
The mechanical rigidity is primarily provided by the two carbon fibre plates, each 2.3~mm thick. They consist of a custom-made epoxy-glued stack of eight prepreg unidirectional plies disposed with a relative angle of, respectively, $0^{\circ}$, $90^{\circ}$, $45^{\circ}$, $-45^{\circ}$, $45^{\circ}$, $90^{\circ}$, $0^{\circ}$. The Divinycell H250 foam, with a density of 250 kg/m$^3$, provides the necessary rigidity to maintain the two carbon fibre plates at the nominal distance (15~mm, 20~mm or 30~mm depending on the panel) while being light enough to allow charged particles to leave the SuperFGD depositing a minimal amount of energy and undergoing low multiple scattering.
The total thickness of the panels varies between approximately 24~mm and 56~mm.

The box plates are fixed to each other with stainless steel M6 socket head cap screws. The stainless steel corner brackets are attached to the aluminium frame by stainless steel M8 socket head cap screws. The box has a total of 111,776 holes to allow both ends of all the WLS fibres to exit, to be coupled on one side with the MPPCs and on the other side with the LED light guide plates. The holes have a conical shape, with a diameter of 5~mm on the inner side of the box and a diameter of 3~mm on the outer side, to account for a possible misalignment with the holes in the scintillator cubes as well as for the maximum WLS fibre bending radius (conservatively, 50~mm).

The glass fibre plates were produced by NEXUS \cite{Nexus_Website}. Then, carbon fibre sandwich plates were produced and glued to the glass fibre plates at CompositeDesign \cite{CompositeDesign_Website}. Finally, the plates were milled, machined and all the holes for the WLS fibres were drilled at CIMFORM \cite{Cimform_Website}. The box weighs about 536~kg without any cubes, lifting lugs, readout, calibration modules, or external electronics and cables.

After the box was fully assembled, a maximum sagging of 3~mm was measured in the centre of the bottom panel which was consistent with results of a finite element analysis (FEA). The FEA was tuned and validated with multiple measurements with different prototypes, including: an experimental modal analysis of the material at CERN, where vibrations were excited acoustically by a speaker placed behind the piece and a set of high-speed cameras viewed the front side of the sandwich and whose digital images were correlated to analyse the vibration amplitudes and normal modes; and the sagging of a 2.3~m long and 15~cm carbon fibre-Divinycell sandwich supporting 56 layers of scintillator cubes at INR in Moscow.

\begin{figure}[h]
\centering
\includegraphics[scale=0.42]{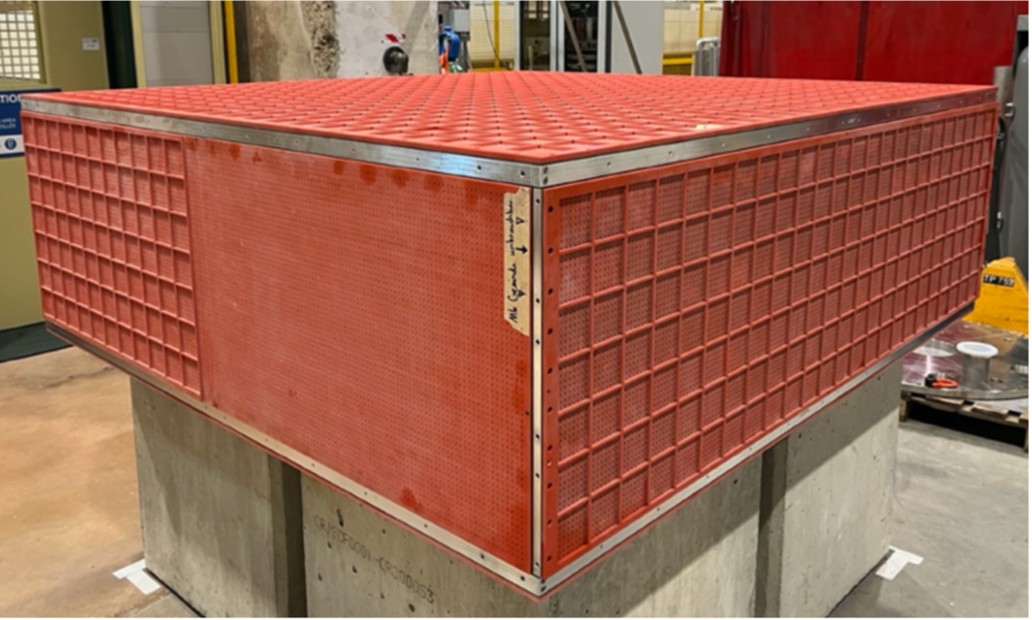}
\includegraphics[scale=0.50]{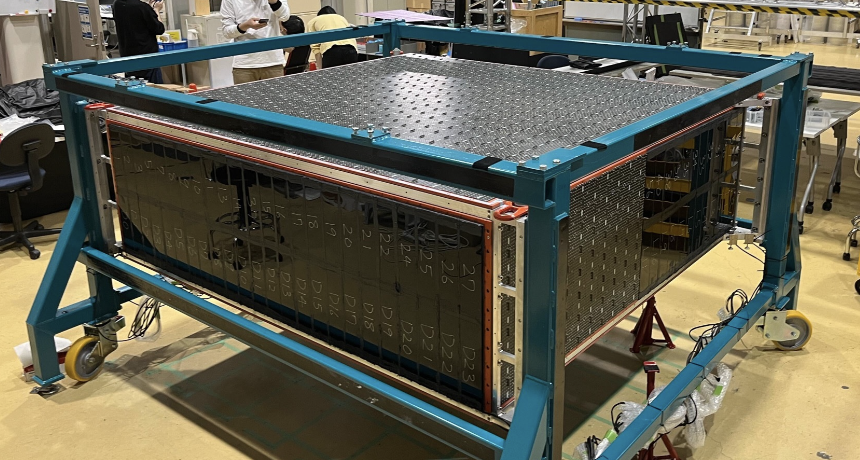}
\caption{
Top: assembled mechanical box placed on concrete blocks. Bottom: mechanical box with all the MPPC-PCBs and the LED light guide plates integrated. The box is fixed to a stainless steel frame.
}
\label{fig:box}
\end{figure}

The mounted mechanical box is shown in Fig.~\ref{fig:box}, alongside a photo showing how the SuperFGD box is fixed to a stainless steel frame analogous to that of ND280 with four corner brackets made of stainless steel AISI 316L. 

The mechanical box must also facilitate the precise integration of the MPPC-PCBs onto the detector. One important constraint is given by the coupling between the WLS fibre and the corresponding MPPC. Any horizontal misalignment would result in part of the scintillation light not reaching the MPPC active area, reducing the light yield of the detector. If such misalignment varies from channel to channel, there would be an additional source of non-uniformity that would require calibration. Given that the MPPC surface area is $1.3 \times 1.3~\text{mm}^2$ and the WLS fibre has a diameter of 1~mm, the maximum misalignment must be smaller than 0.3~mm. Moreover, the polished surface of the WLS fibre must be as close as possible to the MPPC active surface, to maximise its acceptance to scintillation photons. With a nominal numerical aperture of NA=0.72~\cite{Kuraray_Y11_Catalogue} and with a distance of 0.2 mm, a light loss up to 5\% would be expected, increasing up to more than 20\% at 0.5~mm.

Tolerances were validated with a 2.3~m long prototype and found to be consistent with the nominal design: within 0.2~mm between two fibre holes up to a distance of 2~m; within 0.05~mm between the WLS fibre hole and MPPC-PCB integration hole, as well as between two WLS fibres within a single pocket.

The manufacturing process was validated with a $1\times1~\text{m}^2$ box panel prototype, that was also exposed to thermal cycles to test the manufacturing quality and evaluate its sensitivity to thermal expansions between +5 and +35 $^{\circ}$C with LVDT and strain gauge measurements. 

The WLS fibres are fixed with EJ-500 optical cement glue to the inside of cylindrical black plastic optical connectors produced at Uniplast~\cite{Uniplast_Website}. The tolerance on the inner and outer diameters and the top lid thickness of the connectors was contained within $50$~{\textmu}m.

\begin{figure}[h]
\centering
\includegraphics[scale=0.48]{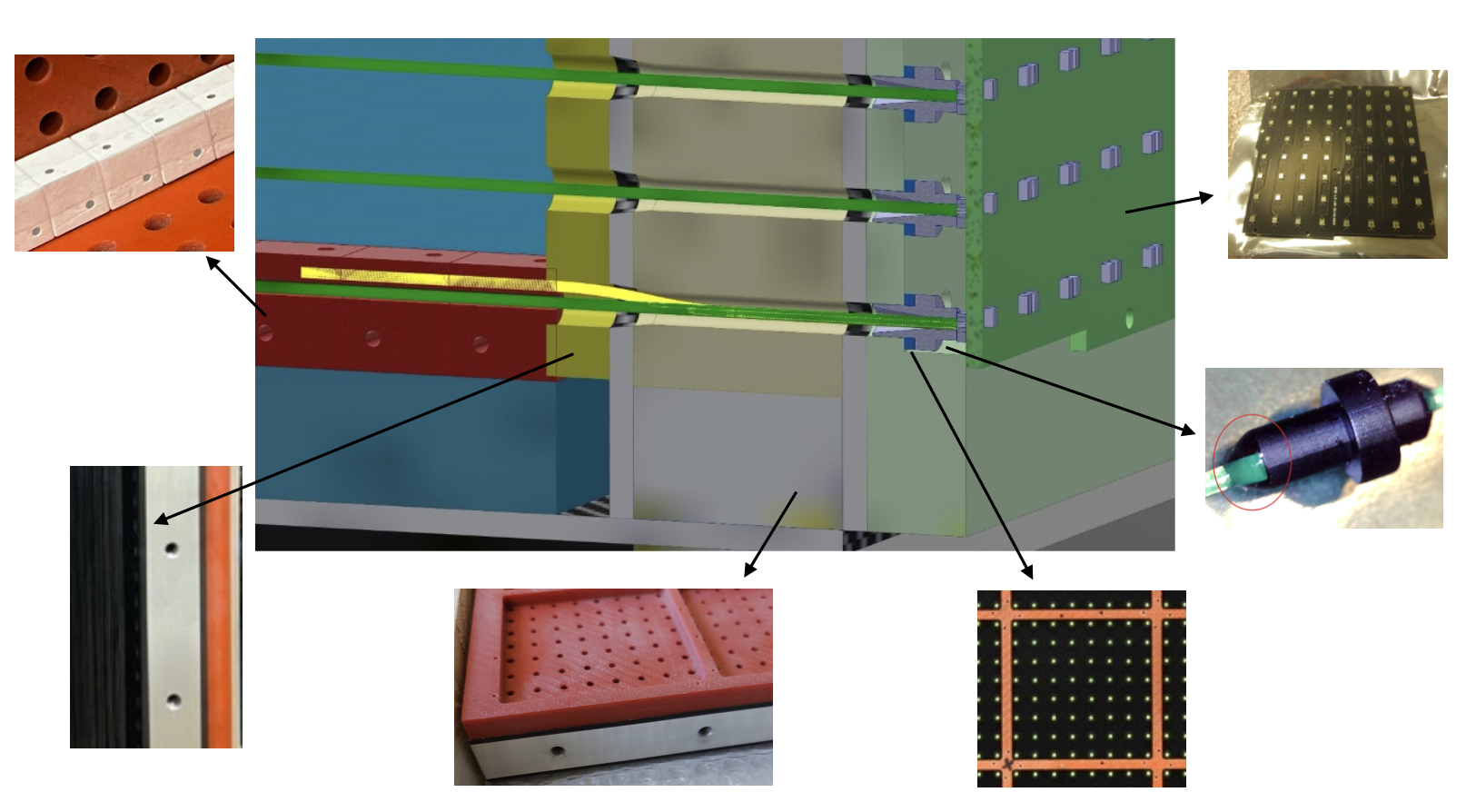}
\caption{
A 3D model of the right box plate is shown together with photographs of various parts. From left to right the following parts are shown: plastic scintillator cubes, soft EPDM foam, two carbon fibre plates sandwiching the rigid foam and the aluminium frame, the fibre glass layer where pockets are machined and drilled, WLS fibres epoxy glued to plastic optical connectors that provide a precise alignment with the MPPCs, an MPPC-PCB.
}
\label{fig:box-fiber-mppc-design}
\end{figure}

Fig.~\ref{fig:box-fiber-mppc-design} shows the design of the box and the integration of the optical connectors and the MPPC-PCBs. The MPPC-PCB is mounted on two aluminium pins and screwed to the fibre glass layer, providing precise alignment between the WLS fibre and the MPPC. A layer of soft foam is placed on the bottom side of the optical connector to push the end of the WLS fibre against the MPPC to optimise the coupling.

\begin{figure}[ht]
\centering
\includegraphics[scale=0.5]{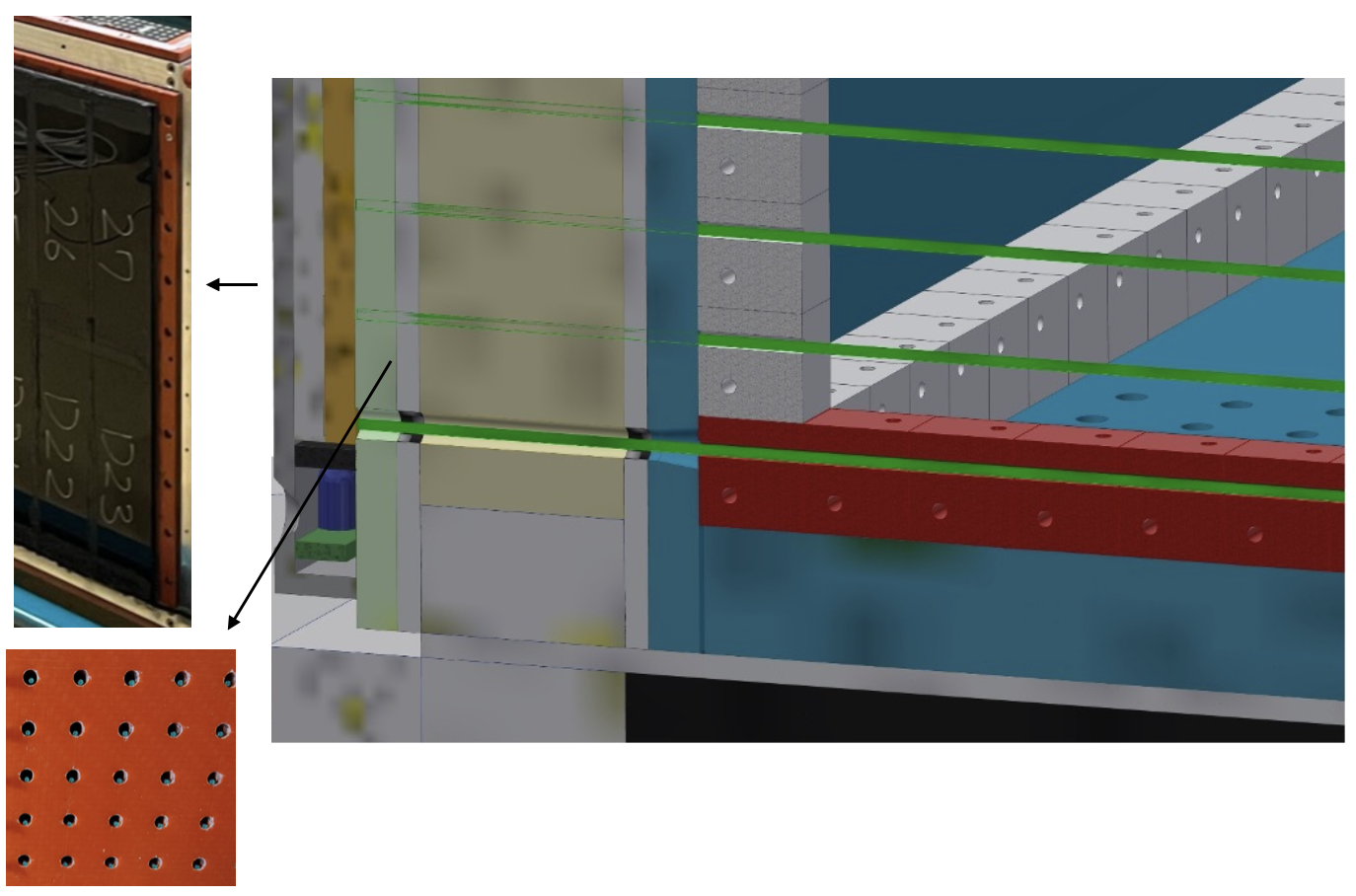}
\caption{
A 3D model of the upstream left corner of the box showing an attached LED calibration module. Photographs of the outer side of the LED calibration module and of the WLS fibres sticking out of the box holes are shown.
}
\label{fig:box-fiber-led}
\end{figure}

Fig.~\ref{fig:box-fiber-led} shows the integration of the LED light guide plates as well as their alignment with the WLS fibres (see Sec.~\ref{sec:led}).

\subsection{Light Shielding}
In order to reduce background noise coming from ambient light and LED light from readout electronics, it was necessary to design an external light shield for the detector. Since this light shield would have to be externally mounted on the detector, the primary design consideration centred on routing electronics cables through the shield without permitting excessive light to leak through. Because the SAMTEC readout cables ended in a broad, flat connector at the MPPC-PCB, and the MPPC-PCBs themselves were largely opaque to light near their centres, it was decided to mount a thin, opaque sheet on the surface of the MPPC-PCBs, with small slits cut for each MPPC-PCB to cable connection. This way, any gap available to light would only exist where secondary light shielding (i.e. the opaque MPPCs and SAMTEC cables) would be available to assist.

The light shield was constructed out of large sheets of 0.254~mm thick black polyester. These sheets then had an array of 8~mm $\times$ 29~mm slits laser cut at the location of each MPPC-PCB to SAMTEC cable connection. The size of the sheets was sufficiently large to cover the beam-left, beam-right, upstream, and downstream sides of the detector in single sheets, however the top and bottom of the detector each required two sheets due to the limited sheet stock size. The cut sheets were mounted to the SuperFGD box using a non-conductive and non-corrosive room temperature vulcanising (RTV) silicone rubber. The performance of the light shielding sheets was tested using a subset of readout electronics and torches to determine any vulnerabilities. After testing, it was concluded that the light shield performed sufficiently to proceed with the cabling of the detector.

%% file: assembly.tex
The assembly of the detector took place in J-PARC from October 2022 to April 2023. First, the mechanical box was half-assembled, then the pre-assembled cube layers were stacked and aligned in the box, and then the box was sealed. The installation of the WLS fibres, MPPC-PCBs, LED calibration system, light barrier, and the readout cables followed in this order. A specialised support system and top access structure were constructed and employed for the detector installation.

\subsection{Assembly of scintillator layers}
The installation procedure began with the assembly of the first three panels of the detector box on the support system.
The alignment of the panels and the alignment of the fibre holes was performed using four 1~cm metal cubes.
Each side of the box and the holes were cleaned with a vacuum cleaner before the installation of the scintillator cubes.

\begin{figure}[h!]
\centering
\includegraphics[width=0.46\columnwidth]{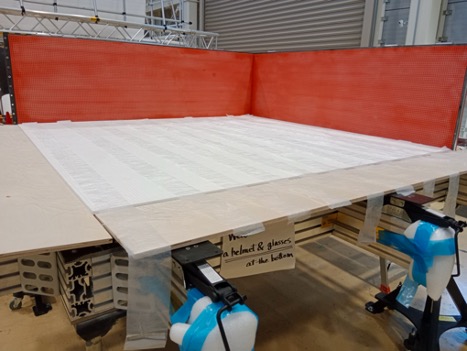}
\includegraphics[width=0.46\columnwidth]{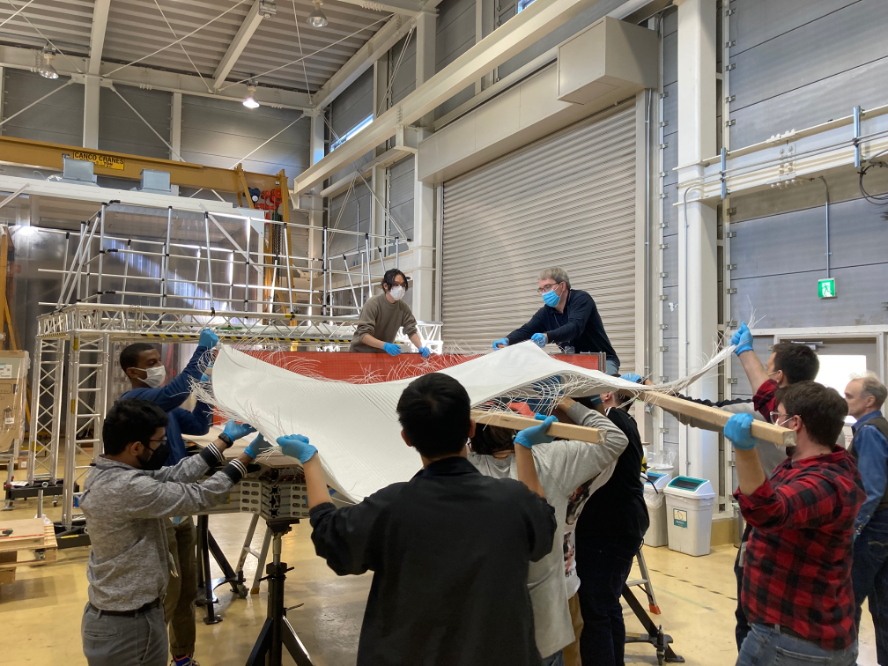}
\caption{Left: Teflon lines to separate the scintillator layer. Right: placing a layer of cubes inside the box.}
\label{fig:cubes_instalation_1}
\end{figure}

\begin{figure}[h!]
\centering
\includegraphics[width=0.46\columnwidth]{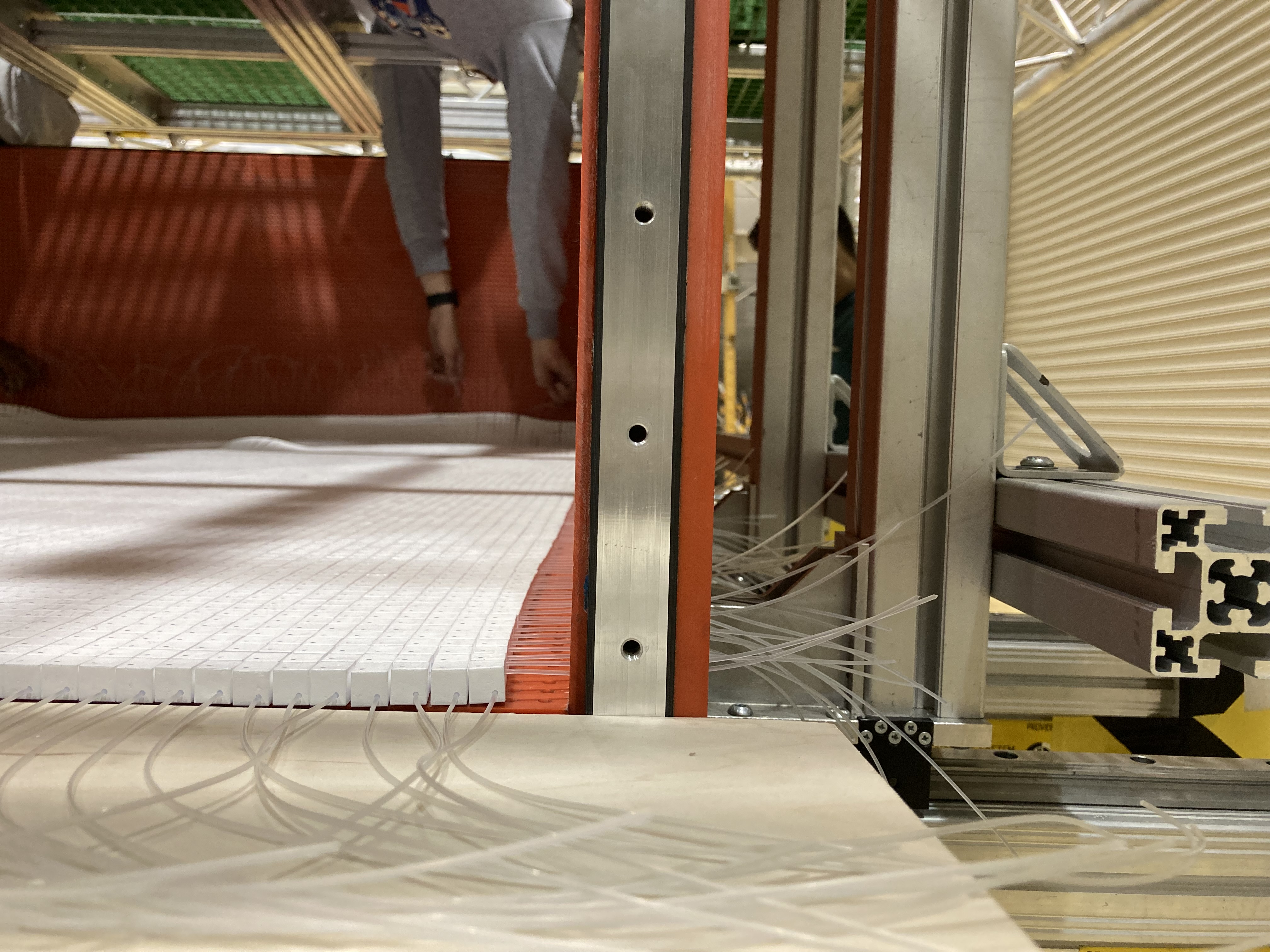}
\includegraphics[width=0.46\columnwidth]{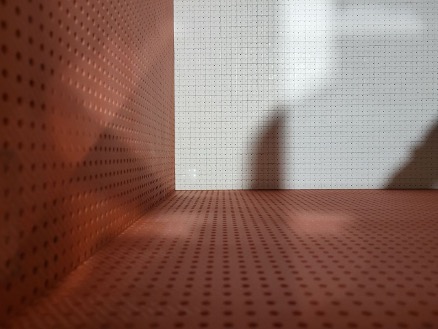}
\caption{Left: insertion of fishing lines in the holes in the mechanical box. Right: the layer of cubes after being pushed into the corner to align the cubes within the box.}
\label{fig:cubes_instalation_2}
\end{figure}

\begin{figure}
\centering
\includegraphics[width=0.46\columnwidth]{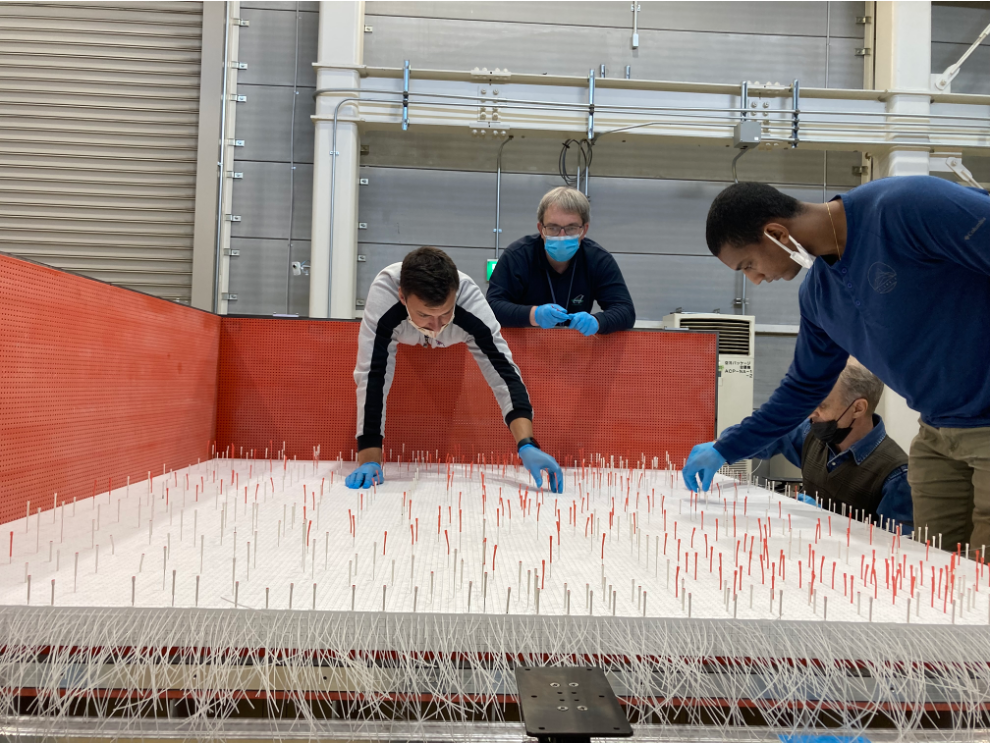}
\includegraphics[width=0.46\columnwidth]{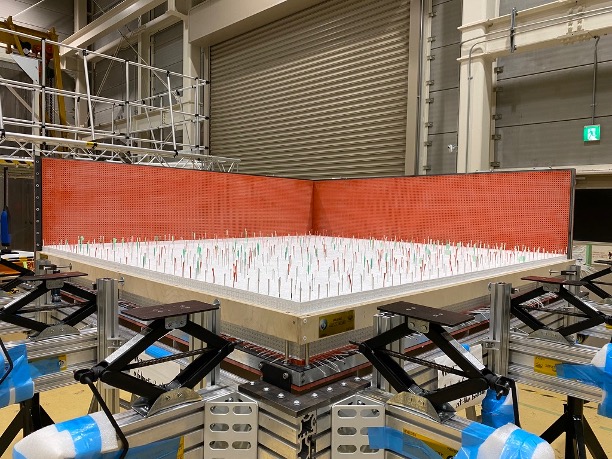}
\caption{Left: checking the alignment of the scintillator cubes using knitting needles. Right: wooden stoppers installed to maintain the alignment of the cubes throughout the installation process.}
\label{fig:cubes_instalation_3}
\end{figure}

The installation of each scintillator layer proceeded as follows: place Teflon strips on the preceding layer of cubes to facilitate mounted layer sliding, place the layer of cubes on top of the previous layer, remove the Teflon strips, insert the fishing lines of the scintillator layer through the holes in the mechanical box, push the layer of cubes towards the assembled box corner to align the cubes within the box, check the vertical alignment of cubes using 1,000 knitting needles. Wooden stoppers were installed to maintain cube alignment throughout the installation process. The procedure is shown in Fig.~\ref{fig:cubes_instalation_1}, Fig.~\ref{fig:cubes_instalation_2} and Fig.~\ref{fig:cubes_instalation_3}.

During the installation geometrical measurements were performed to ensure that the horizontal dimensions met the design specifications. Additionally, after the installation of several layers, the fishing lines of the first scintillator layer were pulled to check the feasibility of replacing the fishing lines with WLS fibres. This was repeated throughout the installation.

The wooden stoppers were removed after the installation of the final scintillator layer to allow for the installation of the second half of the box. The fishing lines from each layer were threaded through the holes of the box before the side panels were aligned and fixed with screws.

\subsection{Fibre installation}

Once assembled, the box was moved to the frame shown in Fig.~\ref{fig:box} prior to the fibre installation.
The box was lifted and held by a movable crane while the assembly table was disassembled and the frame was positioned below. The box was secured to the frame using the four corner brackets so that the box sagged in the same way as in the ND280 basket.

\begin{figure}[h!]
\centering
\includegraphics[width=0.49\columnwidth]{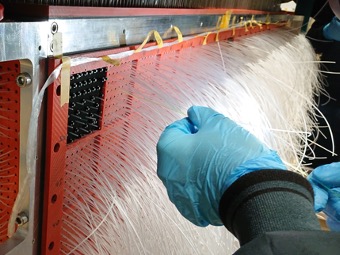}
\includegraphics[width=0.49\columnwidth]{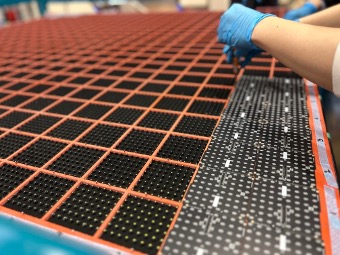}
\caption{Left: installation of fibres in the first $8\times8$ channels. Right: installation of MPPC-PCBs}
\label{fig:fiber_mppc_insertion}
\end{figure}

The fishing lines were then removed and WLS fibres were installed in the horizontal holes, as shown in Fig.~\ref{fig:fiber_mppc_insertion}. To maintain the cube alignment provided by the fishing lines, the fishing line removal and fibre insertion were repeated in units of $8\times8$ channels, corresponding to one MPPC-PCB, rather than removing all fishing lines at once.  The protruding ends of the fibres were initially cut 5~mm from the box surface, postponing the final cut until all fibres were installed out of concern that the box could deform further as the fishing lines and welding rods were removed.

The installed fibres were tested for damage in parallel to the insertion work. Several MPPC-PCBs and $8\times8$ LED arrays were used for the test. By measuring the intensity of the LED light injected from the other end of the fibres, approximately 40 potentially damaged fibres were identified and replaced. Fibres were replaced when the relative light yield was less than 70\% of the average. The LEDs were driven with DC power and the MPPCs were operated in the linear region so that the test could be done quickly without light shielding. The LED array and the MPPCs were calibrated using measurements without fibres such that channel-by-channel differences in the MPPC gain and the LED intensity could be cancelled. Many of the damaged fibres were cracked near the connector and were located at the edges of the $8\times8$ unit, where the connector was most difficult to insert. This information was fed back to the workers to prevent additional damage of fibres. Once the fibres were tested, MPPC-PCBs were attached, as shown in Fig.~\ref{fig:fiber_mppc_insertion}.

The vertical fibres were installed following a similar procedure. However while the horizontal holes had all been aligned by fishing lines, only around one third of the vertical holes were aligned by welding rods. Some holes were not aligned well enough for a fibre to pass through. In these cases alignment was recovered using a welding rod that was roughened at the tip, which in some cases was used to enlarge the hole where the alignment was particularly difficult. In case of damage to the previously installed horizontal fibres by this operation, one third of the horizontal fibres were retested after the vertical fibres were installed.
No additional damage was found.

\subsection{LGP modules, light shield, and readout cables}

\begin{figure}[h!]
\centering
\includegraphics[width=0.48\columnwidth]{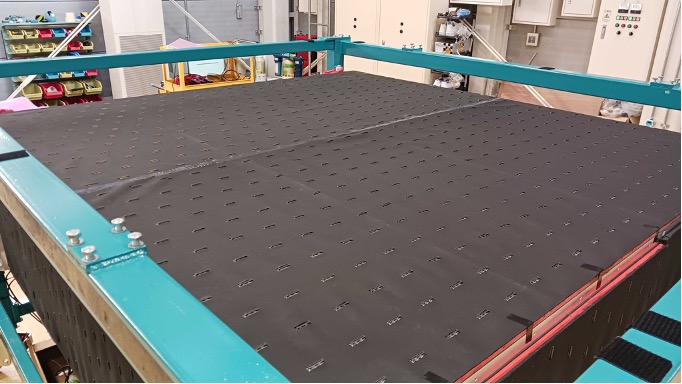}
\includegraphics[width=0.48\columnwidth]{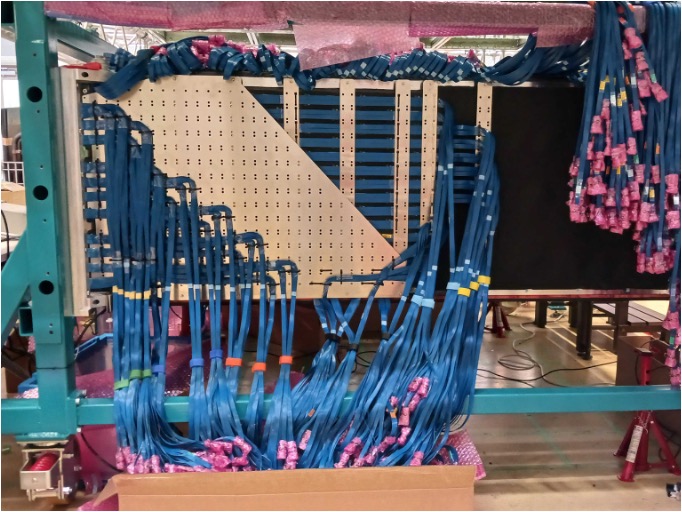}
\caption{Left: light shielding sheets attached on the box. Right: readout cables were attached to the MPPC-PCBs and secured on aluminium supports on the box.}
\label{fig:light_shield}
\end{figure}

After the installation of all fibres and MPPC-PCBs, the protruding fibres were cut parallel to the surface of the box. The LGP modules were then installed. The gaps between the modules and the box surface were filled with RTV silicon rubber for light tightness. The light shielding sheets were mounted using this rubber and then the readout cables were attached to the MPPC-PCBs through specifically prepared holes in the sheets, see Fig.~\ref{fig:light_shield}.

Finally, measurements for all MPPC-PCBs were performed using 32-channel readout boards (CAEN DT5702). LED light from the LGP modules was measured to verify the functionality of the LGP modules, the fibres, the MPPC-PCBs, and the readout cables. As a result of this testing, two MPPC-PCBs were replaced due to a defective channel and a damaged connector pin, respectively. The noise rate while illuminating the detector with LED torches was also measured to identify light leaks. Where such leaks were found, the light shielding was reinforced using RTV rubber.

%% file: photosensors.tex
\subsection{Photosensors}

\begin{figure}[tbp]
\centering
\includegraphics[width=0.42\textwidth]{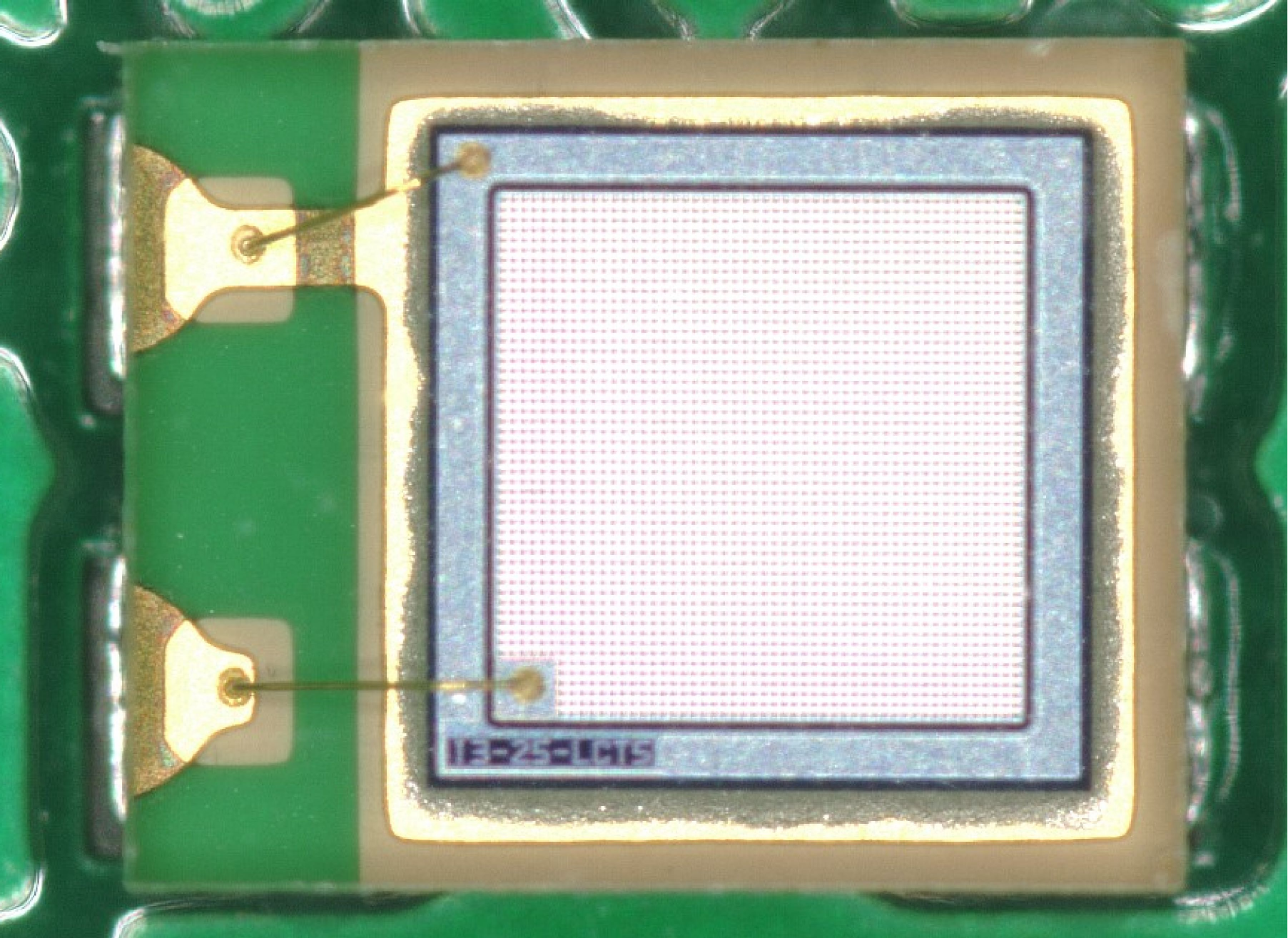}
\includegraphics[width=0.47\textwidth]{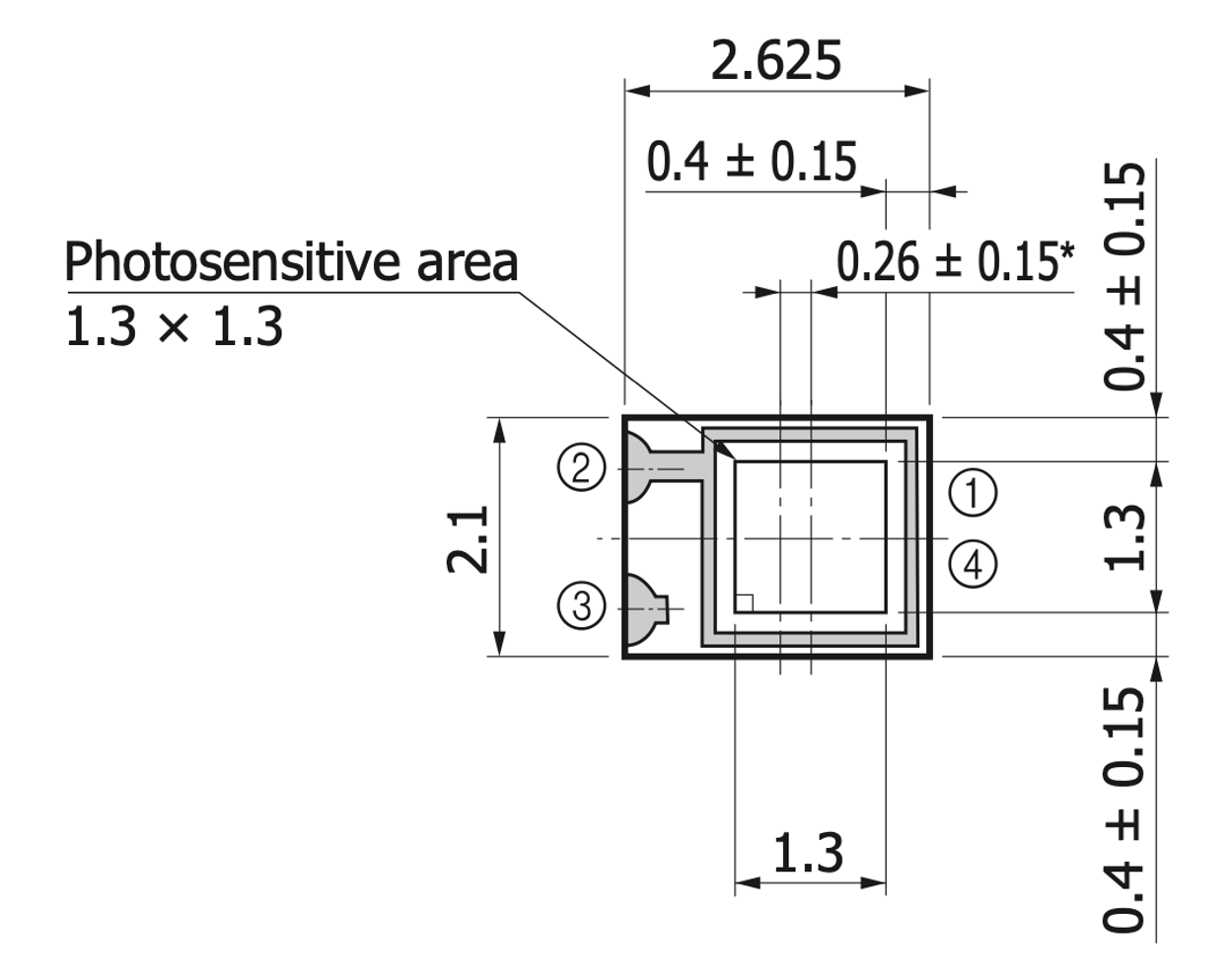}
\caption{Left: picture of the MPPC S13360-1325PE. Right: dimensional specifications of S13360-1325PE from the MPPC catalogue of Hamamatsu Photonics.}
\label{fig:Target-mppc_picture}
\end{figure}

The photosensor is the key device to detect the scintillation light. We adopted Multi-Pixel Photon Counters (MPPCs) produced by Hamamatsu Photonics K.K. The MPPCs have been successfully used in all plastic scintillator detectors of the current near detectors of T2K since 2009~\cite{Abe:2011ks,Yokoyama:2008hn,Vacheret:2011zza}. The MPPC type chosen for the SuperFGD is S13360-1325PE \cite{MPPC_HPK_Datasheet}, which is shown in Fig.~\ref{fig:Target-mppc_picture}.

The specifications of the S13360-1325PE are summarised in Table~\ref{tab:Target-mppc_spec}. Its sensitive area is 1.3~mm $\times$ 1.3~mm, which is designed to match the 1~mm diameter of the WLS fibre. The slightly larger sensitive area ensures a geometrical collection factor for all photons exiting the fibre and sets achievable tolerance requirements on component placements.
  
The pixel pitch is 25~{\textmu}m, resulting in a total of 2668 pixels, in order to attain a larger dynamic range comparing with other types of MPPC used in ND280. The surface mount package was chosen to minimize space requirements and cost. Developments over the 15 years since the construction of the original ND280 mean that S13360-1325PE has about an order of magnitude lower dark noise rate, cross-talk probability, and after-pulse probability compared to S10362-13-050C which were used at that time.

\begin{table}[htbp]
  \centering
  \begin{tabular}{ll} \hline \hline
    Item & Specification \\ \hline
    Effective photosensitive area & 1.3~mm $\times$ 1.3~mm \\
    Pixel pitch & 25~{\textmu}m \\
    Number of pixels & 2668~pixels \\
    Fill factor & 47\% \\
    Package type & Surface mount \\
    Breakdown voltage (V$_{\rm BR}$) & 53 $\pm$ 5~V \\
    Peak sensitivity wavelength & 450~nm \\
    Photon detection efficiency &  25\% \\
    Gain &  $7.0 \times 10^{5}$ \\
    Dark count & 70~kcps (typ.) \\
    Crosstalk probability & 1\% \\ \hline \hline
  \end{tabular}
  \caption{Specifications of the S13360-1325PE MPPC \cite{MPPC_HPK_Datasheet}. The characteristics are measured at (V$_{\rm BR}$+5)~V and 25 $^{\circ}$C.}
  \label{tab:Target-mppc_spec}
\end{table}

\begin{figure}[tbp]
\centering
\includegraphics[width=0.42\textwidth]{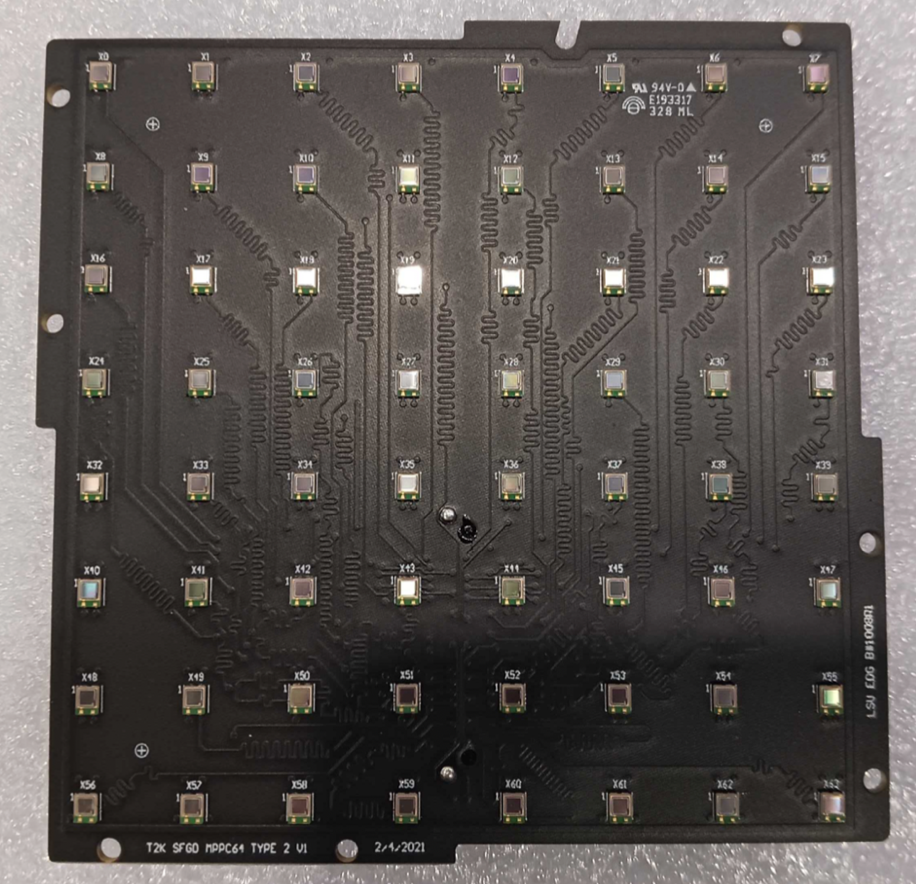}
\hspace*{0.08\textwidth}
\includegraphics[width=0.42\textwidth]{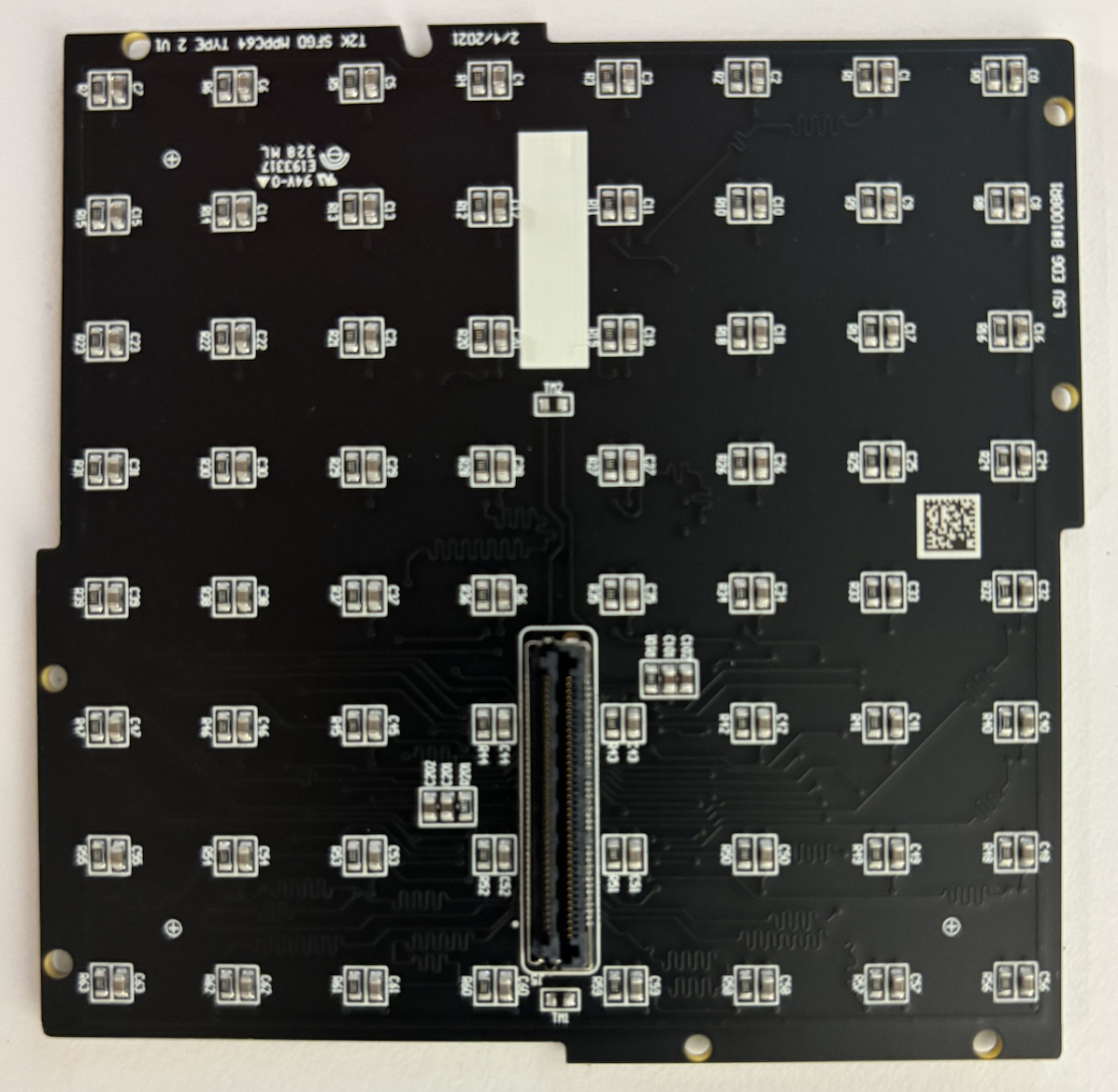}
\caption{Left: picture of MPPC-PCB photosensor side. Right: picture of MPPC-PCB connector side.}
\label{fig:MPPC64-PCB_picture}
\end{figure}

MPPCs with a similar operation voltage ($<\sim$0.15V) were grouped in quantities of 768 and packaged in standard reels to enable automatic mounting onto printed circuit boards (PCBs). Each PCB accommodates 64 MPPCs, which are arranged in a grid of $8 \times 8$ with spacing of 10.3~mm between MPPCs in both directions. The boards measure $83.8 \times 83.8$ mm$^2$ and the outer edges are shaped to allow for close tiling of boards. Photos of MPPC-PCBs are shown in Fig.~\ref{fig:MPPC64-PCB_picture}.

The PCB is a 1.7~mm thick four-layer board and has two separate bias voltage inputs which get distributed to two groups of 32 channels. Within a group of 32 MPPCs, the bias voltage for individual MPPCs can be modulated by $\pm 1.25$ V through the front end readout board. Each channel is equipped with an RC filter and gets read out through an 80-pin SAMTEC connector LSHM-140-02.5-L-DV-A-S from which a ribbon cable brings signals to the front end readout board. The PCB signal traces within a group of 32 channels are matched in length to achieve a spread in signal arrival times better than 24~ps and the difference in the mean arrival times between signals from the two groups of 32 channels amounts to 186~ns. Two versions of the PCB have been designed and built with the main difference in the connector location to optimise routing of the cables. The PCBs are rotationally symmetric with respect to the 64 photosensitive MPPC areas which provides an additional degree of freedom to effectively manage the cable routing. MPPC-PCBs are equipped with two temperature sensors and during the production process all vias are backfilled with black epoxy to prevent background light from reaching the MPPC photosensitive areas.

The SuperFGD requires a total of 881 MPPC-PCBs, with an almost equal split between the two types of PCBs. Quality checks and detailed characterisations of all MPPCs were performed after they were mounted onto PCBs along with checks of the PCBs themselves. The quality assurance and control procedures included checks on the mechanical dimensions of the PCBs, their alignment and mounting features, and the position accuracy of the mounting of the MPPCs on the board. Standard deviations of the MPPCs mounting positions relative to their nominal position were found to be 30~{\textmu}m which is consistent with expected uncertainties due to the reflow soldering process and within the acceptance requirements. All PCBs and MPPC front faces were submitted to rigorous visual inspection, including a detailed inspection with a microscope for a representative number of boards. MPPC-PCBs which exhibited significant surface scratches or damages on the MPPC photosensitive areas were rejected. The functional performance tests and characterisation of all MPPCs consisted of measurements in a temperature controlled environment at 20 $^\circ$C to measure the breakdown voltage, the gain at various overvoltages, the dark rate at threshold values of 0.5 and 1.5 photoelectrons (p.e.) and cross-talk behaviour. For a fraction of MPPCs additional parameters such as the relative photon detection efficiency (PDE) and detailed dark rate versus threshold curves were measured. Table ~\ref{tab:mppc_requirements} shows the acceptance requirements and the measured MPPC performance for a total of around 900 MPPC-PCBs tested. The measured PDE was used as a reference value for comparison with the observed light yield of each channel after installation.

\begin{table}[htbp]
  \centering
  \begin{tabular}{lll} \hline \hline
    Item & Requirements & Measured \\ \hline
    Gain variation &  $\le$ 12 \% & $\le$5\% (1 sigma) \\
    V$_{BR}$  (per PCB) & within $\pm$ 1.25~V  & within $\pm$ 0.6~V (1 sigma)\\   
    Dark count ($>$ 0.5 p.e.) & $\le$200~kcps & $\le$ (50 $\pm$20) kcps\\
    Crosstalk probability & $\le$3\% & $\le$ 1.5 $\pm$0.3 \% \\ \hline \hline
  \end{tabular}
  \caption{Acceptance requirements and measured parameters for MPPCs for use in the SuperFGD. The characteristics are for (V$_{\rm BR}$+5)~V and 20 $^\circ$C.}
  \label{tab:mppc_requirements}
\end{table}

\subsection{Wavelength Shifting Fibres}

Wavelength shifting (WLS) fibres are commonly used to collect light from large area scintillators. The SuperFGD uses the same fibre type as the original ND280 scintillator detectors, Y-11~(200) produced by Kuraray Co., Ltd \cite{Kuraray_Y11_Catalogue}. The performance and quality of this fibre are very well established by many experiments. The main specifications are summarised in Table~\ref{tab:Target-fiber}.

\begin{table}[ht]
  \centering
  \begin{tabular}{ll} \hline \hline
    Item & Specification \\ \hline
    Fibre type & Round shape, Multi-cladding, S-type \\
    Diameter & 1.0~mm\\
    Materials & Core: Polystyrene (PS),  \\
                     & Middle clad: Polymethylmethacrylate (PMMA), \\
                     & Outer clad: Fluorinated polymer (FP) \\
    Refractive index & Core: 1.59, Middle clad: 1.49, Outer clad: 1.42\\
    Density & Core: 1.05~g/cm$^2$, Middle clad: 1.19~g/cm$^2$, \\
    & Outer clad: 1.43~g/cm$^2$ \\
    Absorption wavelength & 430~nm (peak) \\
    Emission wavelength & 476~nm (peak) \\
    Trapping efficiency & $\sim$5.4\% \\
    Attenuation length & $>$3.5~m \\ \hline \hline
  \end{tabular}
  \caption{Main parameters of the WLS fibre Y-11(200) according to Kuraray specification~\cite{Kuraray_Y11_Catalogue}.}
  \label{tab:Target-fiber}
\end{table}

It is a multi-cladding, round shape type fibre with 1.0~mm diameter. The fibre is flexible S-type with molecular orientation that makes it mechanically stronger against clacking than standard (non-S) type. The absorption spectrum, which peaks at 430~nm, is well matched with the wavelength of the blue light emitted by the scintillator cubes. The emission spectrum depends on the fibre length due to attenuation and peaks at $\sim$505~nm for a 1~m long fibre.

We investigated the performance of 60~cm long Y-11 fibres with a different treatment of the open fibre end: fibre broken by hand, cut by a wire cutter, ground by sandpaper shows almost the same reflection at the end, with marginally increase for a polished end. Reflector at the broken or cut end produces no noticeable effect on the light yield. Reflective paint on a flat, polished end increases the light signal by 25\% at the point close to this end.

However it was impractical to prepare the fibre ends in advance of fibre installation because we could not provide the precise position of the open fibre end at the level of the box plane. Misalignment of the open fibre end and reflector at the end also deteriorates the capture of light from the LED calibration system (see Sec.~\ref{sec:led} for more details), so the decision was made to insert the WLS fibres and cut off the protruding part of the fibre with end cutting pliers at the level of the box wall. The opposite fibre end is glued into a plastic ferrule which is fixed tightly inside a box wall providing an optical interface with the corresponding MPPC.

%% file: ledsystem.tex
\begin{figure}[h!]
\centering
\includegraphics[scale=0.35]{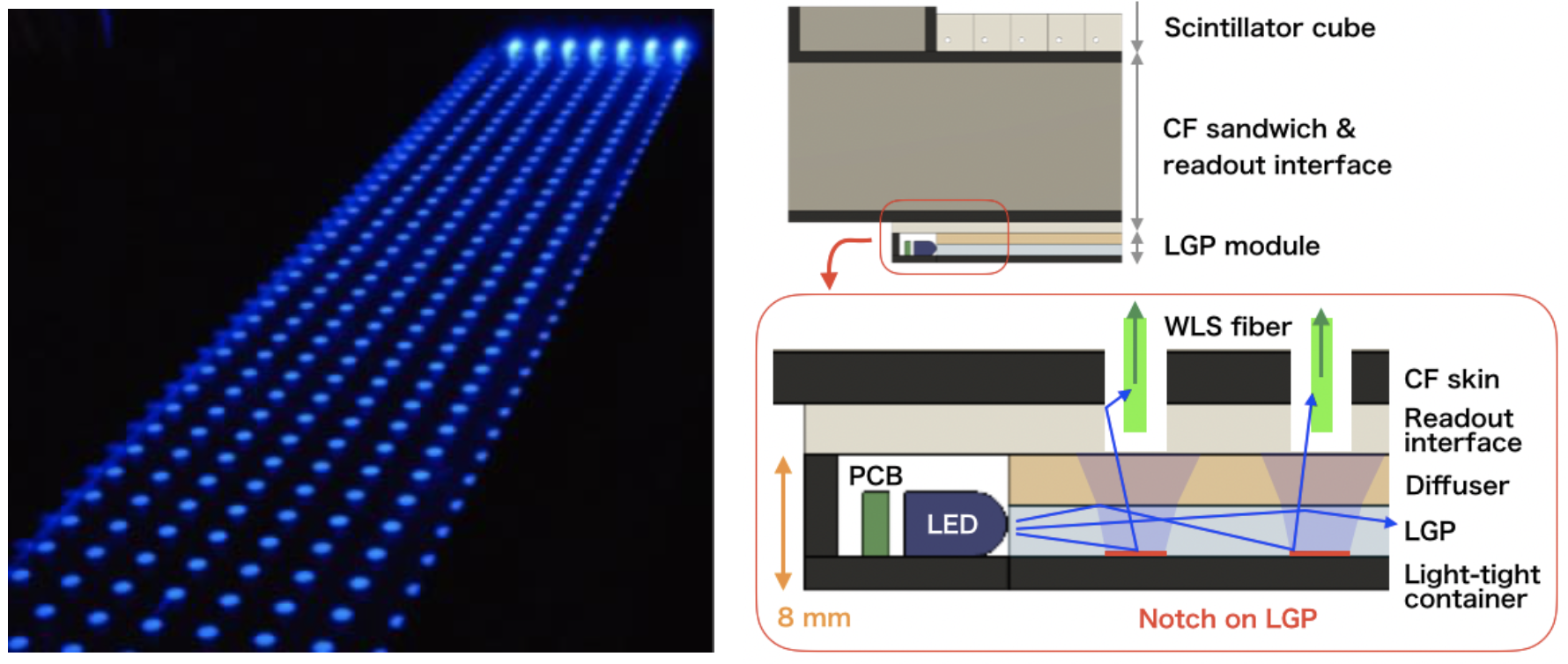}
\caption{A photo of the LGP and conceptual drawing to inject LED light for a large number of MPPCs through fibres simultaneously.}
\label{ledsystem:concept}
\end{figure}

The LED calibration system is useful for detector commissioning, regular MPPC calibration, and long-term stability monitoring. The system is designed to provide appropriate photoelectrons sufficiently uniformly across many channels while being compact due to space constraints. A Light Guide Plate (LGP) was adopted as the optical distributor, an example of which is shown in Fig.~\ref{ledsystem:concept}. The system is composed of the notched LGP, a diffuse plate, a black acrylic container, an LED-array PCB with collimator, and an LED driver. Blue LED light is pulsed through the LGP, scattered by 3~mm diameter notches aligned with fibre ends, and captured by WLS fibres, which then provide re-emitted light to MPPCs. LGP module specifications are summarised in Table~\ref{ledsystem:specification}.

\begin{table}[htbp]
  \centering
  \begin{tabular}{ll} \hline \hline
    Item & Specification \\ \hline
    LGP & KURARAY CO., LTD, COMOGLAS(P), 3 mm thick\\
    Diffuse plate & KURARAY CO., LTD, PARAGLAS(422L), 3 mm thick\\ 
    Container & Assembled black acrylic plates, 2 mm thick\\
    LED & NICHIA CORPORATION, Blue LED (NSPB300B) \\   
    LED array PCB & Custom PCB with a resistance and U.FL connector\\
    LED collimator & Glued acrylic plates with a custom shape\\
    Screw & Hirosugi-Keiki Co.,Ltd., RENY screw (RYF-0312B) \\
    Cables & Coaxial cable (FWS5030) assembled with U.FL plugs \\
    LED driver & Custom device developed for SuperFGD\\\hline \hline
  \end{tabular}
  \caption{Main specifications of the LED system.}
  \label{ledsystem:specification}
\end{table}

\begin{figure}[h!]
\centering
\includegraphics[scale=0.32]{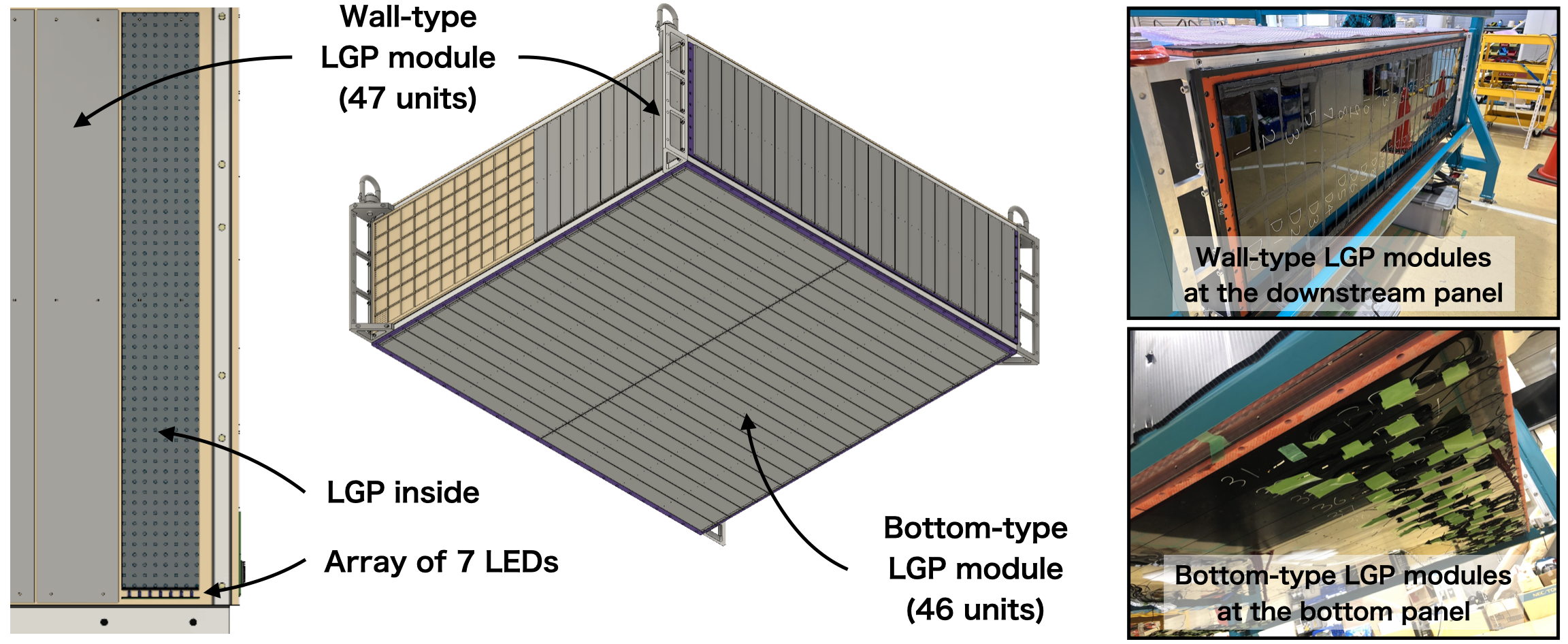}
\caption{Mechanical design of the LGP module on the detector surface (left), wall and bottom module assemblies to the detector (middle), and photos of two types of the modules after the detector assembly (right).}
\label{ledsystem:mechanics}
\end{figure}

As shown in Figure~\ref{ledsystem:mechanics}, two sizes of LGPs were designed to cover all channels: one with $56\times8$ notches for the downstream and side walls (47 modules, $586.4\times81.1\times8.0~$mm each), and another with 96$\times$8 notches for the bottom (46 modules, $999.5 \times 81.1 \times 8.0~$mm each). This modular design improved production, quality control, and assembly. Notch positions and depths were laser-processed precisely. Seven bullet-shaped LEDs with collimators, mounted on a PCB, are aligned between eight notch rows per LGP for optimal uniformity. We optimised the LGP thickness, diameter of the notches and diffuse plate thickness for better light uniformity. The outer black acrylic container prevents light leakage, resulting in an 8~mm module thickness in total. The two module types have 6 and 10 through-holes, respectively, for countersunk screws to limit sag to under 500~{\textmu}m. To minimise the non-uniformity induced by a LGP-fibre gap, WLS fibres were cut flush with the box surface after insertion.

\begin{figure}[h!]
\centering
\includegraphics[scale=0.32]{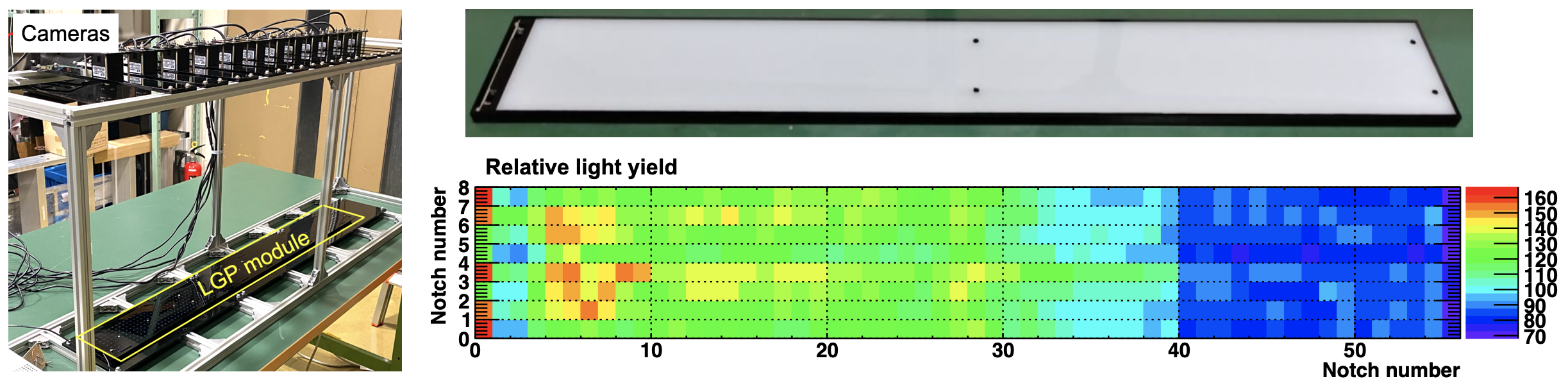}
\caption{An evaluation system for the LGP module (left) and a test result of relative light yield on a wall-type module (right).}
\label{ledsystem:evaluation}
\end{figure}

Quality control involved mechanical and optical inspections. Plates were visually inspected for scratches and stains. Dimensions were carefully checked using an inspection jig to prevent non-uniformity induced by fibre misalignment. An evaluation system with CMOS cameras was developed to rapidly check light intensity variations as shown in Figure~\ref{ledsystem:evaluation}. Twelve cameras captured plane images of all notches for an LGP module with DC-powered LEDs. Light intensities for all notches were obtained by analysing images taken in a dark box. Light yield variation within a module, estimated as the ratio of the standard deviation to the mean, was approximately 22\% (13\%) for wall-type (bottom-type) modules.

\begin{figure}[h!]
\centering
\includegraphics[scale=0.33]{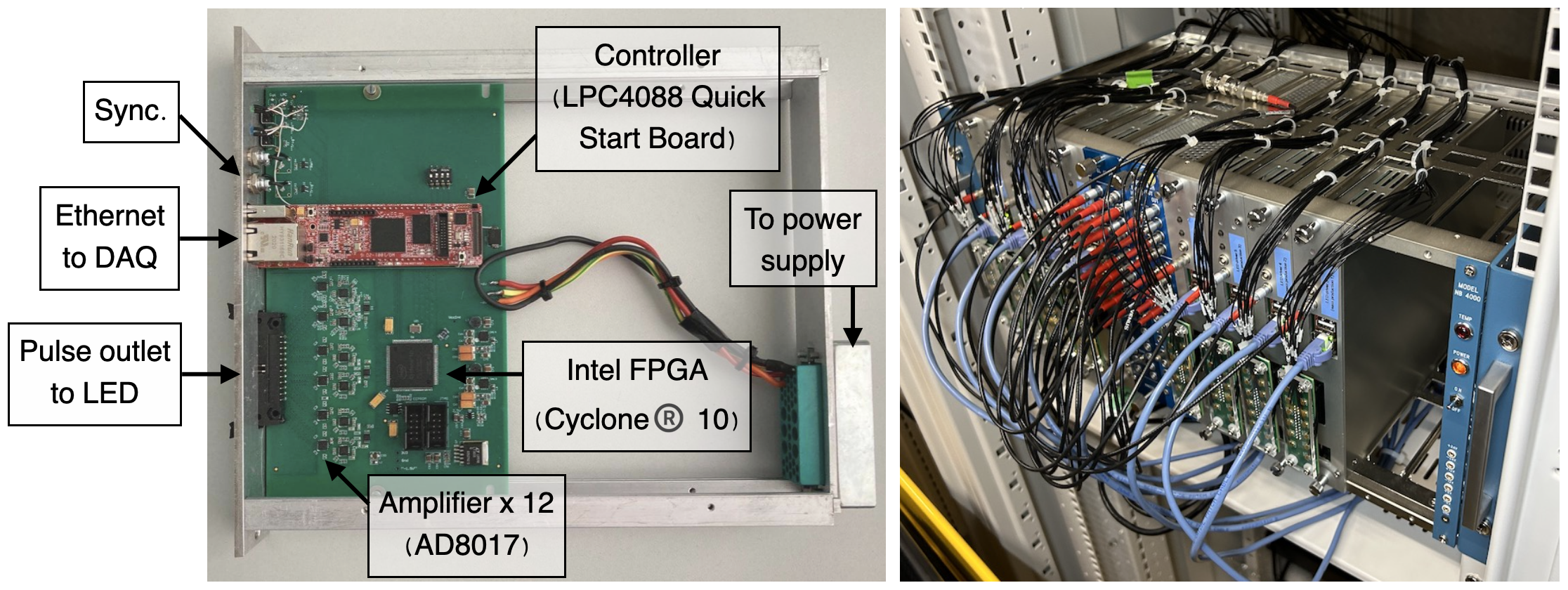}
\caption{Photos of the LED driver module (left) and the eight modules in the NIM crate (right).}
\label{ledsystem:driver}
\end{figure}

The custom LED driver shown in Figure~\ref{ledsystem:driver} generates pulses for the LED. It consists of analogue circuitry, an FPGA, and a microcomputer, powered by a NIM crate. The FPGA-generated digital signal is amplified, outputting pulses up to $\sim$7.0~V. Pulse height and width are adjustable per channel and the standard setting is 6.5~V with a 30~ns width. Eight 12-channel modules cover all 93 LGP modules. The crate is located one floor below the detector and a total of 30~m of coaxial cable bundles with 1.13~mm outer diameter and 50~$\Omega$ impedance connect the drivers and the LGP modules via an intermediate board on a patch panel.

%% file: electronics.tex
\renewcommand{\degree}{${}^\circ$}

The digitisation of the MPPC analogue signals is ensured by dedicated readout electronics boards, organised in 16 crates, eight of which are located on each side of the detector.

\subsection{Architecture of the electronics}

\begin{figure}[h]
    \centering
    \includegraphics[width=1\linewidth]{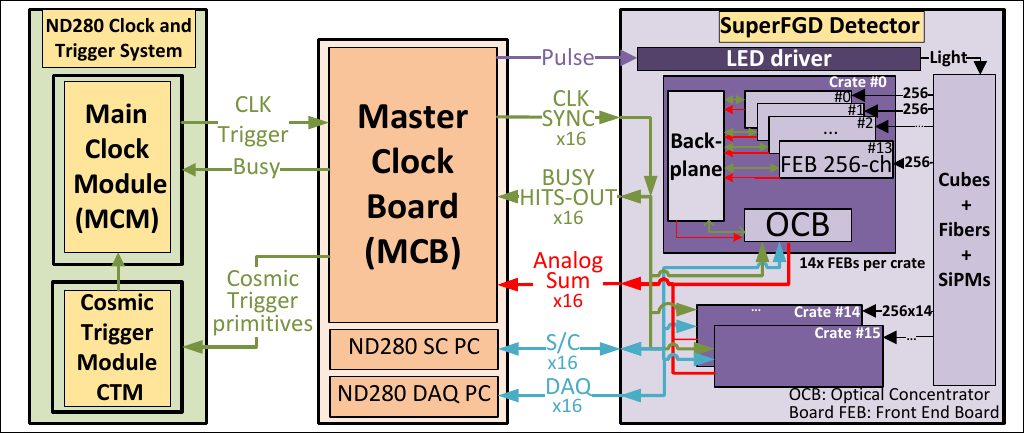}
    \caption{Overview of the DAQ system for the SuperFGD detector.}
\label{fig:sFGD_FEB_DAQ_architecture}
\end{figure}

Each of the 16 Schroff standard crates comprises 14 Front-End Boards (FEBs) and one Optical Concentrator Board (OCB). Four MPPC-PCBs are connected to the front panel of each FEB by four cables and one MPPC Interface Board (MIB) as described in Sec.~\ref{sub:cables}, thus allowing the digitisation of 256 channels per FEB. In each crate, the 14 FEBs and the OCB are connected to a dedicated backplane, which allows the distribution of the low voltage power and MPPC bias voltage, as well as digital FEB-to-OCB links. In each crate, the OCB receives the clock and control signals and sends back the digital data to the Data Acquisition (DAQ) through Ethernet cables. The FEB receives the clock, slow control and synchronisation signals, and the MPPC bias voltage from the OCB, while the main power (12~V) is received directly from the backplane, from a low voltage power supply module located on the lower floor. The FEB also sends back slow control and DAQ data to the OCB through the backplane connections. Each OCB is connected through a CAT-7 Ethernet cable to an Ethernet switch, which is in turn linked to two PCs: one for DAQ and one for Global Slow Control (GSC). The OCB acts as a DAQ data concentrator for all FEBs within the crate and also for the SC commands dispatcher used to configure the FEB. An Ethernet switch is used to access the DAQ PC, from a PC located in the control room. The DAQ PC handles the file storage of the data. The DAQ synchronisation system is composed of 1 Master Clock Board (MCB) driving the FEBs through OCBs via the backplane, and the MCB to OCB connections use CAT-7 Ethernet cables. This general architecture can be seen on Fig.~\ref{fig:sFGD_FEB_DAQ_architecture}.
 
\subsection{Front-end electronics boards (FEBs)} \label{sec:FEB}

\begin{figure}[h]
\centering
\includegraphics[width=0.9\linewidth]{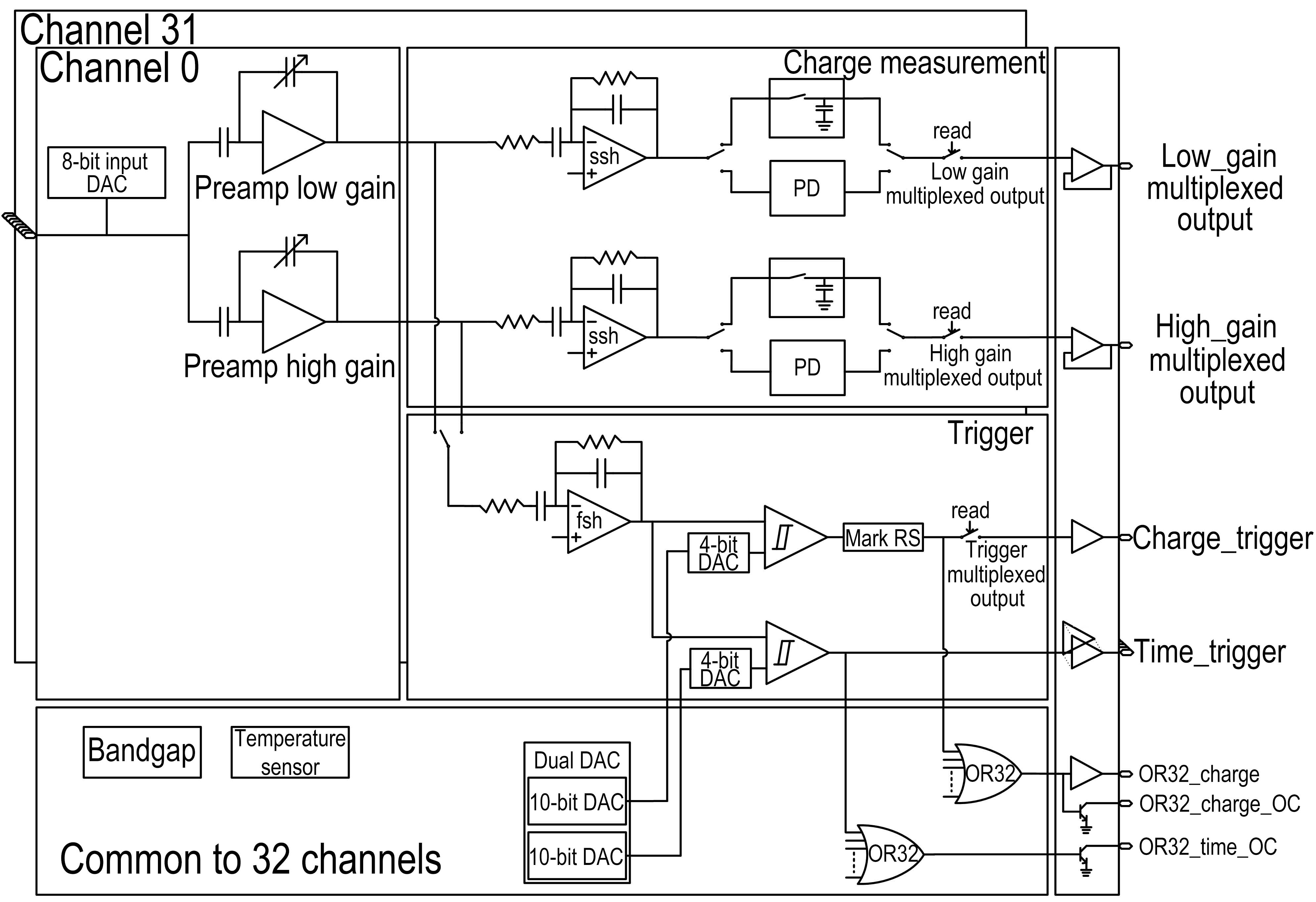}
\caption{Schematic of the CITIROC circuit for amplification, shaping, and triggering~\protect{\cite{Weeroc2019}}.}
\label{fig:citiroc_circuit}
\end{figure}

The baseline design of the FEB is structured around eight CITIROCs (Cherenkov Imaging Telescope Integrated Read Out Chip~\cite{Weeroc2019}) previously used by T2K for the BabyMIND detector \cite{Antonova_2017}. A CITIROC is a 32-channel front-end chip developed by Omega laboratory at Ecole Polytechnique and designed for the readout of a large number of SiPM devices. Each channels charge readout is achieved through two independent programmable preamplification and shaping circuits. The two circuits differ for the input capacitance, being 1.5~pF in low gain (LG) mode, and 15~pF in high gain (HG) mode. The feedback capacitance is user controlled with six-bit precision over a range of 1575~pF (25~pF steps). This provides great freedom in the choice of the amplification and consequently a wide dynamic range. Moreover, the shapers (\textit{slow shaper}) shaping time is programmable with three-bit precision, in steps of 12.5~ns between 12.5~ns and 87.5~ns. Signal sampling can also be selected by the user, choosing between a fixed delay Track \& Hold or a Peak Detector system. The latter was chosen as the default setting and all studies presented here are performed with this setting. All the aforementioned parameters are independently programmed for each CITIROC channel. Analogue low gain and high gain signals are then digitised by ADC chips external to the CITIROC. In parallel, time readout is achieved by latching one of the preamplification outputs (HG or LG, user selected) to a fast shaper circuit with a fixed 15~ns shaping time. A constant threshold discriminator is responsible for generation of the rising edge and falling edge time stamps. The discriminator is controlled by a 10-bit precision DAC common to all the CITIROC channels, while each channels threshold can be fine tuned with an additional four-bit DAC. An analogous circuit provides the CITIROCs trigger signal. Figure~\ref{fig:citiroc_circuit} shows the schematic circuit inside the CITIROC.

\begin{figure}[h]
\centering
\includegraphics[width=0.8\linewidth]{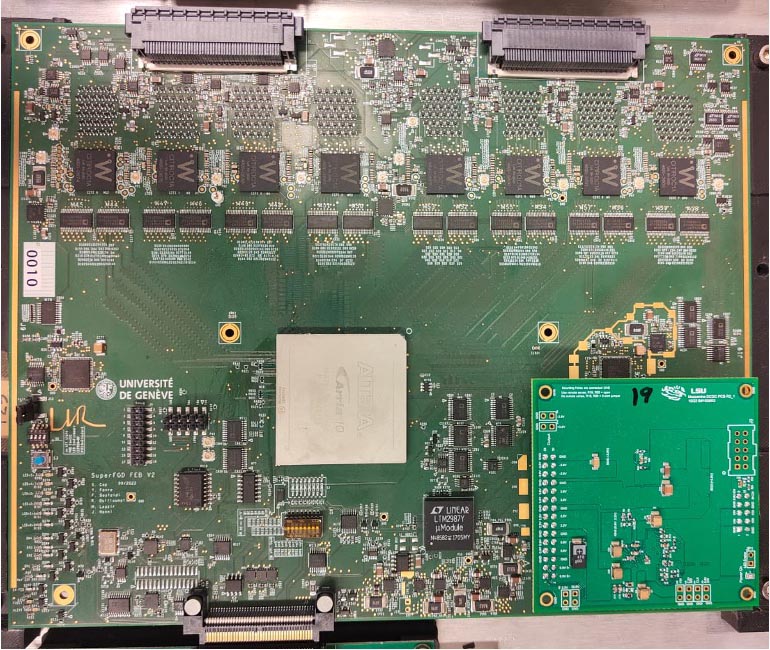}
\caption{The final version of the FEB board with its low voltage mezzanine. One clearly sees the separation between the analogue domain containing the analogue connectors and the CITIROC chips (at the top), and the digital area, containing the FPGA (center and bottom).}
\label{fig:FEBseule}
\end{figure}

\begin{figure}[h]
\centering
\includegraphics[width=0.9\linewidth]{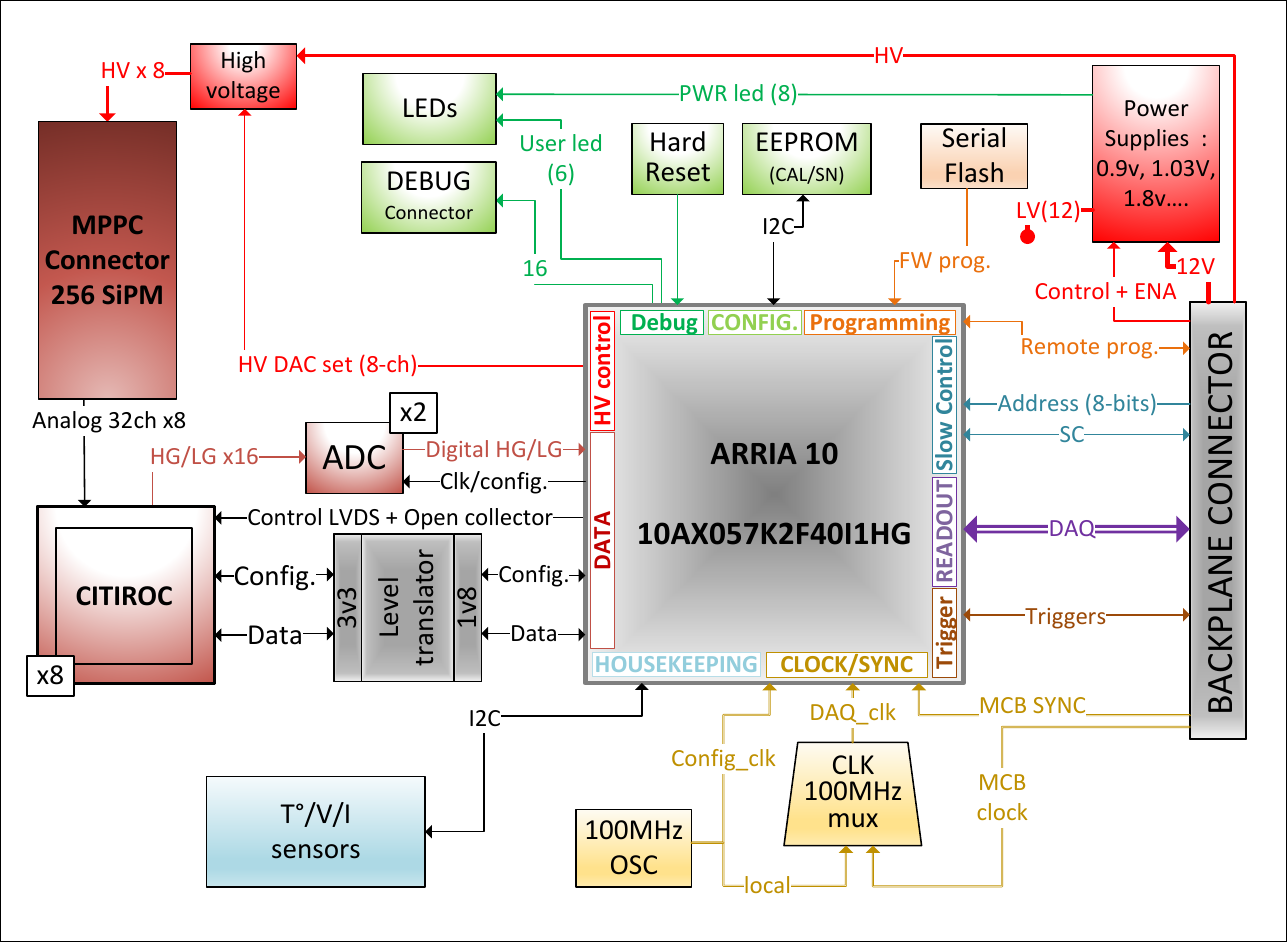}
\caption{Detailed architecture of the FEB V2.}
\label{fig:FEB_V2_architecture}
\end{figure}

Figure~\ref{fig:FEBseule} shows a photo of the final version of the FEB board. Figure~\ref{fig:FEB_V2_architecture} shows the architecture of the FEB, which includes the following components:
\begin{itemize}
    \item MPPC connector: 256 channel connections, through the MIB board, from four MPPC-PCBs. This is located on the front side of the board to ensure a clear separation between the analogue/SiPM signals domain and digital domain.
    \item ASIC: Eight CITIROCs and their 3.3~V to 1.8~V level translators to connect to the FPGA (I/O level = 1.8~V).
    \item ADC: Two 12-bit 8-channel ADCs and their signal conditioning stage for sampling the analogue HG and LG outputs from the eight ASICs. The conditioning stage per HG or LG path includes a single ended to differential level conversion and a baseline subtraction stage in order to optimise the dynamic range of the single ended CITIROC output
    to the ADC full differential input. The 1.25~V baseline of the CITIROC is not accurate 
    so the subtraction stage is tunable and based on a 12-bit DAC allowing a $\sim\pm$200~mV adjustable level set by the FPGA.
    \item FPGA: For digital data processing and interface management.
    \item Clock: A 100~MHz local oscillator for low level FPGA functions and configuration. There is a clock multiplexer setting the clock for DAQ: a local (stand alone) mode or an external 100~MHz from the MCB.
    \item Backplane connector: The backplane connector located on the rear side of the board allows communication between the FEBs and OCB.
    \item Address: The geographical address of the board combining FEB slot in the crate (four bits) and the crate number (four bits).
    \item MCB buffer: The CLOCK and SYNC receivers connected on the backplane.
    \item Trigger buffer: The M-LVDS trigger input, the LVDS trigger output 
    and the TrigOutOR output signal (open drain transistor). This latter signal can be connected to four different sources: the NOR32 or NOR32T coming from the eight CITIROC chips, a specific digital trigger built by the FPGA for the whole FEB, or to an analogue sum from MPPC-PCB.
    \item DAQ buffer: The backplane readout signals (three point to point GTX + one LVDS outputs, one point to point GRX input, one common OCB BUSY input). Only the LVDS output will be used by the SuperFGD, the additional GTXs are for other possible high data throughput applications.
    \item SC: The slow control link with multipoint LVDS transceivers connected on the backplane. These lines are common to all FEBs, and the FEB-TX is enabled from protocol addressing decoding.
    \item FW programming: The FPGA programming stage includes: one Serial Flash for FPGA programming, one JTAG connector located close to the FPGA, one Front panel connector containing JTAG and five debug signals.
    \item Remote upgrade: The FPGA firmware remote upgrade link with multipoint LVDS transceivers (signals common to all FEBs) and one common SEL signal connected on the backplane. If SEL is asserted during FEB power-on the FEB starts up in ``FW upgrade mode'' allowing the writing of the application area in the flash memory; if SEL is deasserted the FEB starts up in SuperFGD application mode.
    \item Housekeeping: Sensors (MPPC or FEB temperature, voltage and current) connected via SPI ADCs to the FPGA.
    \item EEPROM: A I2C EEPROM used for storing dedicated FEB information (Serial Number, calibration parameters, etc.).
    \item Power supplies: The low voltage power supplies block includes: the over/under/reverse voltage protection on the main 12~V coming from the backplane, a supervisor to monitor 16 voltages and sequence the start up of the FPGA power supplies, a mezzanine containing four DC/DCs efficient in magnetic field, 11 LDOs, PMBUS and buffer to monitor the supervisor, and the FEB enable signal to start up the FEB.
    \item HV: Eight DAC stages (one per 32-channel CITIROC) to adjust the HV on each group of 32 MPPCs, eight control loops, and the main HV input from the backplane.
    \item Debug: One 16-bit debug connector and one button for reset.
     \item LED: Six LEDs (three green, three orange) are dedicated to the FPGA application, and three to power status: one green for power, one orange for supervisor warning, and one red for supervisor fault.
\end{itemize}

\subsubsection{Synchronisation}

All FEBs are driven by the same clock system running at 100~MHz and synchronised by a 100~kHz beacon called GTS, allowing the time stamping of the events. These signals are generated by the Master Clock Board (MCB) and sent to all FEBs through the OCB and backplane.

\subsubsection{Analogue path}

If at least one MPPC signal crosses the CITIROC programmable threshold of the analogue path (see Fig.~\ref{fig:citiroc_circuit}), the CITIROC asserts its own OR32 signal. The OR32 signals from the eight CITIROCs are grouped in pairs such that it corresponds to one MPPC-PCB, this signal is called OR64. This means there are four OR64 signals per FEB, each of which are individually connected to an open-drain buffer. The outputs of those four open-drain buffers are connected together via a 5~k$\Omega$ serial resistor and produce the Analogue Sum signal. In this way each FEB counts five current values representing the number of MPPC-PCBs containing at least one hit. These 14 Analogue Sum signals are connected in parallel to the OCB and each OCB sends its Analogue Sum to the MCB. In turn, the MCB rounds up the Analogue Sum in one signal and if this signal is in between the defined thresholds, the MCB sends a SuperFGD-Event word back to the OCB. This SuperFGD-Event word is added to the readout data from the FEB by the OCB and processed offline.
In parallel, the MCB continuously sends Gate signals to the FEB, the frequency and the length of the gate depending on the mode of the detector: beam, cosmic or LED calibration. When the gate is closed, the analogue path is disabled, whereas when the gate is open, the first hit directly starts the CITIROC peak detector of the involved channel and at the same time the FPGA programmable hold delay allows other channels to also be self-triggered on their corresponding peak detector.
At the end of the hold delay, all 256 channels of the CITIROCs are frozen until the end of the gate, at which point ADC conversion starts. The ADC conversion time is seen as an analogue dead time from the system point of view. At the end of the ADC conversion process, the peak detectors of all the CITIROCs are reset and the CITIROCs can be retriggered on the next gate. At the same time, all digitised channels having a value greater than an FPGA programmable threshold are sent to the OCB within their corresponding time-stamped GTS beacon allowing a reconstruction of the event. The ADC conversion time is on the order of the GTS period and will typically be sent on the next GTS beacon. The entire mechanism is applied on the HG and LG paths of the CITIROC during gate opened and the FPGA hold delay is started on the first channel triggered on the HG or LG path.

\subsubsection{Timing path}

In parallel to the analogue measurement, and in order to avoid losing events, the FPGA samples every edge of the digital outputs of the CITIROC at the frequency of 400~MHz. In a future firmware version, a doubling of this frequency is scheduled using Double Data Rate (DDR) sampling on rising and falling edges of the clock. These digital outputs are processed into the CITIROC from another trigger path, containing a faster shaper than the analogue trigger path, taken from either the HG or the LG common front-end amplifier, chosen by means of a slow control bit. In this trigger path, each channel crossing a trigger threshold, which can be chose different from the analogue threshold, can be available on the CITIROC individual channel digital output. This architecture is the same for the 32 outputs of the chip. The FEB FPGA timestamps each channel rising edge and falling edge with respect to the GTS signal. This data is sent to the OCB within its corresponding time stamped GTS beacon, allowing a reconstruction of the event with the same FEB analogue data and the other FEBs. The rising edge provides a precise timestamp of the event, and the Time Over Threshold (TOT) can be computed from the difference between the rising and falling edges and then calibrated with the analogue measurement. This allows a measurement of the charge of the events read during the gate open windows, although less precise than the analogue measurement.

\subsubsection{Working modes}

The different working modes (beam, cosmic, or LED calibration) are sent by the MCB and allow the FPGA to switch to the corresponding mode, i.e. change gate frequency, gate width, thresholds and delays. When the corresponding mode is entered/exited by the FPGA, a dedicated data header/trailer is sent to the OCB allowing the redirection of the different data sent by the FEB with respect to their working mode.

\subsection{FEB characterisation}

Several characterisation tests were carried out to assess the FEB performance and to verify the quality of the FEB design. The linearity of the FEB response, electronics cross-talk and channel-to-channel timing synchronisation were studied, as well as the possible effect of the ND280 magnetic field~\cite{SFGD_FEB_TWEPP_2025}.

\subsubsection{Linearity of the FEB response}

The energy resolution of the SuperFGD is affected by non-linearity effects that might arise in the readout electronics response. It is therefore crucial to finely benchmark any non-linearity effects over a wide range of input signal amplitudes. A FEB prototype, reproducing the final electronics design, was put in a test bench where an electronic signal was injected in each FEB channel. The FEB response was studied in terms of ADC units over the 12-bit ADC upon injection of a signal pulse. Several values of the injected charge were considered, varying the amplitude of the injected signal. The whole range of gain settings was explored, varying the feedback capacitance of the amplification circuit within the CITIROC.

It is expected that the response of the electronics stops being linear for large injected charges, determining the upper limit of the detector charge dynamic range for both the LG and HG readouts. The goal of this test was to characterise the range over which the FEB response is linear for different gain settings. The response is considered linear if the ADC response as a function of the injected charge can be described by a straight line with residuals below 1\% of the ADC range.

The input signal for this study was generated with a pulse generator (TTi TG5012A, used to generate square wave signals) and a charge differentiating circuit (composed of a 50~$\Omega$ resistor in parallel to ensure the impedance adaptation and a 100~pF capacitor in series), in order to reproduce the fast-rising-edge signal expected from the MPPCs. All FEB channels underwent the same test to study possible differences between channel responses.

\begin{figure}
\centering
   \includegraphics[width=\linewidth]{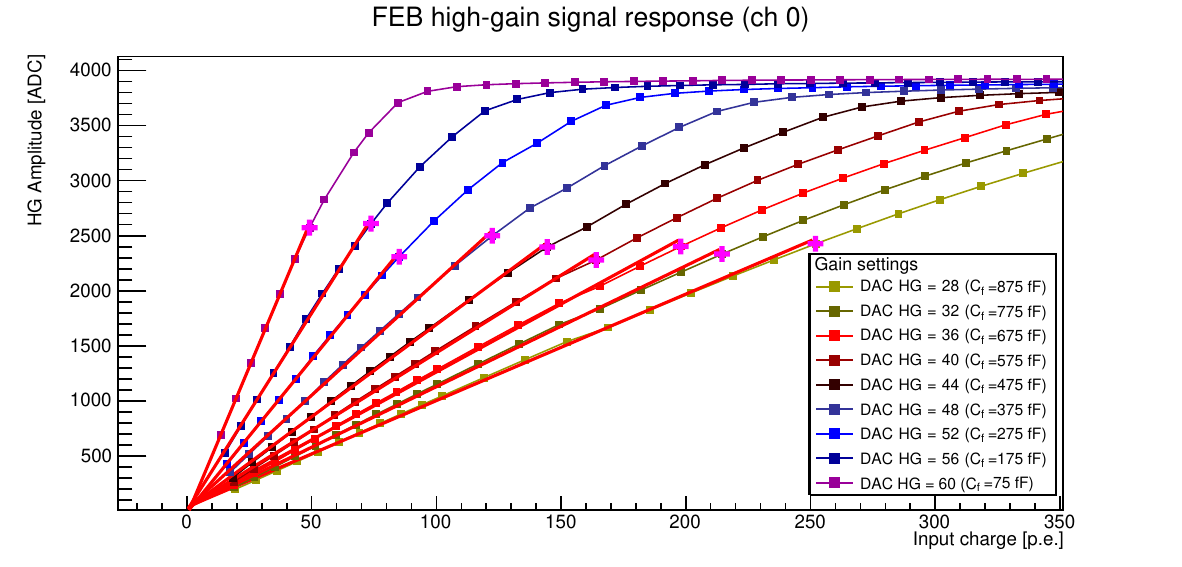}
  \includegraphics[width=\linewidth]{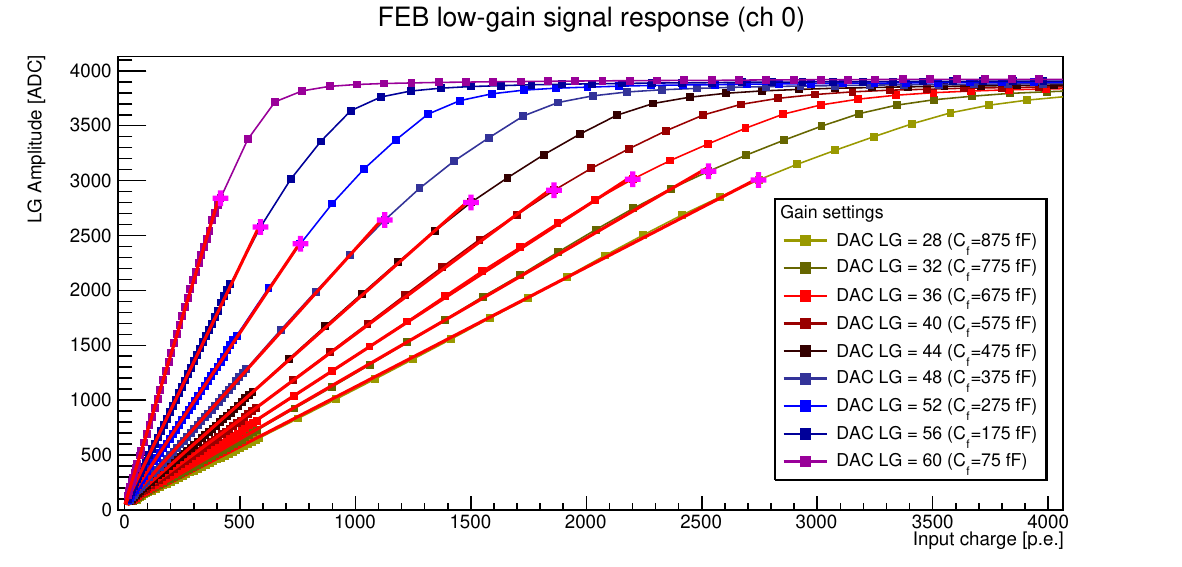}
\caption{Results of the FEB linearity test for channel 0 of an FEB prototype. The ADC response of charge read out for both HG (top) and LG (bottom) is shown for several gain settings and a wide range of input charges. The input charge is estimated from the pulse generator amplitude considering SuperFGD calibration results.}
\label{fig:linearity}
\end{figure}

Figure~\ref{fig:linearity} shows the results of the linearity test for both HG and LG charge readout. The differences between different channels were found to be below 2\%, satisfying FEB performance requirements of uniformity.

\subsubsection{Electronics cross-talk}
\label{sec:electronics_xtalk}

\begin{figure}
\centering
\begin{subfigure}{.5\textwidth}
  \centering
  \includegraphics[width=\linewidth]{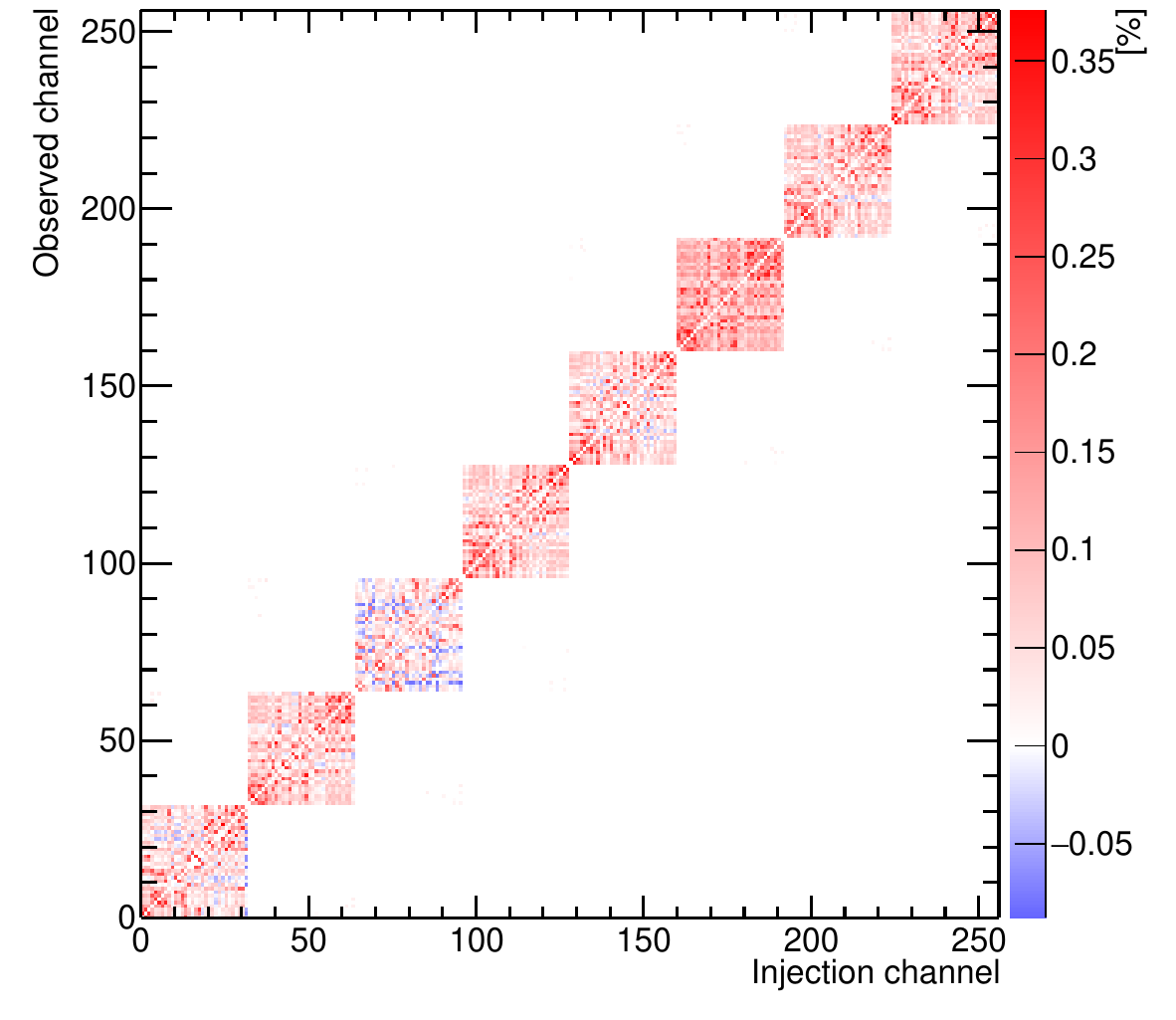}
  \caption{Results for MIB version 1.}
  \label{fig:xtalk1}
\end{subfigure}%
\begin{subfigure}{.5\textwidth}
  \centering
  \includegraphics[width=\linewidth]{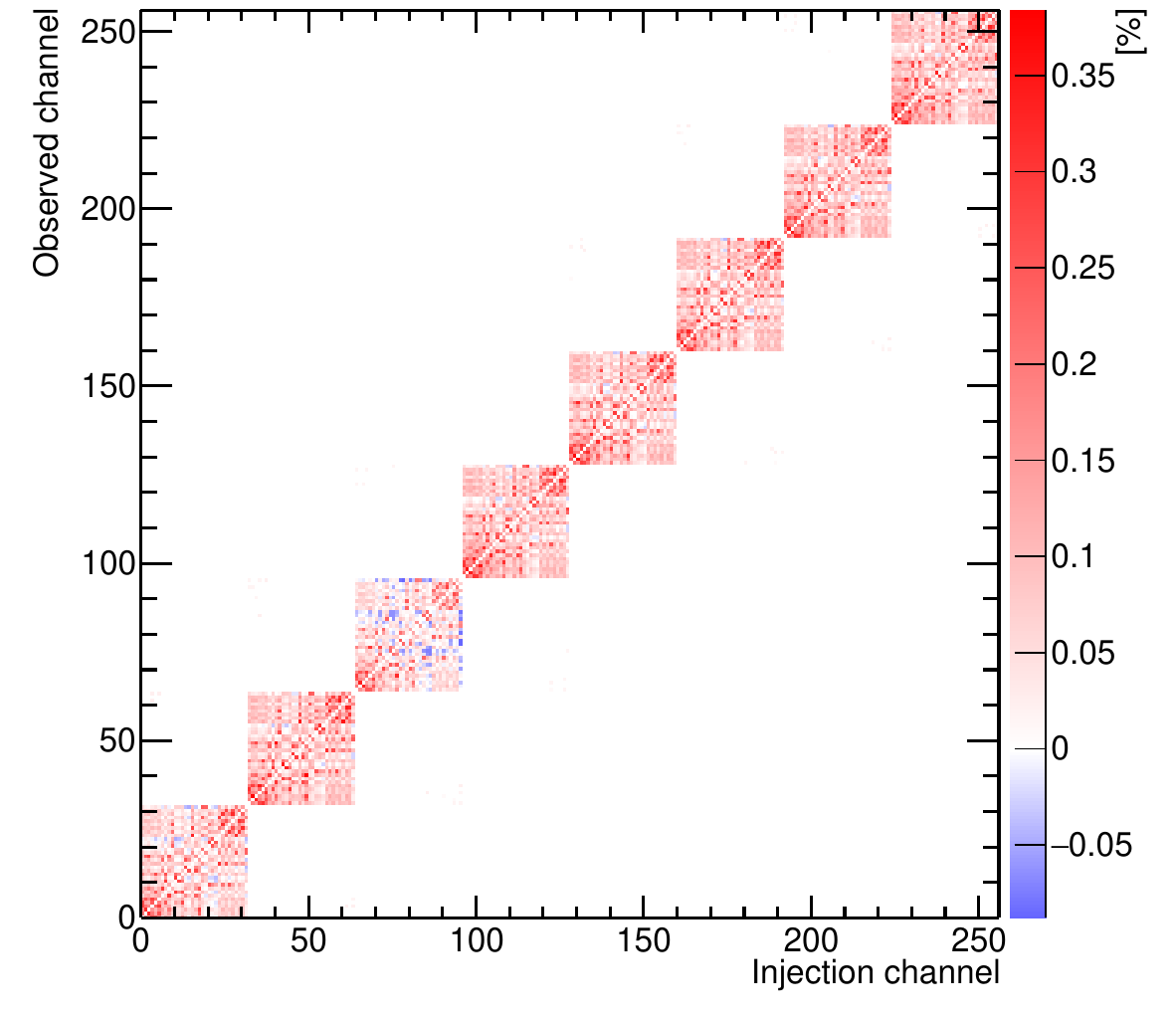}
  \caption{Results for MIB version 2 (final version in use).}
  \label{fig:xtalk2}
\end{subfigure}
\caption{Cross-talk test results shown as a percentage of the injected signal observed in non-injected channels.}
\label{fig:xtalk}
\end{figure}

A test was performed to evaluate the electronics cross-talk induced by neighbouring traces in the FEB and MIB PCBs. The test consisted of injecting one of the 256 channels of the FEB with a large signal ($\sim$1200~p.e.) coming from an MPPC, and observing the average signal recorded in the other channels of the board. This test was performed for two different versions of the MIB, where the specific positions of traces relative to the analogue signal coming from the MPPC differ. Fig.~\ref{fig:xtalk} shows the results of the tests. Cross-talk is observed only within CITIROCs, and not across them. This is expected since the signal traces of different CITIROCs are very far apart, thus there is little chance that interference occurs between them.

The MIB is expected to be the main source of electronics cross-talk since it is a small PCB featuring more than 256 analogue signal traces spread over nine very small areas (typically $2\times2$~cm$^2$). The maximum level of electronics cross-talk observed in this test is around 0.35\%, which is well below the level of the optical cross-talk between neighbouring scintillating cubes, expected to be larger than 1-3\%. In addition, a small ``negative'' effect is observed (see blue channels in Fig.~\ref{fig:xtalk}), which is possibly due to the voltage drop caused on the preamplifier power supply lines in the CITIROC chip, due to the large amplitude of the pulse. These results show that the FEB performance is compliant with the original requirements.

\subsubsection{Timing measurement}

A measurement of the timing difference between FEB channels was performed by simultaneously injecting identical pulses into two separate channels. This was accomplished by using an ordinary ``T'' connector to split the pulse generator output. The signals are then injected into two FEB channels through a high-pass filter, with the same cable lengths.

As described in Section~\ref{sec:FEB}, the timing data is acquired through a fast shaper and a discriminator with an adjustable threshold, applied to one of the preamplification outputs; specifically the HG path for this study. The timing information is calculated using the rising edge of the signal, as it generally offers a more accurate timestamp for the event compared to the falling edge.

\begin{figure}[h] 
    \centering
    \scalebox{0.7}{\includegraphics{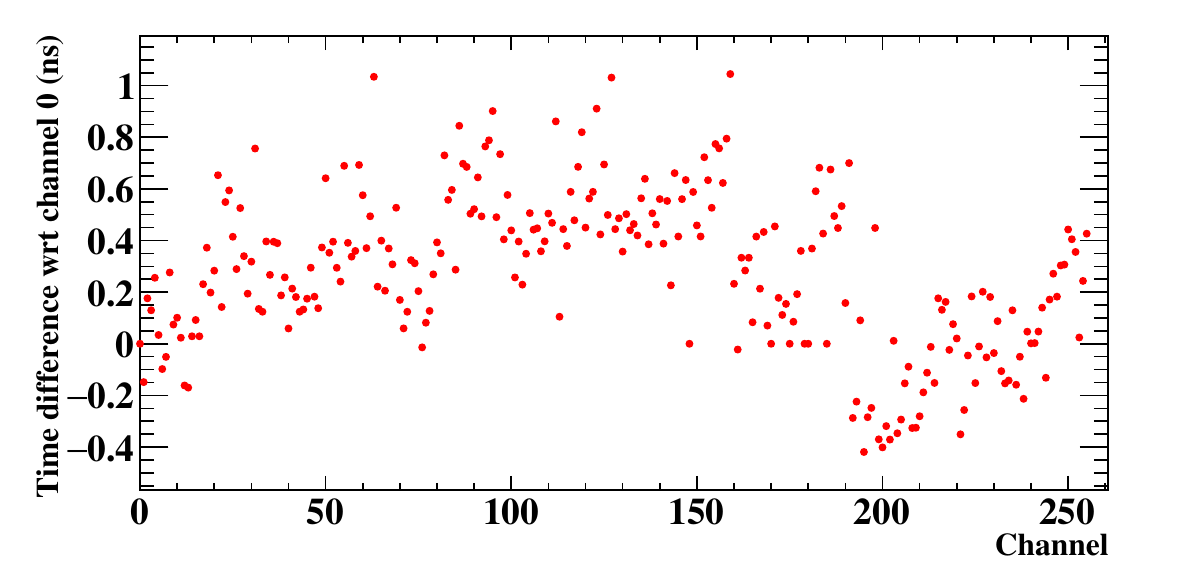}}
    \caption{Timing difference between all the channels with respect to channel 0 in an FEB.}
    \label{time_diff_FEB}
\end{figure}

Figure~\ref{time_diff_FEB} shows the timing difference of the rising edge between each channel and channel 0, which was taken as a reference. The maximum timing difference was around 1.5~ns and could vary up to 2~ns in different FEBs. The observed delay mainly comes from the CITIROC chip itself. It varies between different CITIROC chips and cannot be corrected via the FPGA firmware. Consequently, each CITIROC chip has its own unique timing distribution. These time delays are expected to be fixed for a given channel and can therefore be corrected via calibration, as described in Sec.~\ref{sec:time_resolution}.

\subsubsection{Effect of a magnetic field}

\begin{figure}[h] 
    \centering
    \scalebox{0.8}{\includegraphics{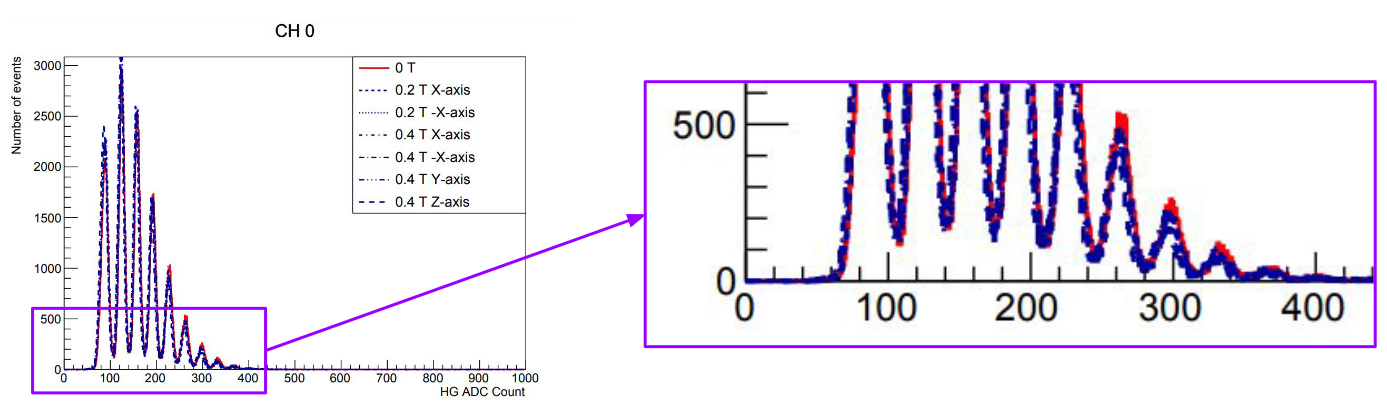}}
    \caption{The MPPC signal digitised by the FEB clearly shows the different peak corresponding to different numbers of photoelectrons (up to 9 in this picture). A zoom on this plot (right) allows to see that no significant effect is seen in these photoelectron peaks recorded with 0.2~T and 0.4~T magnetic fields in different orientations.}
    \label{fig:Finger-Champ-B}
\end{figure}

As the crates containing the electronics boards are installed inside the ND280 magnet, it was important to check that the 0.2~T magnetic field has no influence on the FEB board. Tests were performed at CERN in a dipole magnet at 0.2~T and 0.4~T, using different orientations of the magnetic field with respect to the board.
No significant effect of the different orientations and amplitudes of the magnetic field is observed in the photoelectron peaks recorded, as shown in Figure~\ref{fig:Finger-Champ-B}.

\subsection{Electronics mechanical integration and cooling system}

\begin{figure}[h]
\centering
\includegraphics[height=4cm]{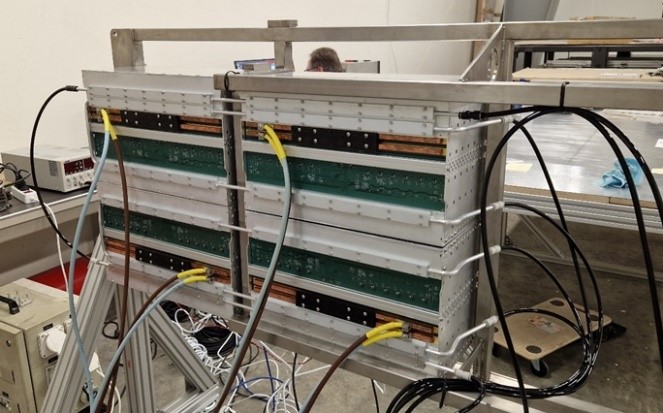}
\includegraphics[height=4cm]{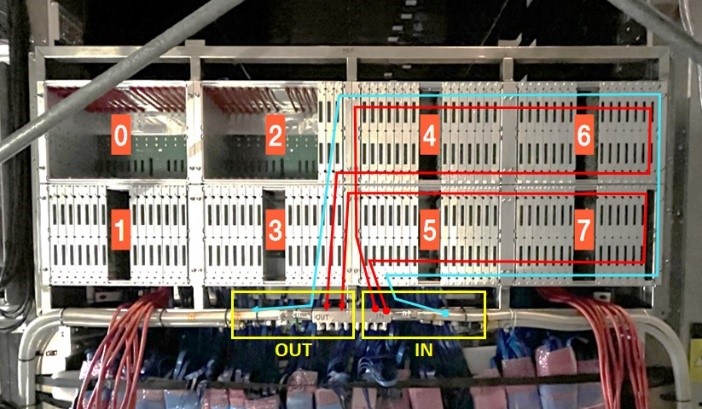}
\caption{Left: crates shown from the back with cooling lines and bars. Right: crates shown from the front before the installation of the FEBs in crates 0 and 2, such that the cooling bars are visible.}
\label{fig:CrateAssemblies}
\end{figure}

\begin{figure}[h]
\centering
\includegraphics[width=0.45\linewidth]{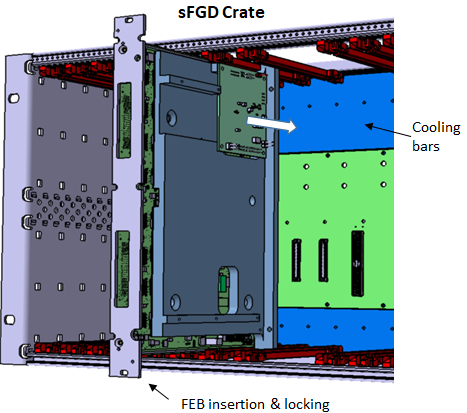}
\includegraphics[width=0.45\linewidth]{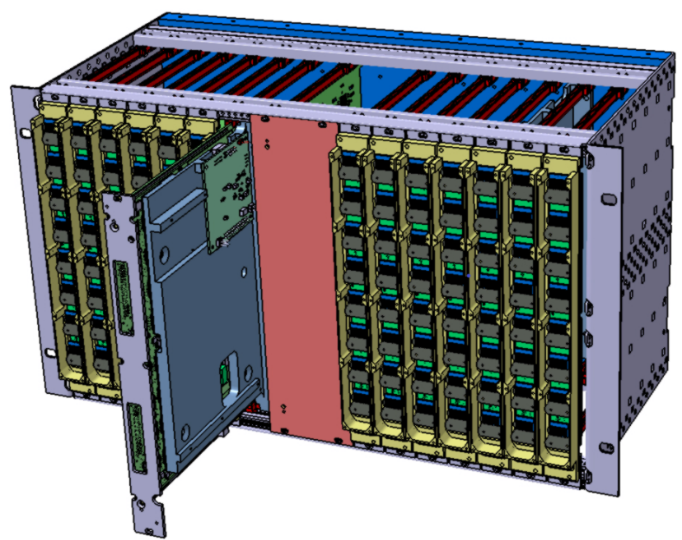}
\caption{FEB insertion into a crate.}
\label{fig:FEB-Insertion}
\end{figure}

A dedicated water cooling system ensures that the hottest components of the FEB, including the FPGA, are kept at a maximum temperature of $\sim$50~$^{\circ}$C. The main component is an aluminium cooling plate conducting the heat up to cooling bars/pipes placed at the rear side of the crate. This geometry allows insertion and extraction of FEBs in the crate without connecting or disconnecting any pipe. A FEA was performed to check the thermal performance of this cooling system
by optimising some parameters (material, shape, cooling pipe size, etc.) to limit the temperature rise. The FEA was compared to tests performed at CERN on the real system which showed very similar performance and validated the thermal design. The crate system is shown in Fig.~\ref{fig:CrateAssemblies} while insertion of an FEB is shown in Fig.~\ref{fig:FEB-Insertion}. The coolant at inlet is set at approximately 18~$^{\circ}$C, using an under-pressurised system to prevent leaks onto the electronics. The maximum water flow depends on both the maximum $\Delta$P (about 0.8~Bar) and the total pressure drop along the lines and manifolds. The total power to dissipate through the cooling system is about 2 $\times$ 2.8~kW. 

\begin{figure}[h]
\centering
\includegraphics[width=\linewidth]{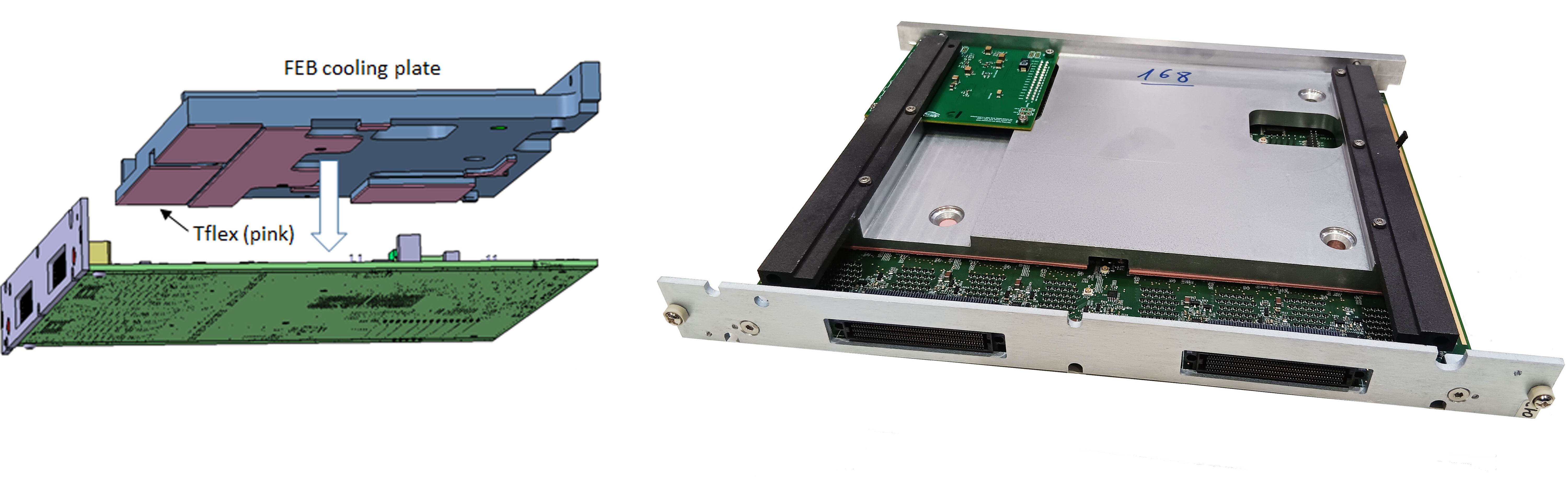}
\caption{Left: model of the aluminium cooling plate (blue) and T-Flex (pink). Right: an FEB equipped with its cooling plate and its low voltage power mezzanine.}
\label{fig:coolingplate}
\end{figure}

The thermal contact between the plate and the cooling bars at the back is facilitated using the thermal paste HTCP 20S. The thermal connection between the cooling plates and the components of the FEB is done by using either the thermal paste or a soft thermal conductive material (TFlex, 1~mm thick) to limit mechanical stresses on the PCB and the components. Fig.~\ref{fig:coolingplate} shows a model of a cooling plate with soft thermal conductive material attached, as well as an FEB equipped with its cooling plate as installed in the SuperFGD detector at J-PARC in 2024. The FEB cooling plates are locked onto the cooling bars using two M3 screws, guided and secured by 3D printed black plastic pieces that are also visible in Fig.~\ref{fig:coolingplate}.

\subsection{Analogue cables}
\label{sub:cables}

\begin{figure}[h]
    \centering
    \includegraphics[width=0.9\linewidth]{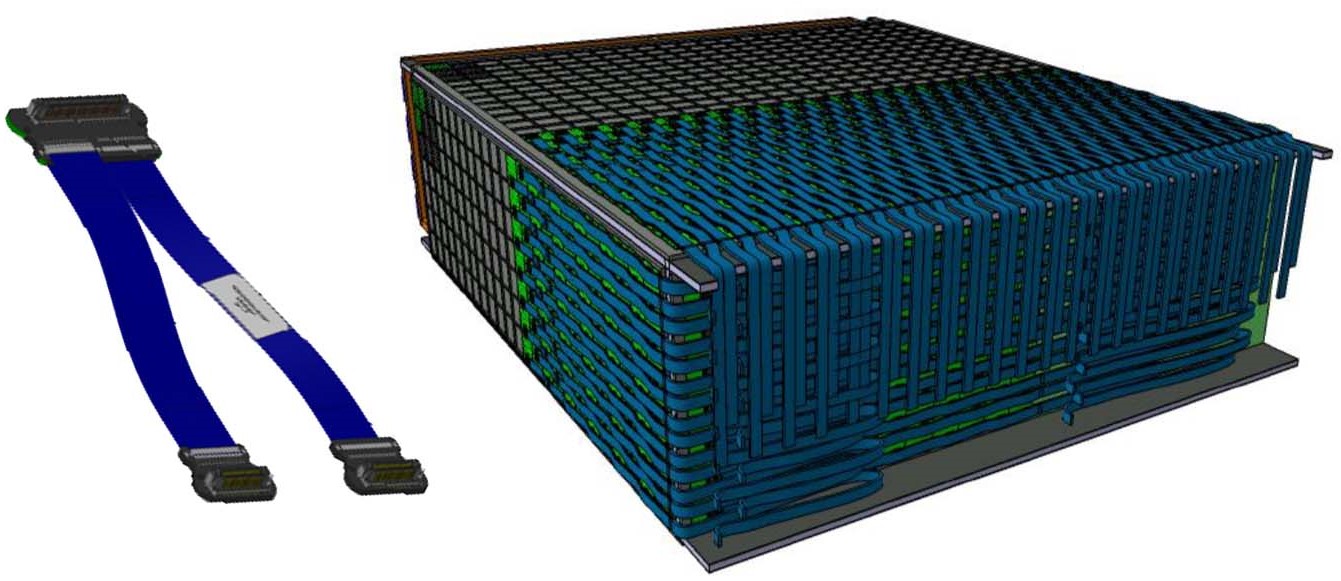}
    \caption{Left: model of the SAMTEC HDR-216294-XX-HLCD custom assembly. Right: model of the cables connecting all MPPC-PCBs to the electronics crates.}
\label{fig:cables}
\end{figure}

\begin{figure}[h]
    \centering
    \includegraphics[width=\linewidth]{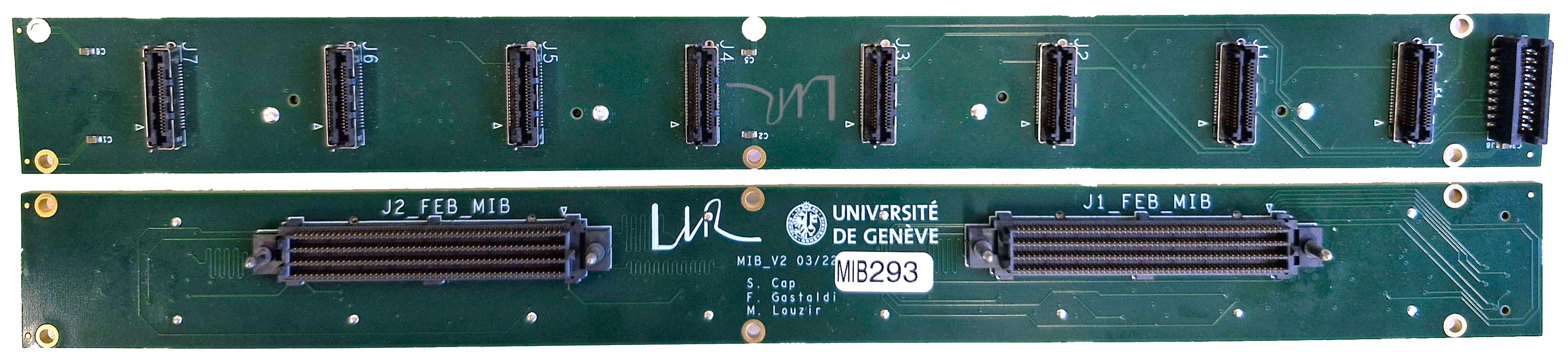}
    \caption{The MPPC Interface Board (MIB) which allows the connection of the 40-channel connectors from four MPPC-PCBs to one FEB.}
\label{fig:MIB}
\end{figure}

The MPPC analogue signals are delivered to the electronics crates using custom 80-channel flat-ribbon SAMTEC micro-coaxial cables~\cite{Samtec_Website}. They are connected by means of one 80-channel connector to the MPPC-PCBs and with two 40-channel connectors to the MIBs. Fig.~\ref{fig:cables} shows a model of the cable assembly. It also shows the cables that are run outside of the light shielding layer along the SuperFGD surface to reach the electronics crates on both sides of the detector. The eight connectors from four MPPC cables are connected to one side of a MIB (see Fig.~\ref{fig:MIB}), the other side of which is directly connected to an FEB.

\subsection{Optical Concentrator Boards (OCBs)}

\begin{figure}
    \centering
    \includegraphics[width=0.8\linewidth]{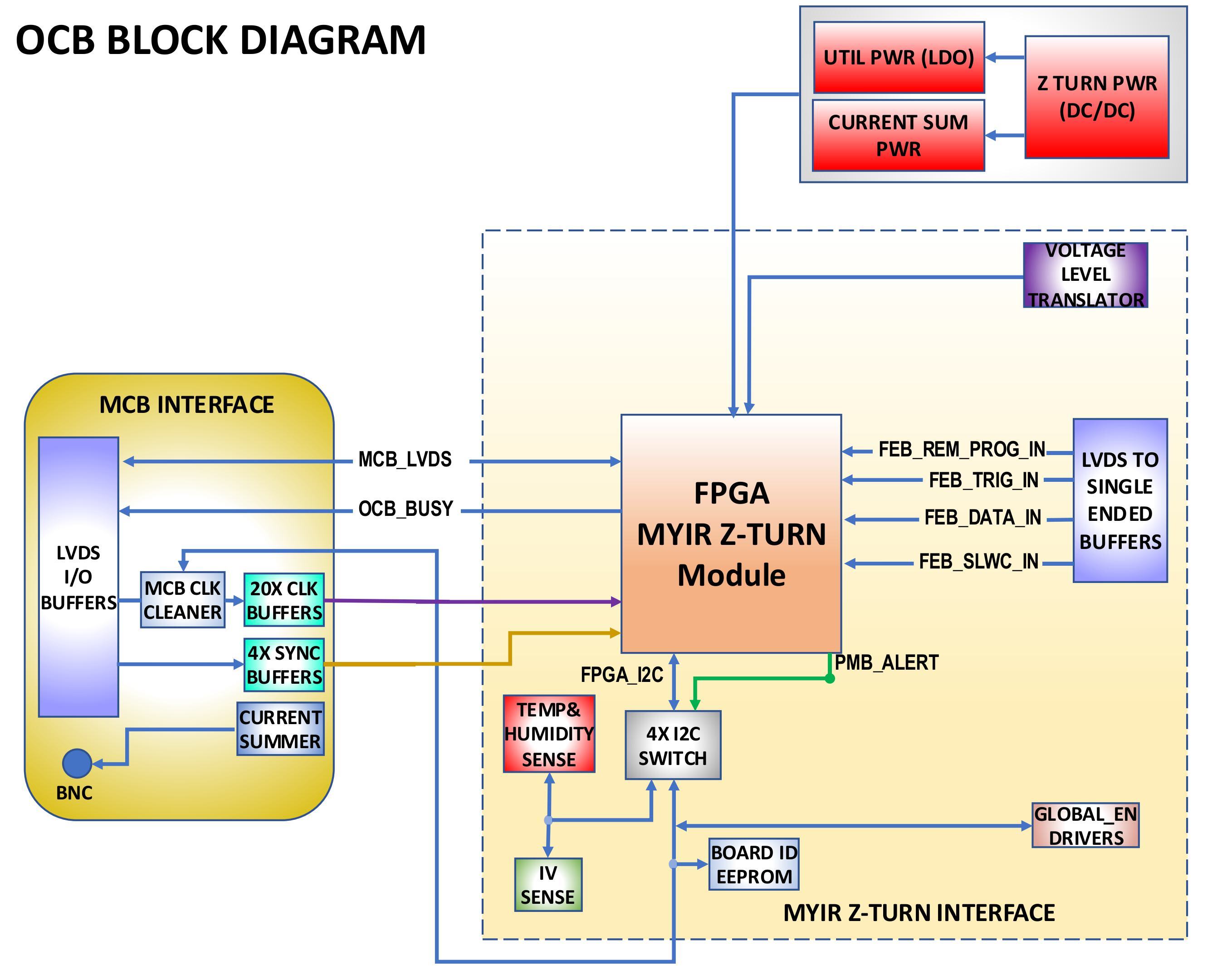}
    \caption{OCB hardware block diagram.}
    \label{fig:ocb-hw}
\end{figure}

\begin{figure}[ht]
    \centering
    \includegraphics[width=0.8\linewidth]{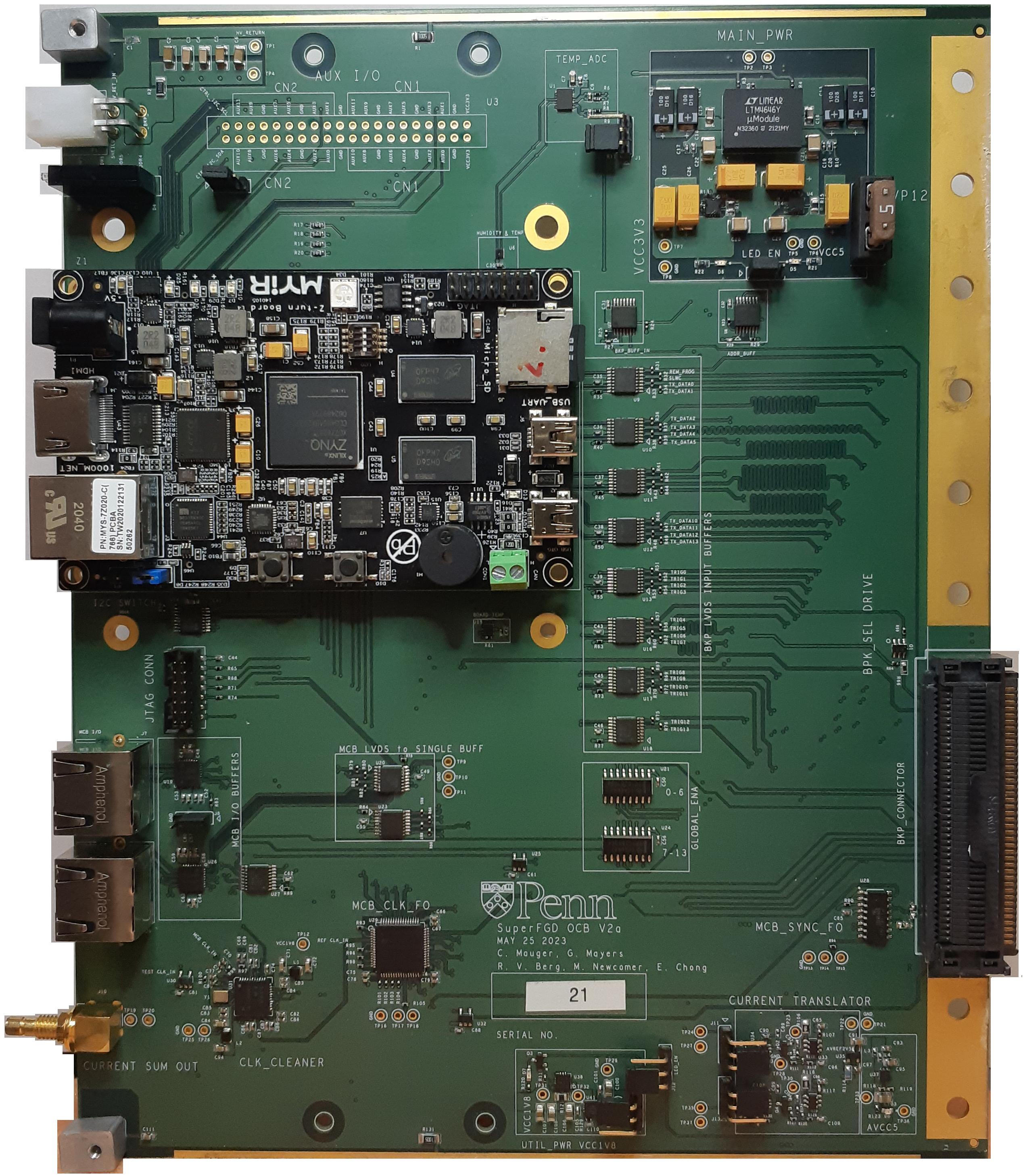}
    \caption{Photograph of an OCB.}
    \label{fig:ocb}
\end{figure}

The Optical Concentrator Board (OCB) is a largely digital device whose main role in the SuperFGD electronics system is to move and organise digital data and commands. In the heart of the OCB is the Zynq-7020 SoC (System-on-Chip) known as the Z-turn board, which has an integrated Artix-7 class FPGA and a dual-core ARM processor. A block diagram and a photograph of an OCB are shown in Fig.~\ref{fig:ocb-hw}
 and Fig.~\ref{fig:ocb},
 respectively.

\begin{figure}
    \centering
    \includegraphics[width=0.9\linewidth]{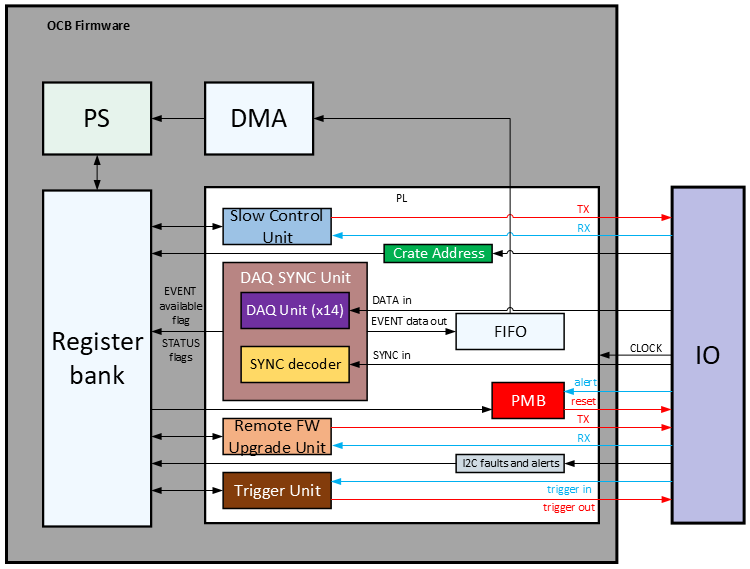}
    \caption{OCB firmware block diagram.}
    \label{fig:ocb-fw}
\end{figure}

The FPGA runs firmware that mainly handles communication protocols between the OCB and 14 FEBs and digital data organisation that is sent to the DAQ system. The embedded processor on the other hand hosts a standard Linux OS running the ND280 DAQ and slow control software. The firmware is loaded onto an FPGA from a binary file saved in an SD card that is inserted into the Z-turn board. Meanwhile, the entire Linux OS is booted through NFS (Network File System), which is hosted on a separate machine. A block diagram of the OCB firmware is shown in Fig.~\ref{fig:ocb-fw}.

\begin{figure}
    \centering
    \includegraphics[width=0.8\linewidth]{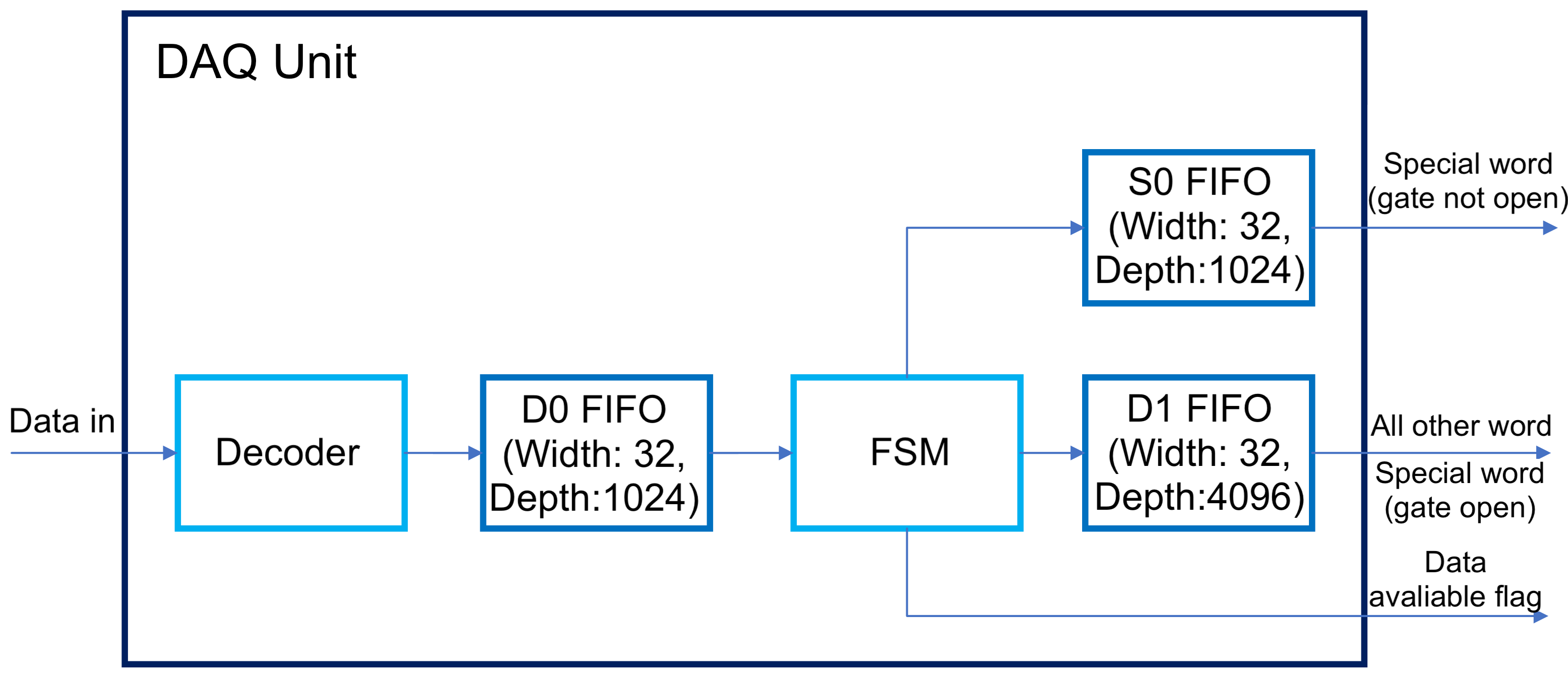}
    \caption{FEB data packet builder.}
    \label{fig:feb-event-builder}
\end{figure}

\begin{figure}[h]
    \centering
    \includegraphics[width=0.8\linewidth]{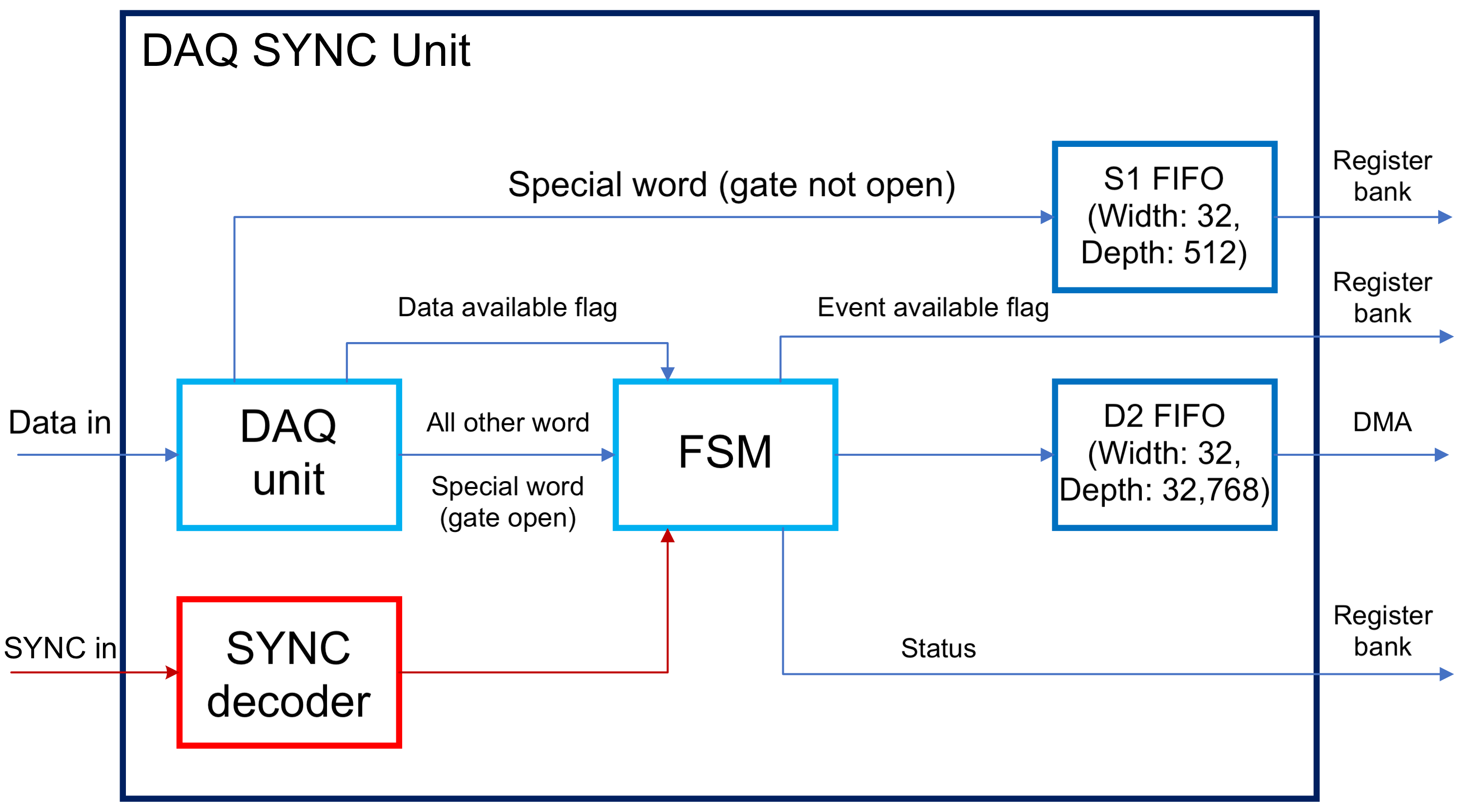}
    \caption{OCB data packet builder.}
    \label{fig:ocb-event-builder}
\end{figure}

Using timing, gate, and trigger information received from the MCB, the OCB acts as a data concentrator that builds local events using data received from 14 FEBs. Diagrams of the FEB and OCB data packet builders are shown in Fig.~\ref{fig:feb-event-builder} 
and Fig.~\ref{fig:ocb-event-builder}, 
respectively. Events built in the OCB are then read out by the DAQ software. The OCB also serves as the interface between the SuperFGD electronics and the global ND280 system. It is through the OCB that ND280 software can communicate with the hardware, such as to configure the FEBs before a run or to monitor slow control parameters.

\begin{figure}[ht]
    \centering
    \includegraphics[width=\textwidth]{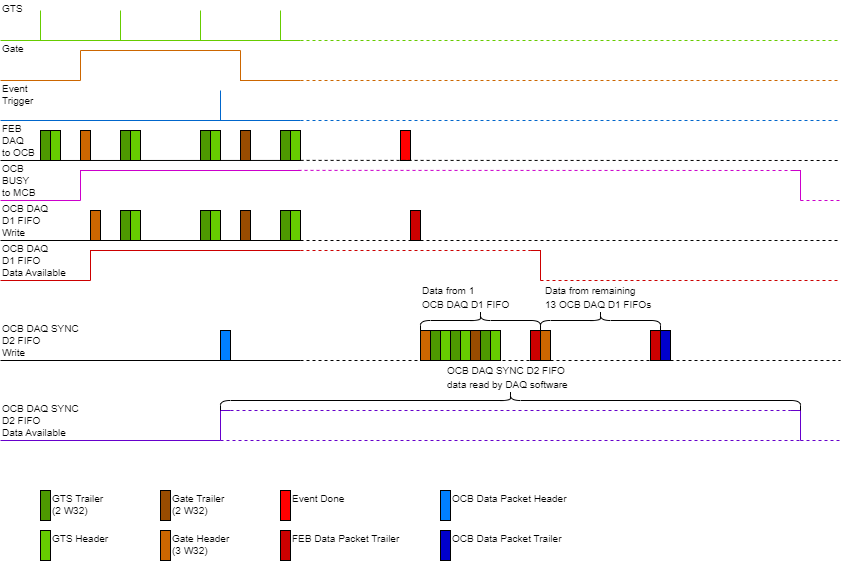}
    \caption{Timing diagram of DAQ SYNC unit event building.}
    \label{fig:event-building-timing-diagram}
\end{figure}

The main logic block in the OCB firmware is the local event builder known as the DAQ SYNC unit. Within the DAQ SYNC unit, a SYNC decoder and 14 DAQ units are instantiated. To start building an event, a gate open followed by an event trigger (before gate close) will need to be received from the MCB. Such information can be obtained by decoding the serial SYNC line from the MCB, which is the function of the SYNC decoder. Depending on the information obtained from the SYNC decoder, the finite state machine (FSM) in the DAQ SYNC unit decides whether to flush all or part of the data received from each FEB or to concatenate the data received from all the FEBs and build an event. On the other hand, data received from each FEB is first built into an FEB data packet by the DAQ unit. When the event trigger is received from the SYNC decoder, the FSM in the DAQ SYNC unit then concatenates 14 FEB data packets into one large OCB data packet. The FSM then triggers a flag to notify the DAQ software of the availability of an OCB data packet to be read out. A busy signal is sent from the OCB to the MCB to signal the readiness of the electronics to receive another gate and trigger from the MCB. Timeouts and error reporting are also implemented in the OCB firmware to take into account any bit error that can occur in the transmission of signals from the MCB and FEBs to the OCB. A timing diagram of the DAQ SYNC unit event building procedure is shown in Fig.~\ref{fig:event-building-timing-diagram}.

%% file: daq.tex
\subsection{Overview}

The software of the SuperFGD DAQ system is built on top of the ``Maximum Integration Data Acquisition System'' (MIDAS)~\cite{midasDAQ}, and integrates into the ND280 DAQ system~\cite{Thorpe:2010iea} which is also based on MIDAS. Due to a large part of the event building process being performed by the OCBs, the SuperFGD DAQ software implementation is relatively simple, consisting of only a single Linux PC running the MIDAS back-end system, and the MIDAS front-end programs running on the Linux OS of the 16 OCBs and the MCB. The final building of complete ND280 events, online monitoring, long-term data storage, and distribution is handled using the existing ND280 DAQ infrastructure.

\subsection{Front-end implementation}

Each OCB front-end program handles the communication with the MIDAS back-end, the configuration of the electronics, the readout of the data stream from the on-board FPGA (``Z-Turn'' board \cite{ZTurnManual}), and the online data-compression for LED calibration data. Upon receiving the run start signal from the MIDAS back-end, the FPGAs are put into readout mode. MIDAS does not provide a way to ``trigger'' events or synchronise the data-readout between the different front-ends, so this is handled at the electronics level. The MIDAS front-end polls the FPGA until data becomes available for readout, and reads out a single ``OCB event'', corresponding to a single trigger received through the OCB. Fast data readout is achieved through direct memory access onto the FPGA. This data is then sent to the MIDAS back-end, where it is accumulated into dedicated buffers for each OCB. Information about readout errors by the FEBs or OCBs is also placed into the same data stream. The MIDAS front-end program scans the data stream online to check for the presence of these error signals and shows alarms to the user if necessary through the MIDAS web interface.

For LED calibration triggers, a large fraction of the MPPCs will send hit data, resulting in very large data streams. To reduce the overall event size, the high gain amplitudes are filled into in-memory histograms within the OCB MIDAS front-end programs, and only a minimal amount of information is pushed to MIDAS to detect potential readout errors. To reduce the maximum size of one MIDAS event, these per-MPPC histograms, comprising about 60~MB in total, are gradually sent to the MIDAS back-end over the course of about 15~minutes. Currently, 15 out of the 3584 total number of histograms are sent every 20 triggers, offset between the OCBs to ensure that one ``SuperFGD event'' contains histograms for one OCB only.

The configuration of the FEB electronics is performed as part of the start-up of the MIDAS front-end programs. The per-MPPC adjustments of the CITIROC and ADC configurations are first constructed using XML files, which are then compiled into binary configuration commands. These are stored in a folder with standardised subdirectory structure on the SuperFGD DAQ machine. This centralised repository is mounted on the OCB's Linux OS using the network file system protocol (NFS). Each OCB MIDAS front-end program then configures each FEB using the FEB-specific configuration files from this directory during start-up.

\subsection{Back-end implementation and integration with ND280}

The back-end PC runs an event builder (EB), a data logger, a user-interface web server, and a ``cascade'' server to communicate with the ND280 ``global'' DAQ system. The EB builds upon a standard MIDAS EB, assembling the ``OCB events'' into a single ``SuperFGD event'' once all 16 buffers have been filled by the OCBs. The default MIDAS implementation is to simply assemble these events according to the order of the received events. This has been extended for the SuperFGD by several checks for the consistency of event IDs among the assembled events, as well as deviations from the incremental changes of the event ID signalling bit flips in the communication line, where the event IDs are issued by the ND280 Main Clock Module (MCM) and electronically distributed through the Master Clock Board (MCB) developed for the upgrade detectors to each OCB.

Integration into the ND280 ``global'' DAQ system is handled similarly to the other sub-detectors of ND280 through a ``cascade'' connection which works in the following way. Both SuperFGD and ``global'' ND280 have their own MIDAS back-ends. The SuperFGD events built by the SuperFGD EB are passed by a cascade ``server'' program ---connected to the SuperFGD MIDAS back-end as a data consumer--- to a cascade ``front-end'' program through a MIDAS-independent TCP/IP connection. The cascade front-end program then feeds these events into the ``global'' ND280 back-end, where they are built into an ND280 event and logged to disk through ND280 back-end programs. The run start/stop messages in turn are transferred in reverse-direction from the ``global'' MIDAS to the SuperFGD MIDAS back-end using the same cascade programs.

\subsection{Triggers}

The ND280 trigger structure controlled by the ND280 MCM consists of a periodic sequence of calibration and cosmic triggers which are interrupted by accelerator-generated beam triggers. The calibration and cosmic triggers are issued at approximately 6~Hz and are stopped before the beam trigger arrives to ensure the electronics are ready to accept the beam trigger. The SuperFGD currently sends empty events to calibration triggers intended for other sub-detectors, whereas for the dedicated ``SuperFGD LED'' calibration trigger, multiple LED gates are opened as described above. For beam triggers, the MCB opens a $40$~{\textmu}s long gate for SuperFGD data-taking after the trigger is received from the ND280 MCM.

To increase the statistics of LED calibration data, the MCB opens three LED gates for each LED calibration trigger received from the ND280 MCM. Each gate corresponds to a single LED flash and the digitisation of charges detected by the MPPCs. However, the buffer for event-building within the OCB's FPGA cannot hold data for multiple LED gates due to its capacity. This data is therefore read out by the MIDAS front-end program for each gate, and used to fill the histograms, while only generating a single MIDAS event to ensure consistent event building within the MIDAS back-end.

The cosmic triggers differ from the other triggers in that they are issued by the ND280 MCM in response to trigger primitives raised by different sub-detectors within a cosmic data-taking window. During this cosmic data-taking window, the MCB opens a continuous sequence of gates. If no cosmic trigger is received from the MCM while the gate is open, the gate is closed and re-opened without digitisation and data readout. If a cosmic trigger is received, the hits recorded by the MPPCs are digitised and sent to the DAQ. Switching between these two operating modes is done in response to command words issued by the MCM to the MCB, signalling the beginning and end of a cosmic data-taking window.

Typical SuperFGD event data-sizes are 15--50~kB per event for a beam trigger. With repeated LED triggers, a data-throughput of about 80~MB/s at approximately 300~Hz can be reached if the data is passed through the system without filling histograms. By filling histograms and only transmitting minimum error information, the actual network throughput and final storage data size during both beam operation and dedicated calibration runs can be reduced as much as necessary. The filling of histograms introduces a performance penalty that reduces the attainable event rate to about 120~Hz. Since during beam operation LED gates are only opened at a frequency of about 20~Hz, this does not cause problems during regular data taking.

\subsection{Performance monitoring}
\label{sec:performance_monitor}

The performance of the SuperFGD is monitored during its operation. An online monitoring program reads the data output stream of the SuperFGD and displays plots which are analysed by detector experts to identify issues and ensure good quality data is being recorded. The plots are filled over time during a run, and are reset every two hours or when a new run is started. They are monitored during data-taking in shifts to ensure 24 hour coverage so that detector experts can be notified immediately to rectify any issues. Offline monitoring is also performed to determine flags for data quality which are input into physics analyses.

Hit maps are used in live monitoring of the different outputs from the CITIROCs: the high gain, low gain, rising edge and falling edge (see Sec.~\ref{sec:FEB}), as well as the number of times a channel was activated from any readout.

The number of channels without any hits during the monitoring period, or ``empty channels'', is displayed along with the number of channels with more than five times as many hits as the mean over the whole detector, or ``hot channels''. Large deviations in these are indicative of an error in the data readout and will prompt a reconfiguration of the relevant FEBs.

Live monitoring of the LED calibration system and the high gain performance is also performed using LED data saved to histograms as described above. A rapid version of the high gain measurement described in Sec.~\ref{sec:led_calibration} is applied to the histogram data, using smoothing and local maxima instead of fits to estimate the photoelectron peak positions.
Regions of the detector where the LED system, electronics, or mechanics is malfunctioning will be apparent in a plot of the measured high gain averaged in an ASIC by unexpected blank regions or regions with gain shifts, respectively.

Other live monitoring of the SuperFGD includes the data size from each event, the ADC spectra in the high and low gain, and the number of hits per event.

\begin{figure}[t]
    \centering
    \includegraphics[width=0.7\textwidth]{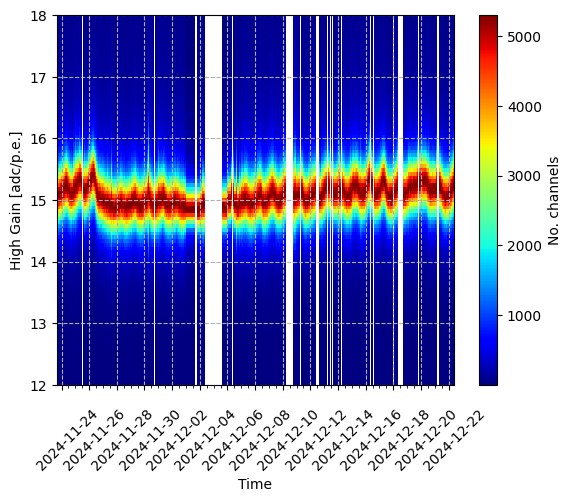}
    \caption{The high gain distribution of the SuperFGD as a function of time is shown for part of the November-December 2024 run period, where diurnal fluctuations in the gain are seen due to changing temperatures inside the ND280 magnet.}
    \label{fig:SFGD_gainMon}
\end{figure}

Data quality checks are performed offline with an analysis of the high gain, and OCB error codes. The high gain data saved in histograms for each channel is analysed every two hours and a channel is classified as problematic if the gain measurement procedure that is described in Sec.~\ref{sec:led_calibration} fails. If more than 30 channels in an FEB are classified as problematic, then that period of data taking is flagged as bad and will not be used in physics analyses. The data taking period in November and December of 2024 was flagged as good for $99.6\%$ of the beam time. These checks also provide a history of the gain fluctuations in the detector, as shown in Fig.~\ref{fig:SFGD_gainMon}.

%% file: slowcontrol.tex
To ensure the proper functioning and maximise the lifetime of the 16 OCBs, 222 FEBs, and 881 MPPC-PCBs that make up the SuperFGD electronics, a set of more than 7,000 sensors is installed on these boards, allowing the measurement of monitoring variables such as bias voltage, current, power, temperature, and relative humidity on different components, including DC-DC converters, power mezzanines, FPGAs and others.

The sensors are read every 10 seconds by two front-end programs running on the processing system of the OCB's SoC ``Z-turn'' board, one front-end for the OCB and one for FEB variables. To communicate with the Z-turn, the sensors on the OCB use an I$^{2}$C bus, while the ones on the FEBs and MPPCs, do it through the two Slow Control lines. In both cases the data is sent using the UART protocol.

Using an RPC protocol, the front-end programs send the data from the sensors to be written to the Online Database (ODB) \cite{midasDAQ} of the MIDAS back-end, which is installed on the Global Slow Control (GSC) PC, also known as the ``MIDAS GSC''. This data is logged and can be retrieved at will. 

The alarm system in the MIDAS GSC handles one warning and one alarm for each of the variables reported. If any of the variables goes outside their set operation limits, an audiovisual alert is displayed indicating the board and variable of origin so appropriate actions can be taken. When critical variables in the FEBs exceed their ``alarm limit'', an interlock mechanism is triggered, which automatically cuts off the HV and LV to all the FEBs in the crate to avoid potential damage to the hardware.

MIDAS also offers a visual representation of all the monitored parameters, called ``history plots''. These give a quick and easy view over any custom time interval, which is helpful when performing diagnostics of the electronics or when looking for variations or patterns in general. This is particularly useful for calibration purposes, where variations in MPPC HV and temperature are important.

The Slow Control system includes the control of the HV and LV supply for all the boards, reset of DAQ readout parameters, reset of a POSIX semaphore that synchronises the access of a slow control register shared between the DAQ and Slow Control front-end programs, and a ``clock cleaner'' that locks the clock of the OCB to that of the MCB.

%% file: detector_performance.tex
The response of the SuperFGD to charged particles was commissioned and validated with cosmic data both on surface and once installed in ND280. The detector was also commissioned with neutrino data in Autumn 2023 and Spring 2024. The results reported in this section are obtained with the data collected during the first physics run in May and June 2024. They consist of LED calibration data, cosmic rays, and neutrino beam data.

\subsection{MPPC gain and pedestal calibration}
\label{sec:led_calibration}

The precise calibration of both the electronics pedestal and the MPPC gain, as well as their stability over the run, is of crucial importance to ensure a precise measurement of the particle energy deposit in each readout channel. This relies on the LED-based calibration system described in Sec.~\ref{sec:led}, which is used in dedicated calibration runs during accelerator maintenance periods and also provides calibration data between beam spills for continuous monitoring.

Both the gain and pedestal are measured from the charge signal spectrum obtained by exposing the ends of the WLS fibres to the LED light. The raw data are recorded in ADC units and converted into p.e. using the relation
\begin{align}
    \text{[ADC]} = g \times \text{[p.e.]} + p
\end{align} 
where $g$ is the MPPC gain in units of [ADC/p.e.] and $p$ is the pedestal.

\begin{figure}[h!]
\centering
\includegraphics[width=0.45\columnwidth]{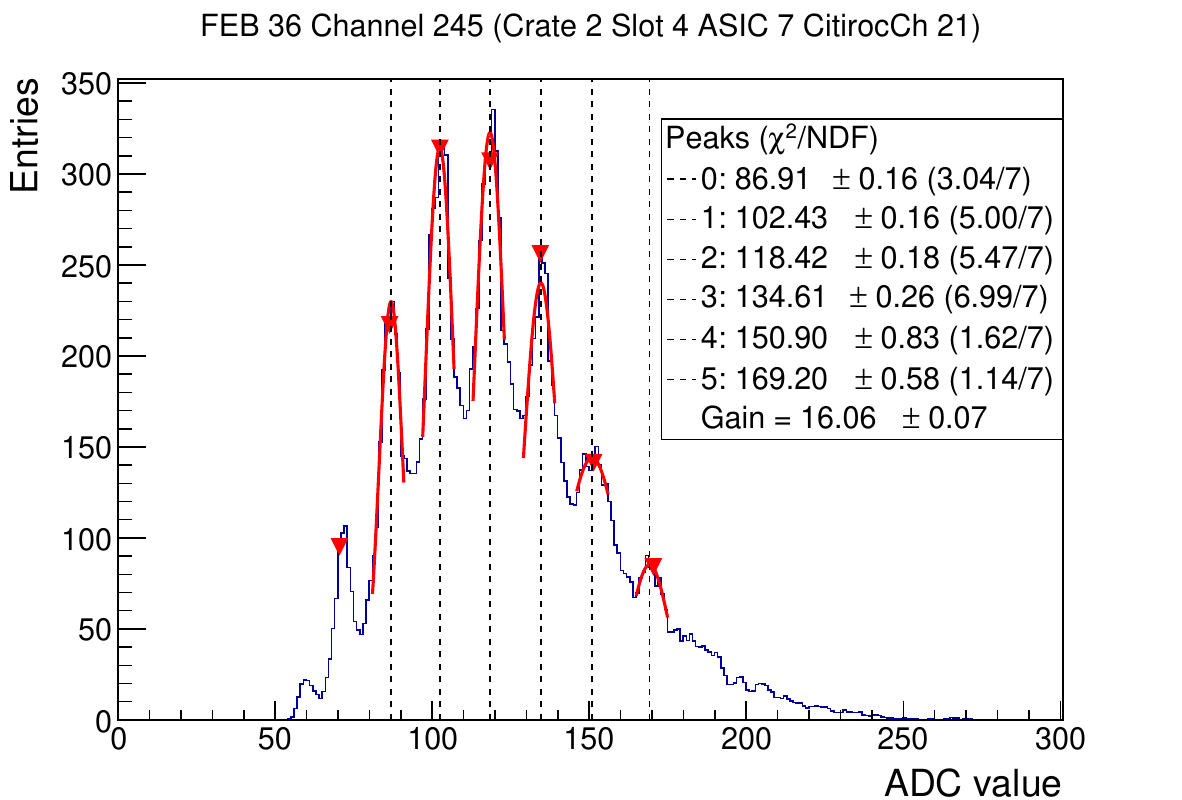}
\includegraphics[width=0.5\columnwidth]{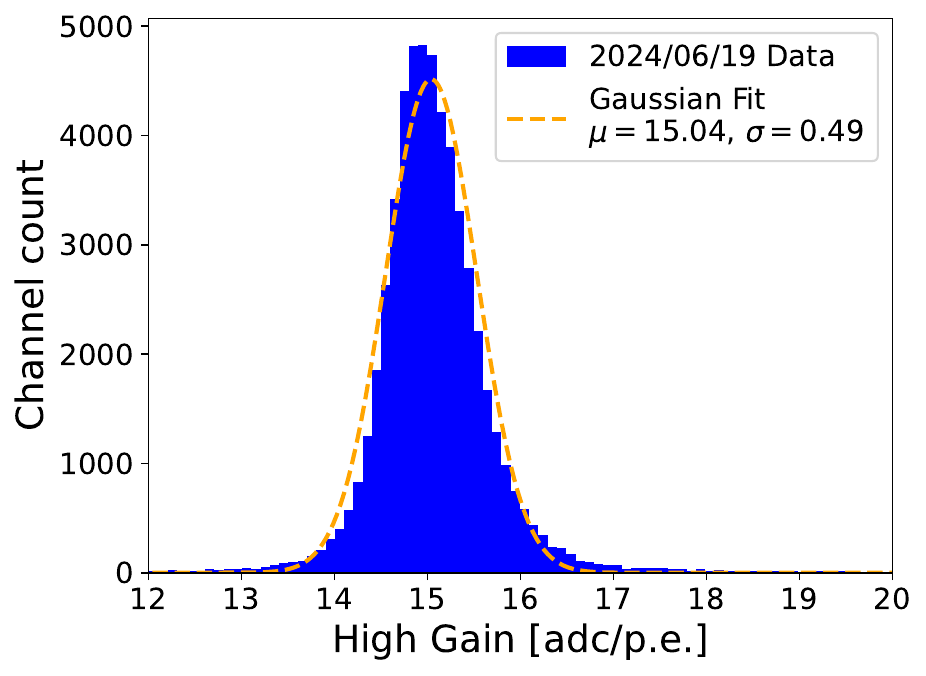}
\includegraphics[width=0.5\columnwidth]{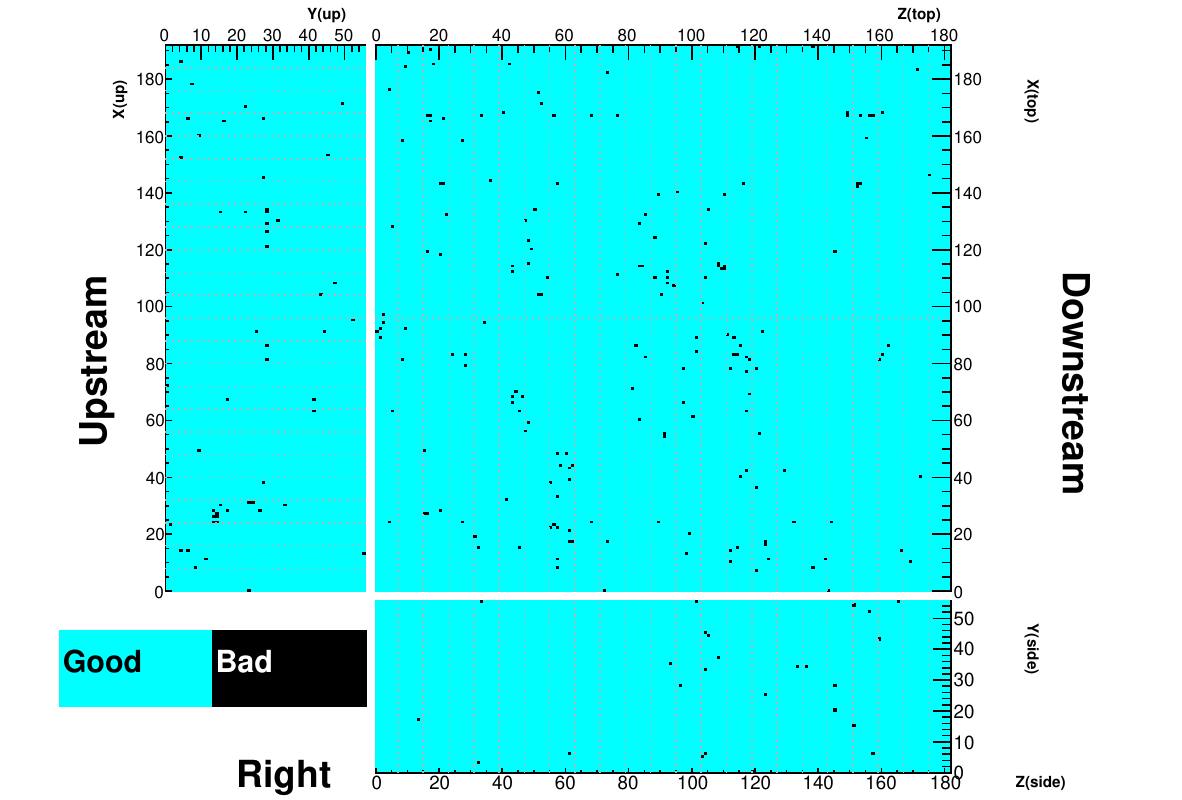}
\includegraphics[width=0.45\columnwidth]{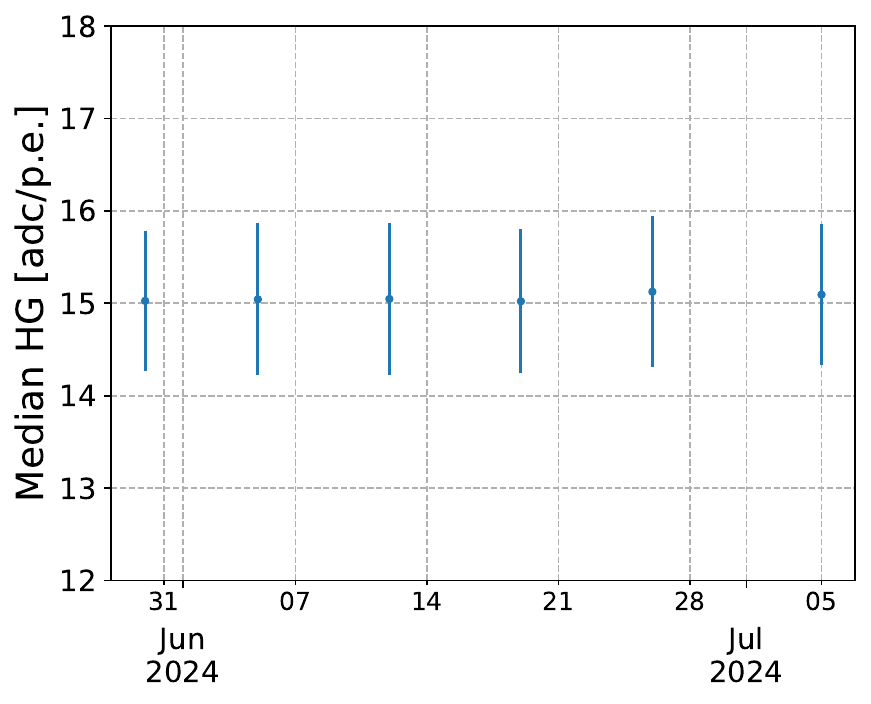}
\caption{
Results of MPPC gain calibration and monitoring performed during June and July 2024.
Top left: example of ADC distribution for one channel obtained during a dedicated LED calibration run.
Top right: distribution of the MPPC gain in units of ADC per p.e. from the data collected during a dedicated LED calibration run. A Gaussian function (dashed orange line) is overlaid and the fitted parameters are shown in the legend.
Bottom left: per-channel quality classification map shows the positions of readout channels that were successfully calibrated (blue) and those where the gain measurement failed (black). Channels where the gain measurement failed account for only 0.45\% of the channels.
Bottom right: stability of the median MPPC gain as a function of time measured in June and July 2024.
}

\label{fig:Calibration_results_mppcgain}
\end{figure}

The LED signal spectrum is shown in Fig.~\ref{fig:Calibration_results_mppcgain}. The spectrum features a distinct multi-peak structure, with each peak corresponding to a specific number of p.e. The MPPC gain is computed as the distance between adjacent peaks, which is approximately constant across the spectrum. To increase the precision the gain is extracted by performing a linear fit to at least three consecutive peaks. Fig.~\ref{fig:Calibration_results_mppcgain} also includes a map showing which channels were successfully calibrated during a dedicated calibration run. The fit of the gain fails for 0.45\% of channels. Since the channels for which the gain fit fails are not localised in one area of the detector, this is an acceptable number of failures that will not have a significant impact on data analysis.

\begin{figure}[h!]
\centering
\includegraphics[width=0.5\columnwidth]{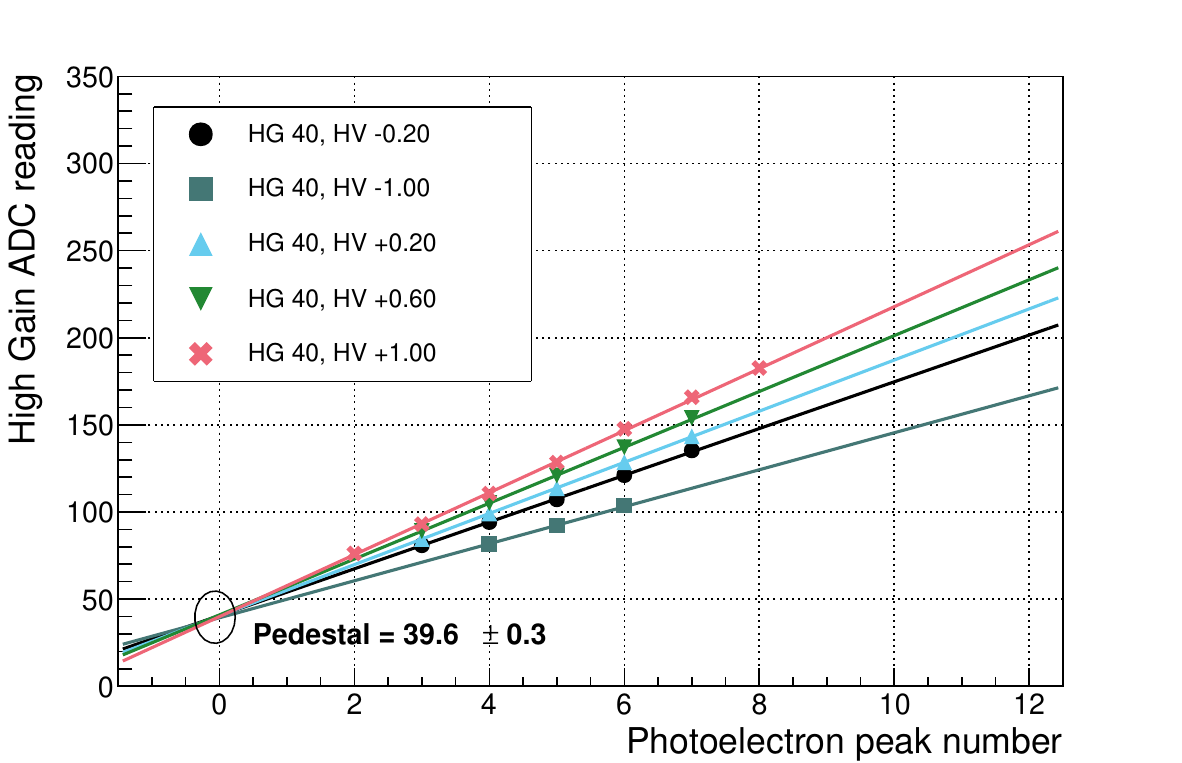}
\includegraphics[width=0.45\columnwidth]{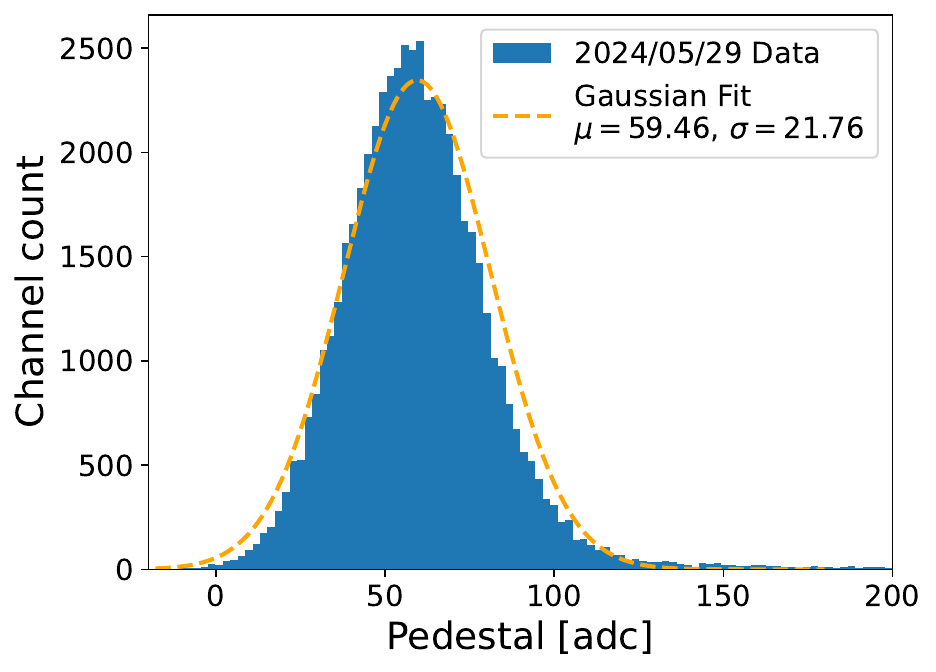}
\caption{
Results of electronics pedestal calibration performed during June and July 2024.
Left: example of extrapolated pedestal value in units of ADC and p.e. peak number obtained using the method described in the text.
Right: distribution of the electronics pedestal in units of ADC from the data collected during a dedicated LED calibration run. A Gaussian function (dashed orange line) is overlaid and the fitted parameters are shown in the legend.
}

\label{fig:Calibration_results_pedestal}
\end{figure}

The electronics baseline is set to be below the discriminator threshold, hence outside of the visible ADC range, to avoid reading out all channels each time a single channel is triggered. This keeps the data rate at an acceptable level. Therefore the pedestal cannot be measured directly and is instead computed by extrapolating straight lines obtained by fitting the p.e. peaks from LED ADC spectra obtained with different electronics HV bias settings. An example is shown in Fig.~\ref{fig:Calibration_results_pedestal}.

The measurements of the MPPC gain and the pedestal performed during June and July 2024 are shown in Fig.~\ref{fig:Calibration_results_mppcgain} and Fig.~\ref{fig:Calibration_results_pedestal}, respectively. The MPPC gain distribution was fitted with a Gaussian function. A mean of 15.0 ADC and a standard deviation of 0.5 ADC were measured. The Gaussian fitted mean pedestal value is 59.5 ADC with a standard deviation of 21.8 ADC. The stability of the median pedestal and MPPC gain values during June and July 2024 are shown in Fig.~\ref{fig:Calibration_results_mppcgain} and Fig.~\ref{fig:Calibration_results_pedestal}, respectively.

\begin{figure}[h!]
\centering
\includegraphics[width=0.45\columnwidth]{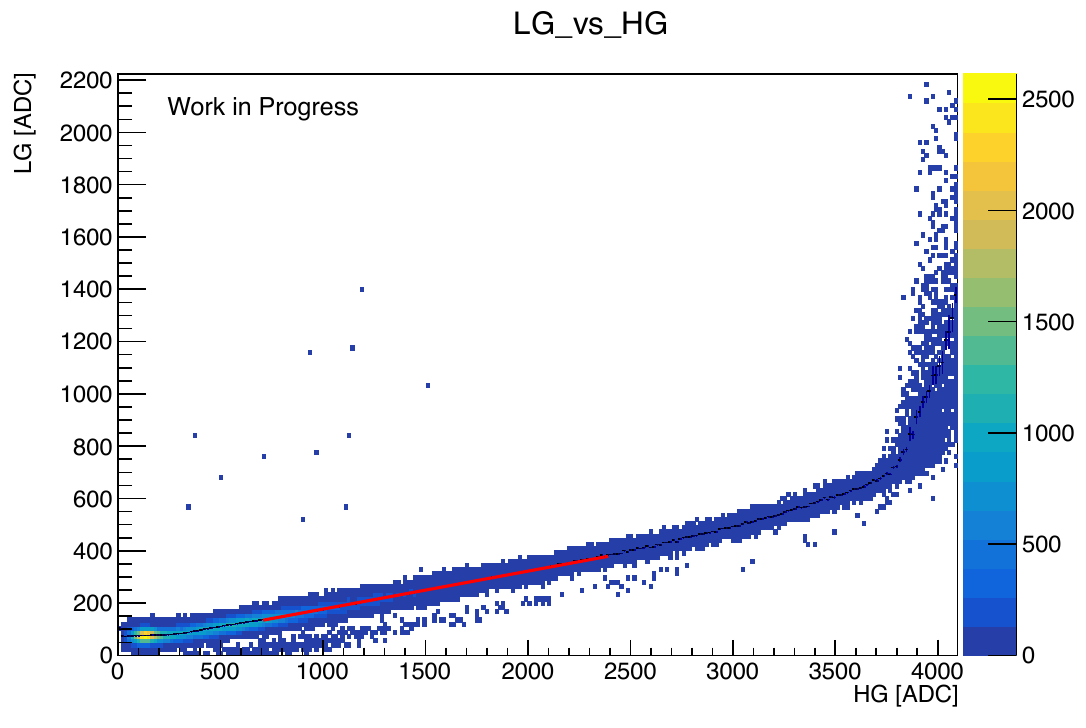}
\includegraphics[width=0.45\columnwidth]{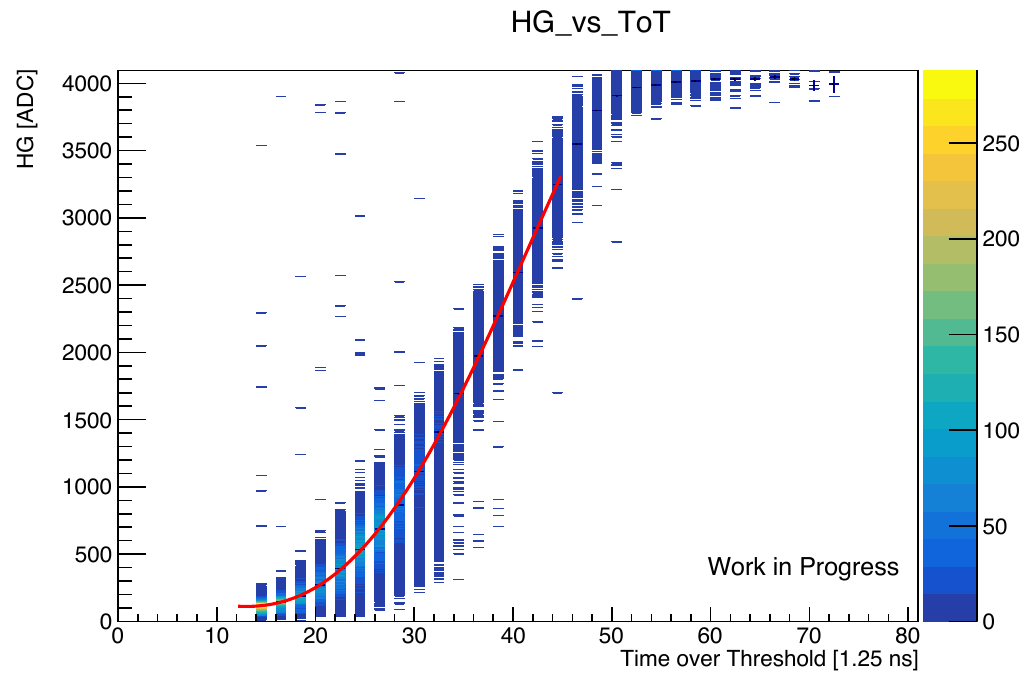}
\includegraphics[width=0.45\columnwidth]{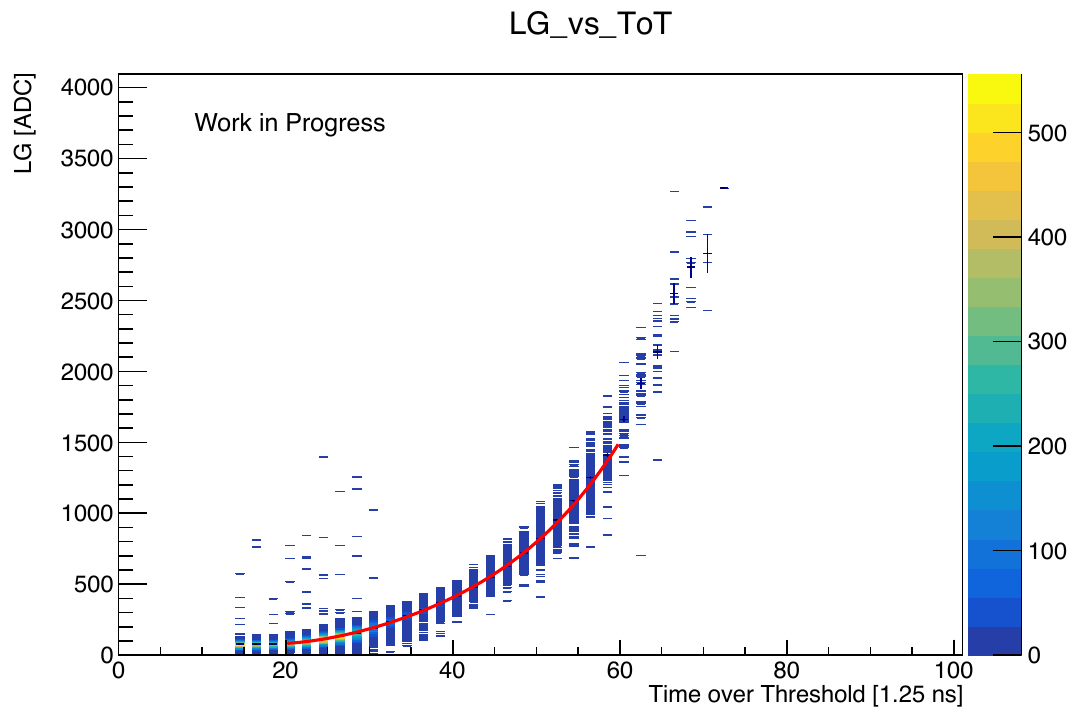}
\includegraphics[width=0.5\columnwidth]{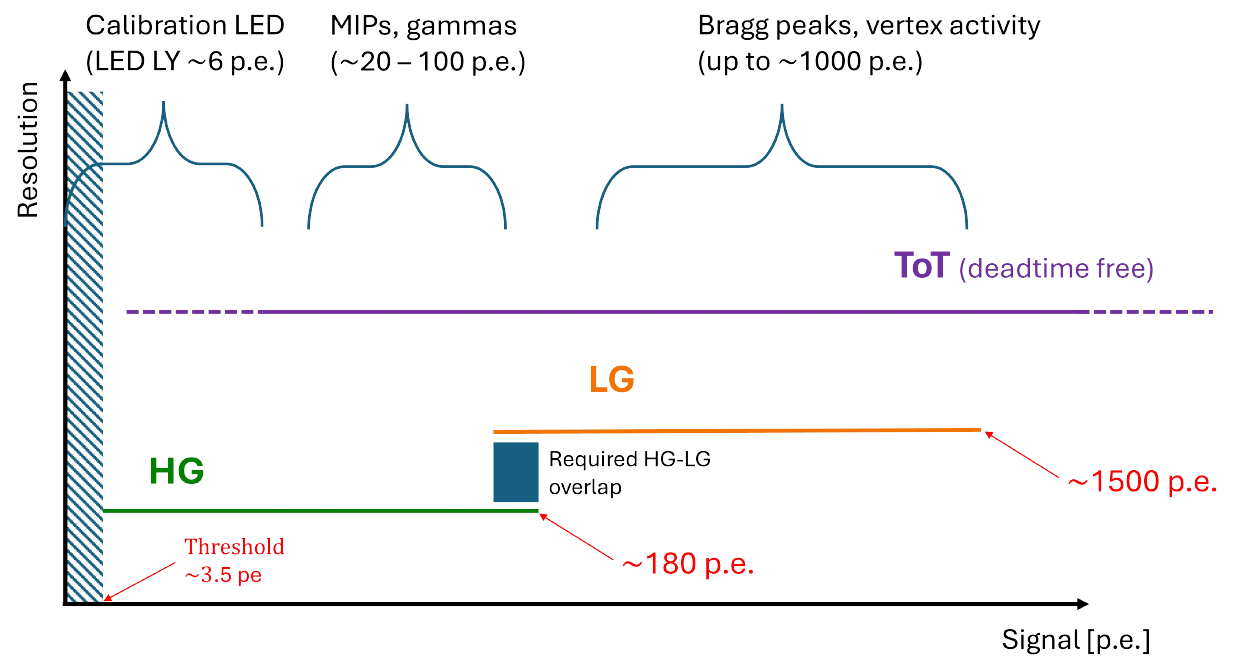}
\caption{Cross-calibration of the LG, HG, and TOT signals.
Top left: HG vs LG in ADC units.
Top right: HG in ADC units vs TOT in timing ticks where each tick is 1.25~ns.
Bottom left: LG in ADC units vs TOT in timing ticks.
}
\label{fig:calibration_charge_cosmics}
\end{figure}

The response of the low gain (LG), high gain (HG), and time over threshold (TOT) electronics signals were cross-calibrated. Fig.~\ref{fig:calibration_charge_cosmics} shows the S-curves obtained for the three combinations. The LG gain is calibrated with respect to the HG gain up to around 180~p.e. In this region, the linearity between the HG and LG is better than 2\%.

\begin{figure}[h!]
    \centering
    \includegraphics[width=0.8\textwidth]{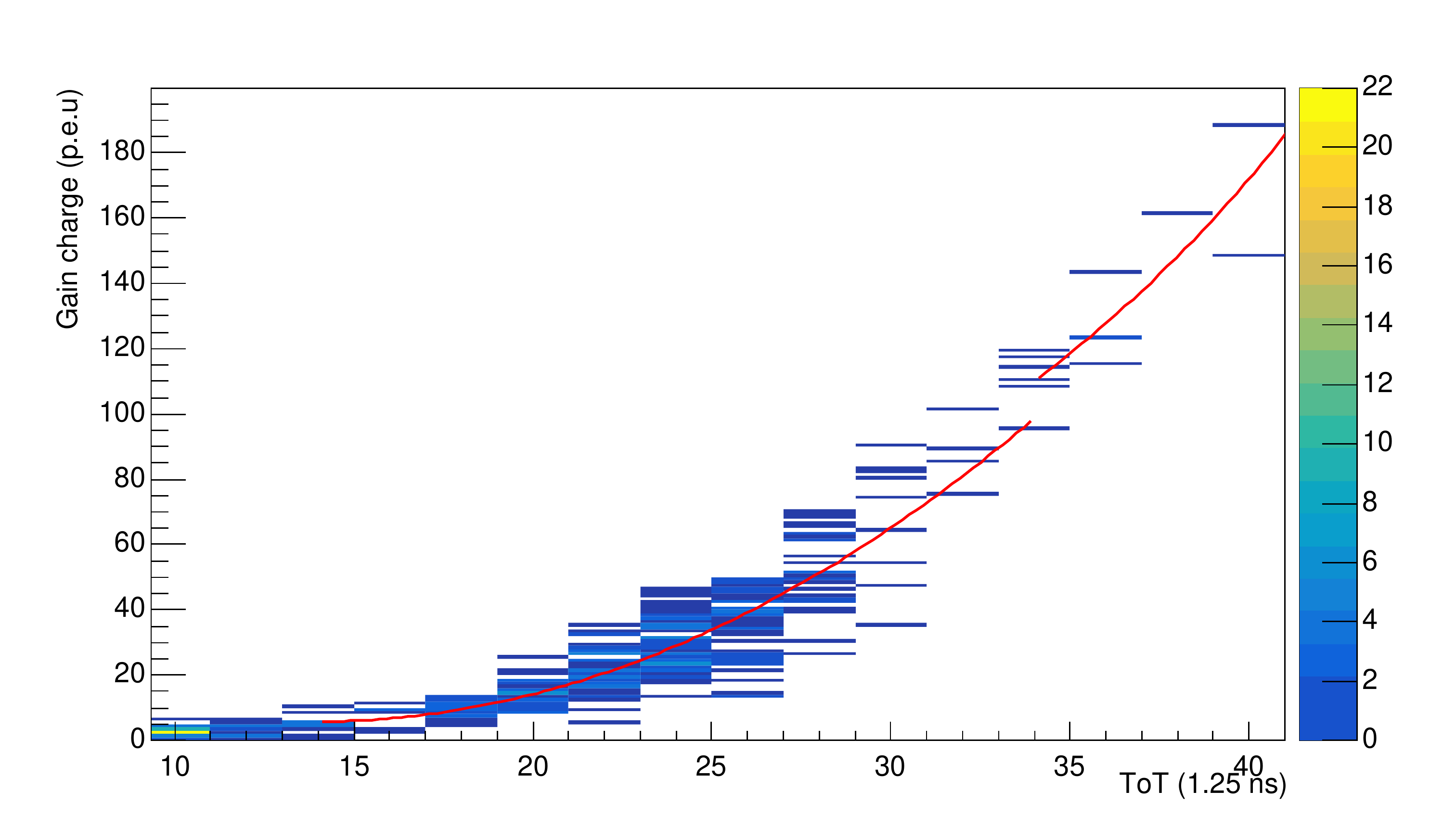}
    \caption{Hit charge from gain calibration vs TOT from June 2024 beam data. The red line is the fitted calibration curve.}
    \label{fig:totsinglechannel}
\end{figure}

The TOT signal is calibrated with respect to the hit charge obtained by the gain calibration. After the HG and LG calibration, the hit charge that has matched TDC and ADC can be calculated. TOT is then used to calculate the charge for hits that do not have matched ADC and TDC (where HG and LG calibration fails). In the high-statistics, low charge region (17.5~ns~$\leq$~ToT~$<$~42.5~ns), the relation between ADC charge and ToT is modelled as a quadratic function $q=a(\text{ToT}-14)^2+b$, fitting parameters $a$ and $b$ to cosmic and beam data samples for each readout channel. In the high charge region (ToT~$\geq$~42.5~ns), we assume an exponential relation $q=\exp[c(\text{ToT}-42.5~\text{ns})+d]$, extracting $c$ and $d$ globally for all channels. An example of the ToT calibration is shown in Fig.~\ref{fig:totsinglechannel}.

\subsection{Detector light yield and WLS fibre attenuation length}
\label{sec:light_yield}

\begin{figure}[h!]
\centering
\includegraphics[width=0.45\columnwidth]{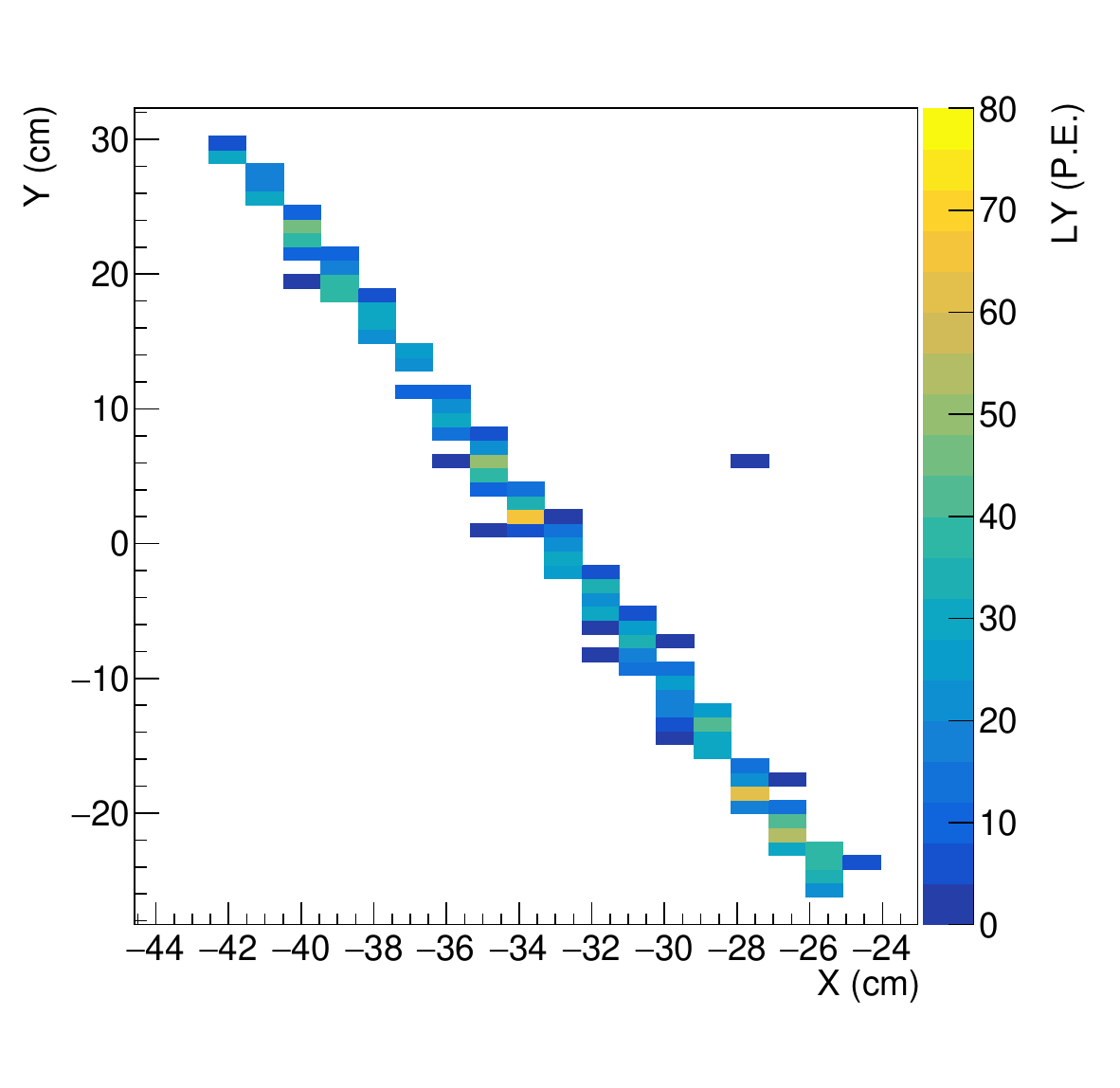}
\includegraphics[width=0.45\columnwidth]{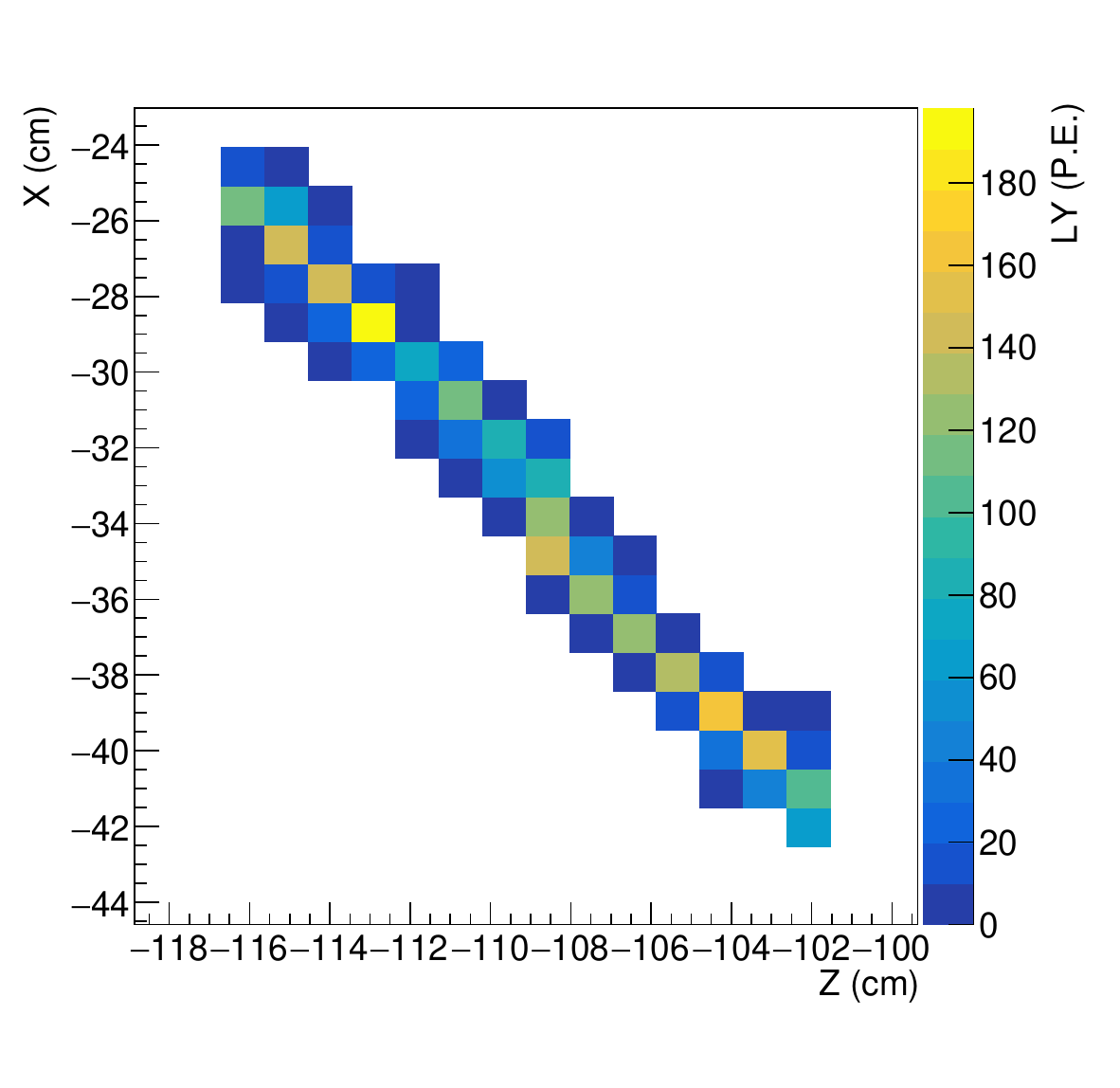}
\includegraphics[width=0.45\columnwidth]{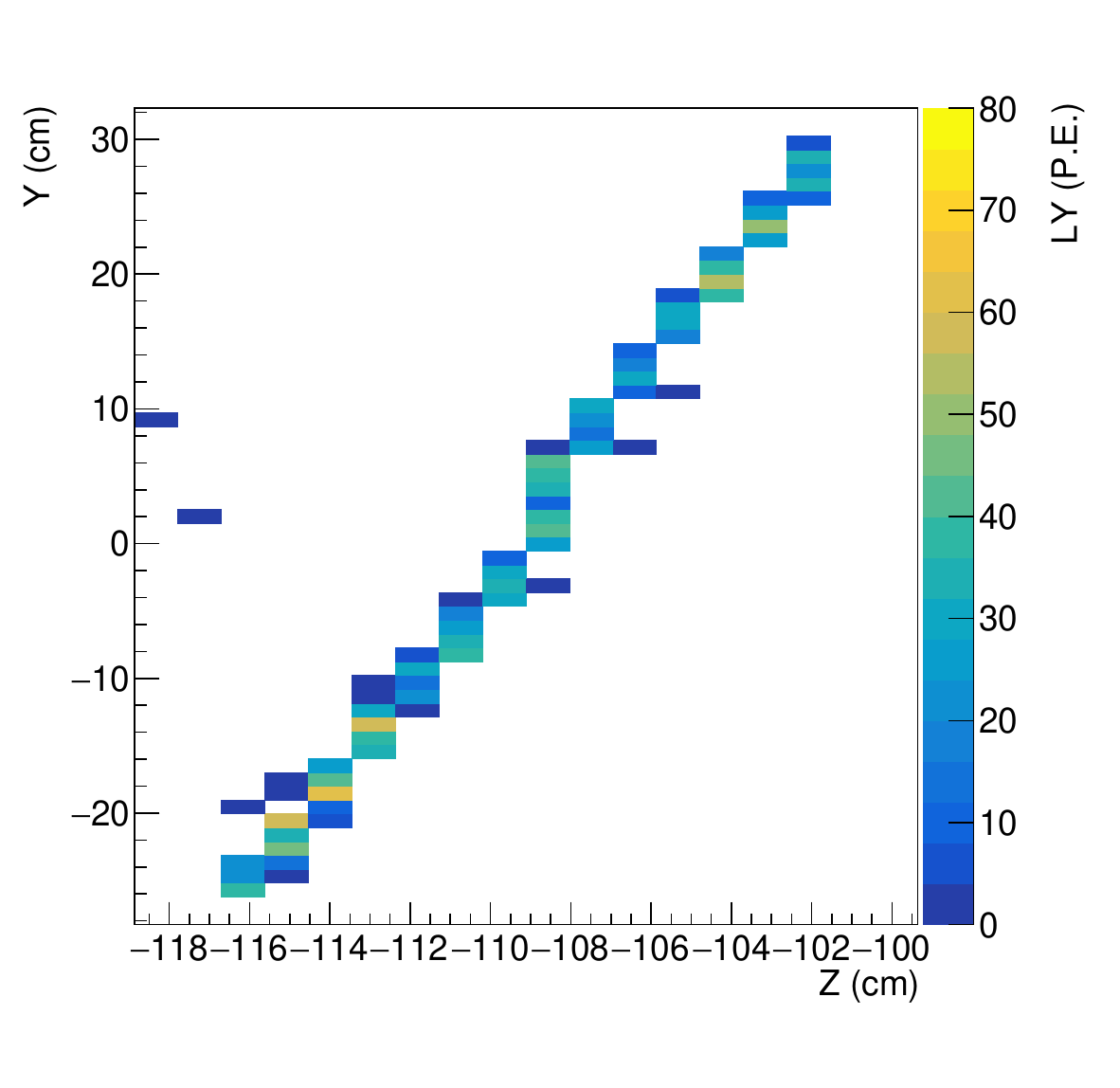}
\includegraphics[width=0.45\columnwidth]{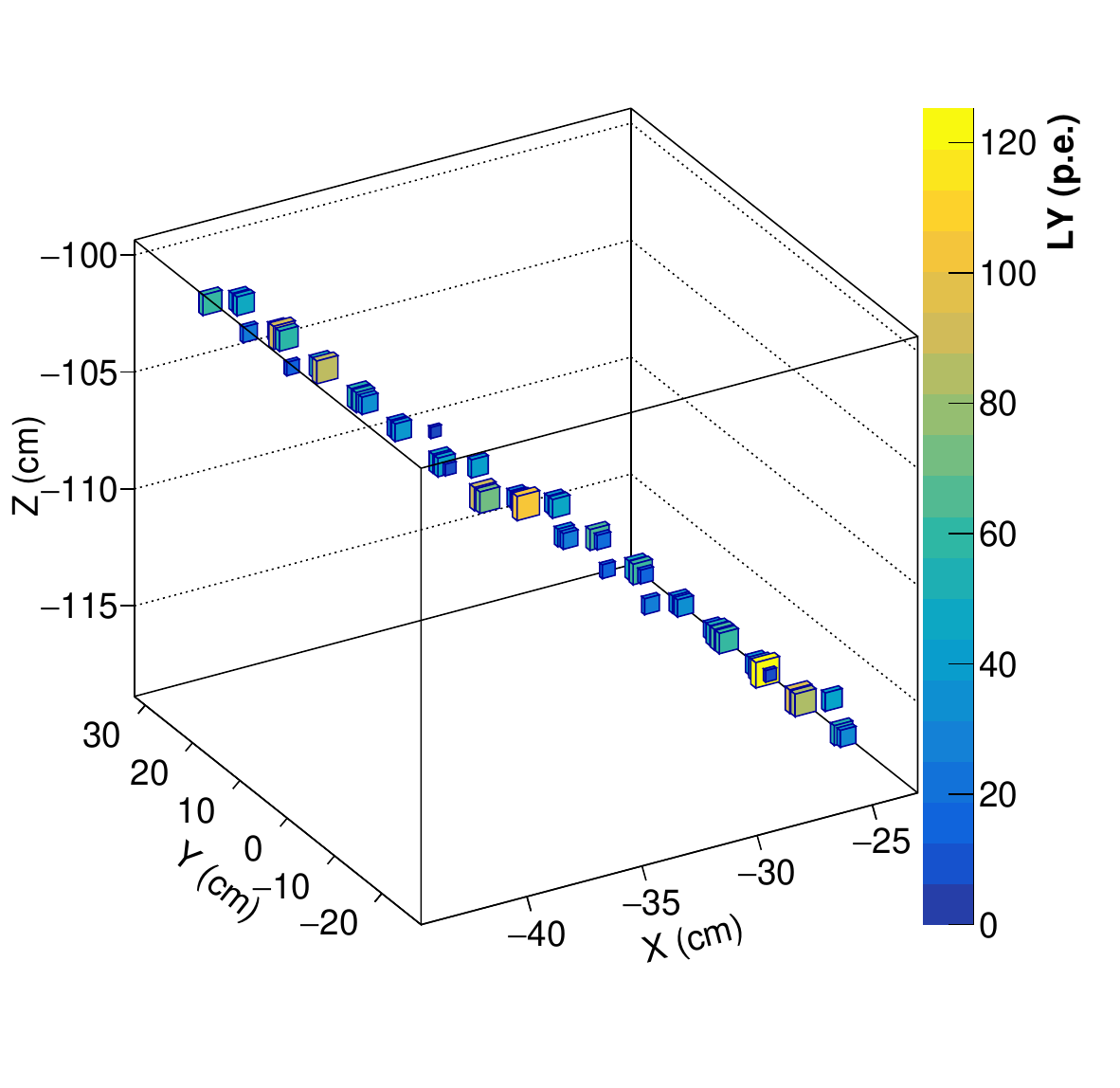}
\caption{
Example of a cosmic ray event selected for the SuperFGD response analysis. The 2D hits in the three readout views are shown. The reconstructed 3D event is shown on the bottom right. It is obtained after rejecting hits with a charge lower than 3.5~p.e. as well as hits with a charge smaller than 10\% of the adjacent highest charge one. The cube size in the 3D event display is proportional to the light yield.
}
\label{fig:event_cosmic_ray}
\end{figure}

The SuperFGD absolute scintillation light yield was characterised by analysing the selected sample of cosmic ray data collected in July 2024. An example of a selected cosmic ray event is shown in Fig.~\ref{fig:event_cosmic_ray}. 
A hit on a single fibre gives a position in two coordinates and is therefore referred to as a 2D hit. Hits that are matched in all three dimensions are referred to as 3D hits.

\begin{figure}[h!]
\centering
\includegraphics[width=0.8\textwidth]{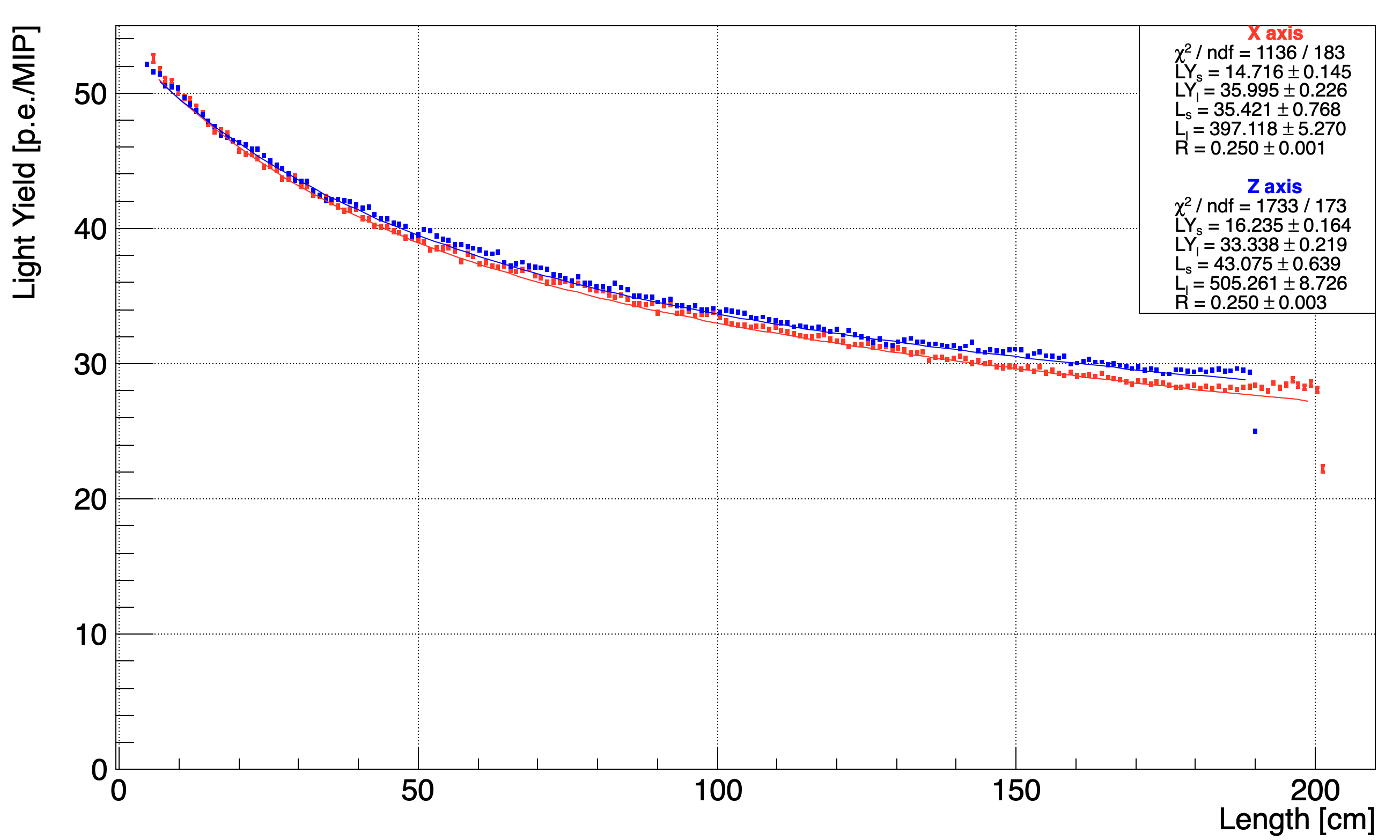}
\caption{
Light yield distribution from cosmic muons measured by all the WLS fibres along the X and Z axes (across horizontally and along  the main neutrino beam axis).
}
\label{fig:attenuation_length}
\end{figure}

Owing to the 3D segmentation and the three 2D projections in the SuperFGD, the light yield of each scintillator cube can be measured from the reconstructed track of single cosmic rays. The light yield is measured independently for each cube along a certain WLS fibre. Its position is provided by the other two orthogonal fibres crossing the same cube. Therefore, it is possible to measure the attenuation length for each WLS fibre of the SuperFGD. This allows potential non-uniformities between the response of different readout channels to be visualised, and possible failures to be identified. An example of an attenuation curve is shown in Fig.~\ref{fig:attenuation_length}.

The attenuation length is measured for each WLS fibre by fitting the light yield ($LY$) as a function of the cube position $x$ along the fibre from the MPPC with:
\begin{align}
     LY &=  LY_S\times exp{(-\frac{x}{L_S})} + LY_L\times exp{(-\frac{x}{L_L})} \\ \nonumber
     &+ R\times[LY_S\times exp{(-\frac{2L - x}{L_S})} +LY_L\times exp{(-\frac{2L - x}{L_L})} ],
    \label{eq:attenuation_length}
\end{align}
where $L$ is the fibre length, $L_S$ and $L_L$ are the short and long attenuation lengths, respectively.

\subsection{Signal crosstalk}
\label{sec:crosstalk_signal}

Although optical isolation is provided by the outermost white surface of the scintillator cubes, there is still a small probability for scintillation photons to leak to an adjacent cube. The result is the presence of hits surrounding tracks produced by ionising particles, as can be seen in Fig.~\ref{fig:event_cosmic_ray}.

Another possible source of apparent cube-to-cube crosstalk can be induced by channel-to-channel electronics crosstalk. Measurements performed on a test bench showed that this is limited to a maximum of 0.35\% (see Sec.~\ref{sec:electronics_xtalk}). Although this contribution to the signal crosstalk is negligible in the cosmic ray sample used to characterise the response of the SuperFGD, its impact can be non-negligible in neutrino interactions with the presence of highly-ionising particles such as low-momentum protons.

\begin{figure}[h!]
\centering
\includegraphics[width=0.48\columnwidth]{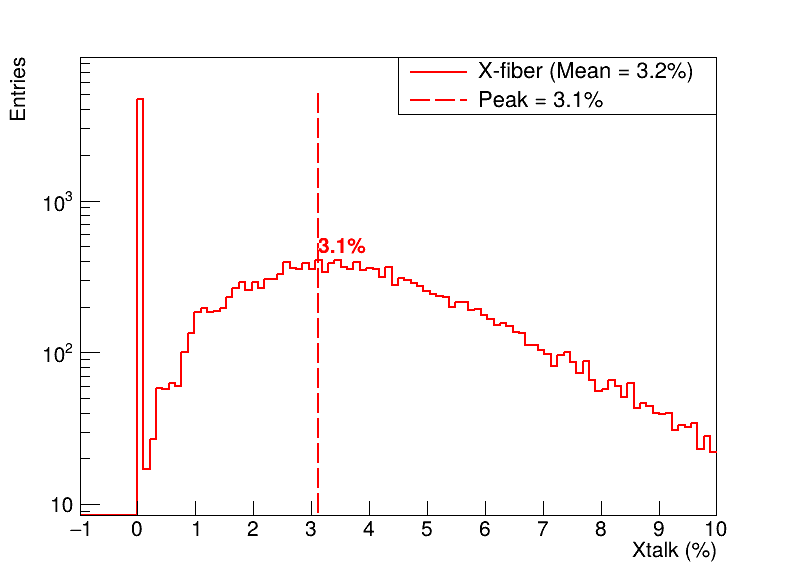}
\includegraphics[width=0.48\columnwidth]{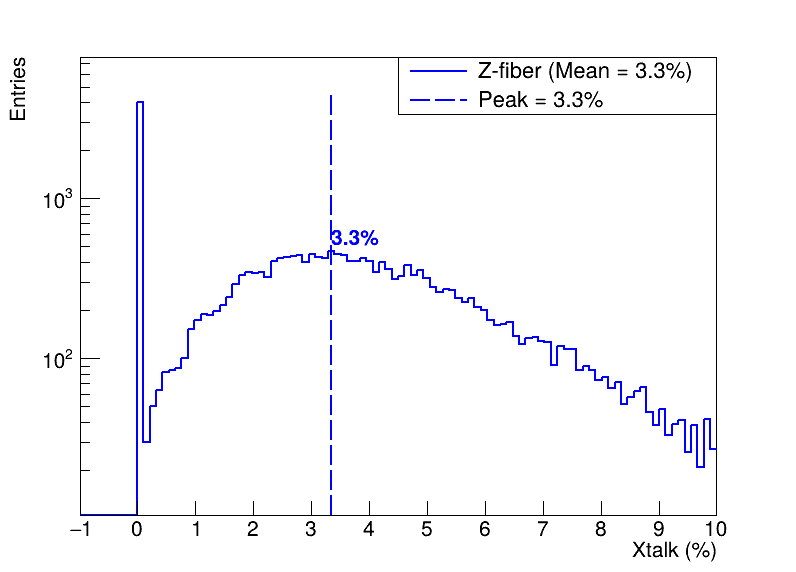}
\caption{
Percentage of hits with crosstalk signals in the X (left) and Z (right) dimensions. The two dimensions show a consistent crosstalk value of around 3\%.
}
\label{fig:cosmic-rays-crosstalk}
\end{figure}

The signal crosstalk has been estimated using a selection of vertical tracks in cosmic data. Crosstalk hits were selected by looking for hits in three consecutive Y layers with the same X or Z coordinate in a track. Signals in the neighbouring cubes of the second hit are assumed to come from crosstalk. Fig.~\ref{fig:cosmic-rays-crosstalk} shows the distribution of crosstalk signals selected. We find an average crosstalk of around 3\%, consistent with earlier prototype measurements.

\subsection{Time resolution}
\label{sec:time_resolution}

Since the SuperFGD active medium is a plastic scintillator, it can provide sub-nanosecond time resolution on the single-cube response, if instrumented with proper readout electronics. This allows, for the first time, the possibility of detecting neutrons produced by neutrino interactions and reconstructing their kinetic energy by measuring the time of flight.

A crucial step for the optimisation of the timing performance of the detector is the time calibration. The two main contributions to the deterioration of the time measurement are time walk effects and the mutual mis-synchronisation of readout channels (\textit{time offsets}). Both of these effects are corrected using datasets of pairs of hits highly correlated in time, or \emph{matching hit pairs}. A matching hit pair is a pair of MPPC hits sharing the origin cube of the triggering photon. Thanks to the peculiar geometry of the SuperFGD, it is possible to estimate the expected time difference between two hits, knowing the distances between the scintillating light emission point and the MPPC that collects the light. We can then define the difference between the measured and expected time differences $\Delta t_{12}$:
\begin{equation}
    \Delta t_{12} = \Delta t_{measured} - \Delta t_{expected}  = (t_1 - t_2) - (s_1 - s_2)/v, 
    \label{eq: deltat_ij}
\end{equation}
where $t_1$ and $t_2$ are the measured times, and $s_1$ and $s_2$ are the distances from the cube to the MPPCs for hit $1$ and $2$ respectively. We denote the speed of light in the fibres by $v$, which is taken as 16.7\,cm/ns in this analysis. Using this definition, the channel time offset calibration is performed using a newly developed Markov chain-based procedure \cite{JINST:2025timecalib}.

\begin{figure}[h!]
\centering
\includegraphics[width=0.45\columnwidth]{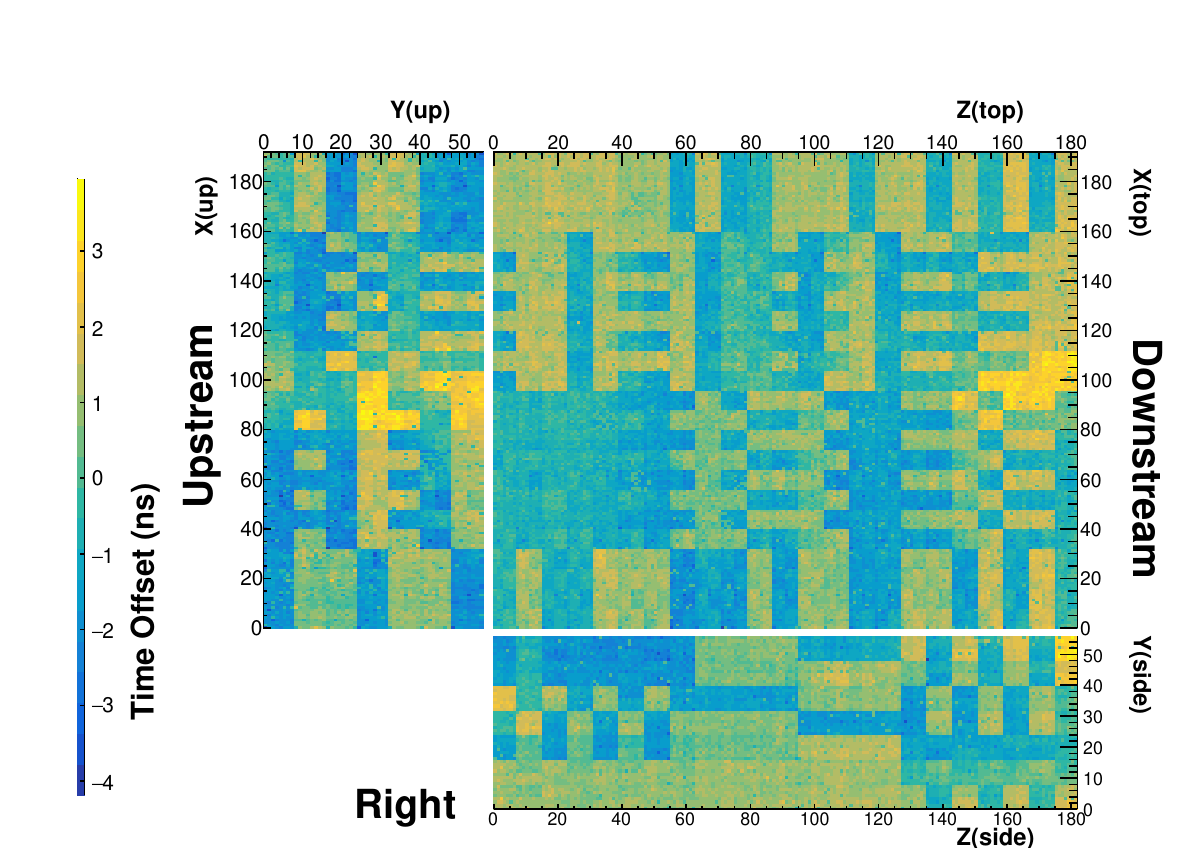}
\includegraphics[width=0.5\columnwidth]{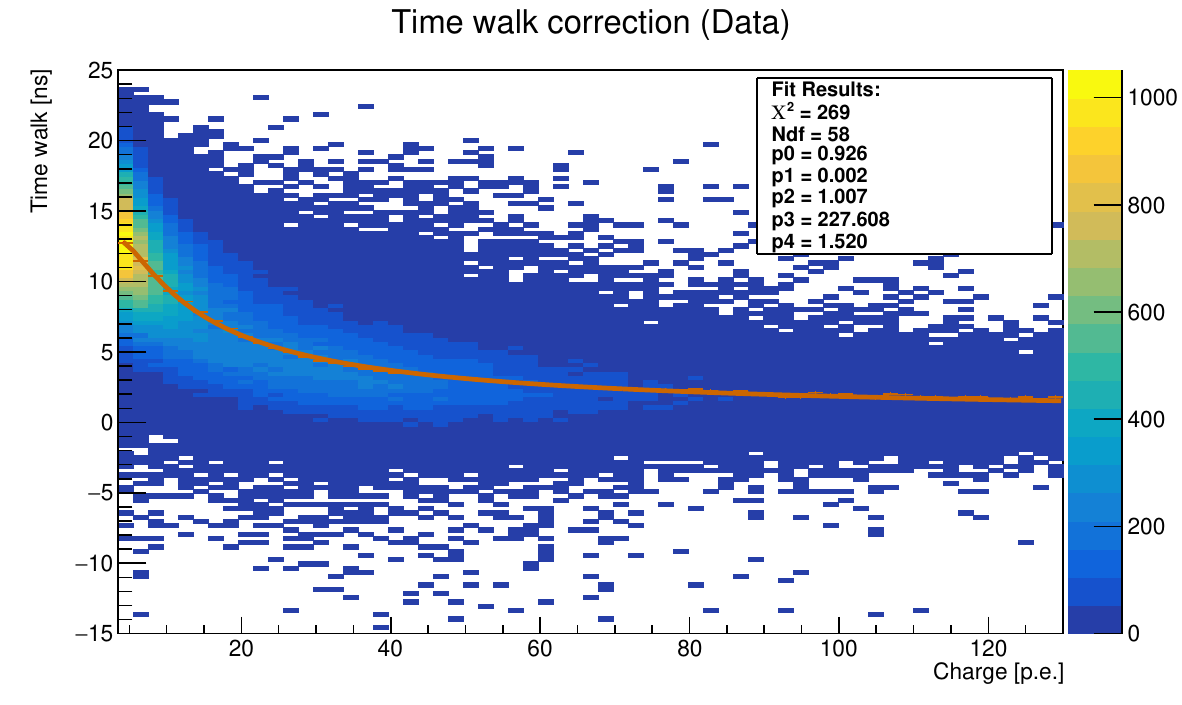}
\includegraphics[width=0.55\columnwidth]{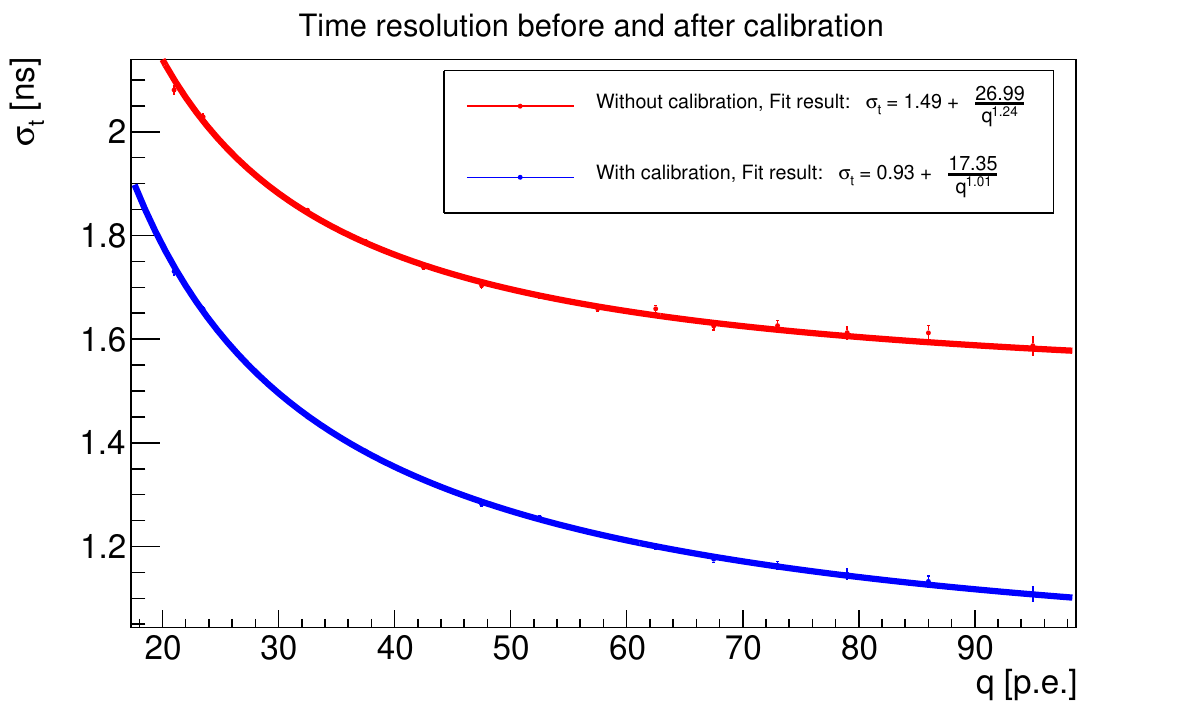}
\caption{
Top left: the time offsets found during calibration.
Top right: the distribution of the time difference between two simultaneous but independent 2D hits as a function of the lowest 2D charge. Events with a charge difference between the 2D hits larger than 120 p.e. were selected.
Bottom: the time resolution as a function of the lowest 2D hit charge before (red) and after (blue) calibration.
The time calibration was performed using the selected cosmic ray data sample. The time propagation of the cosmic rays in the detector as well as of the scintillation light from the cube to the SiPM was accounted for.
}
\label{fig:time-resolution-channel}
\end{figure}

Fig.~\ref{fig:time-resolution-channel} 
shows the obtained time offsets estimated for each channel with this procedure, highlighting that the main contribution to the mis-synchronisation comes from FEB-to-FEB variation.

The time walk has been characterised as a function of the signal charge measured with the CITIROC peak detector, modeling the charge-dependent time delay with a custom function. The time difference between two matching hits is plotted as a function of the lowest deposited charge, for a charge difference larger than 120 p.e.

The function used to fit the data and calibrate the time response of each channel is 
\begin{align}
    \Delta t (q) = \frac{p_0}{\log (p_1 \cdot q + p_2)} - \frac{p_3}{q+p_4}.
\end{align}
The fitted values of the parameters are shown in Fig.~\ref{fig:time-resolution-channel}. The time resolution for a single readout channel is measured from the width of the distribution of time difference $\Delta t_{12}$ between two 2D hits from a sample of cosmic rays. Under the assumption that the spread of $\Delta t_{12}$ is dominated by the time resolution, we can assume that $\sigma_t\simeq\sigma_{\Delta t_{12}}/\sqrt2$. The impact of the calibration on the time resolution is shown in Fig.~\ref{fig:time-resolution-channel} as a function of the charge of the 2D hit recorded in the MPPC. Since this plot considers pairs of 2D hits with a similar deposited charge, the time walk effect is expected to be underestimated. After the calibration, the time resolution for a single channel is about 1.5~ns for an energy loss typical of a MIP crossing a cube at about 1~m from the MPPC ($\sim 30~\text{p.e.}$).

At the time of this analysis, not enough statistics were available in the cosmic ray data sample to correct for the time walk for single or sub-group of WLS fibres, further reducing the time smearing introduced by this effect. This will be performed in future when sufficient statistics are available. Moreover, a more efficient sampling of electronics time, applied also on the clock falling edge, will be implemented in the firmware. Hence, the electronics time resolution is expected to be improved by a factor two.

The precise determination of the time of a particle track is important for the measurement of the time of flight of neutrons in the SuperFGD as well as for charged particle identification (PID) in combination with the TOF detector, to reject background hits by coincidence with respect to the neutrino interaction time, to determine the sense of a particle with respect to its origin.

\begin{figure}[h!]
\centering
\includegraphics[width=0.45\columnwidth]{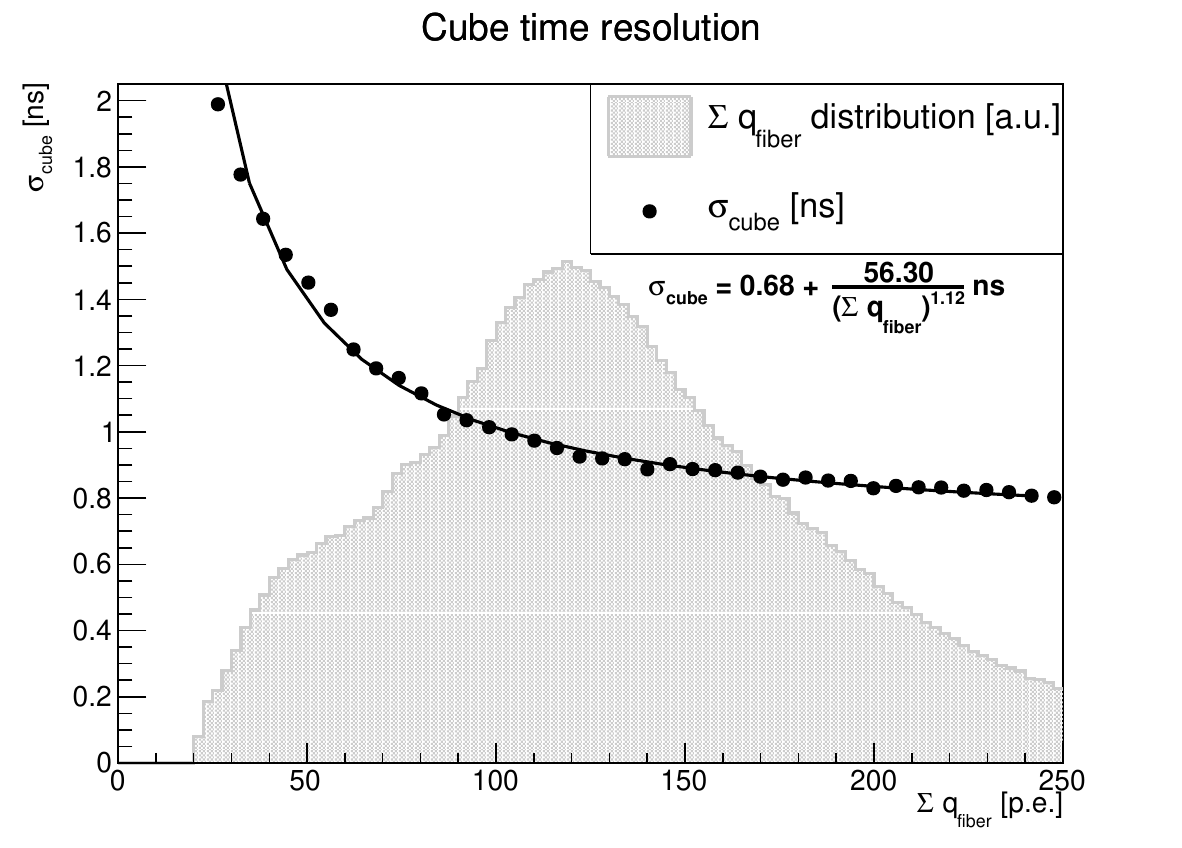}
\includegraphics[width=0.45\columnwidth]{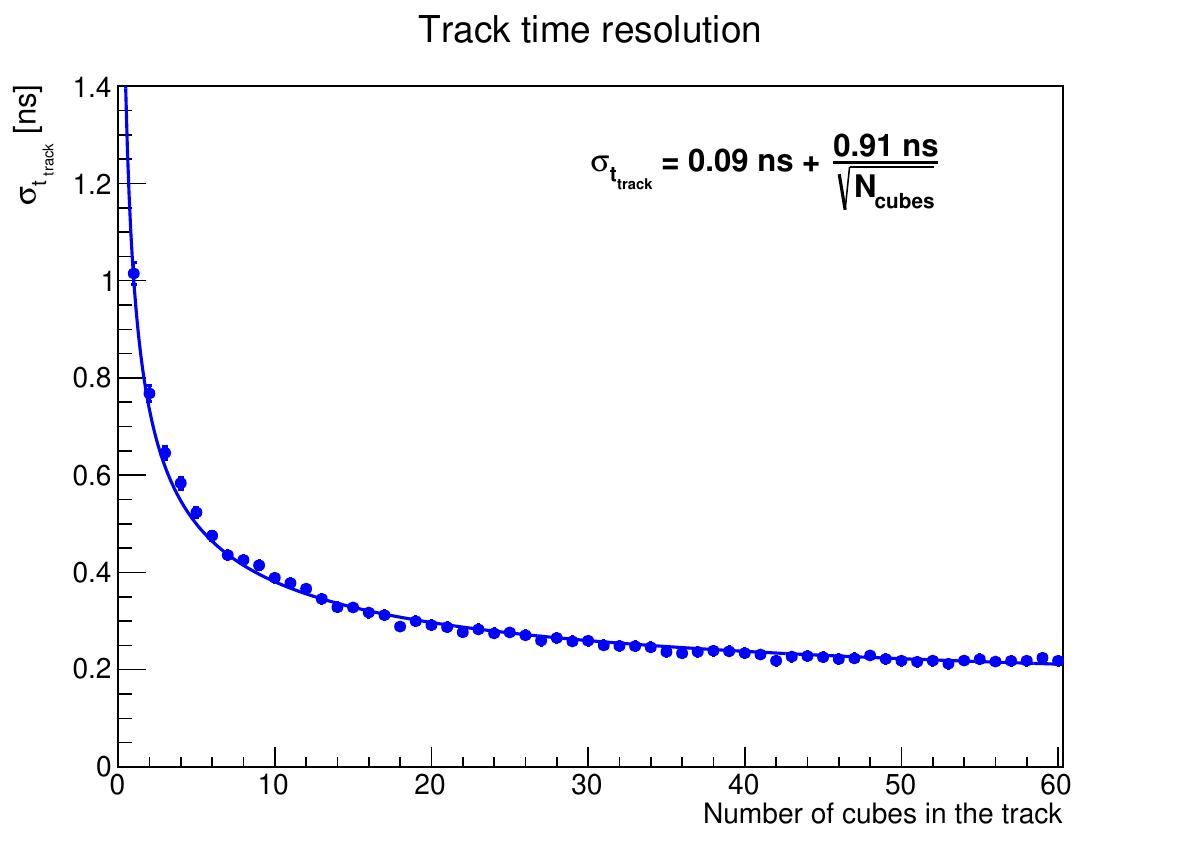}
\caption{
Left: in black, the time resolution of a single cube obtained from a sample of cosmic tracks from a February 2025 cosmic run; in grey, the distribution of the sum of the fibre charges.  Right: the time resolution of a single particle track as a function of the number of cubes forming the track, obtained from a selected sample of cosmic rays.}
\label{fig:time-resolution-single-cube}
\label{fig:time-resolution-t0}
\end{figure}

The time resolution of a single cube as a function of the cube charge is shown in Fig.~\ref{fig:time-resolution-single-cube}. It was obtained from a sample of cosmic tracks from the February 2025 operations period. Cube hits are formed exclusively when the three WLS fibres crossing a cube overcome the charge threshold of 3~p.e. The cube time was computed as the weighted mean of the times measured by the corresponding fibres, while the cube charge was computed as the sum of the fibre charges. The cube time resolution is obtained from the distribution of the difference between the reconstructed times $t_1$ and $t_2$ of two random cubes belonging to the same track, corrected for the speed of the particle. By selecting cubes having similar charges, the cube time resolution is estimated as the standard deviation of the distribution $t_1-t_2$, divided by $\sqrt{2}$. The cube time resolution for MIPs (approximately 120 p.e.) is about 0.95~ns.

The time resolution of single tracks produced by selected cosmic rays was studied as a function of the number of cube hits forming the track. The track 3D hits are sub-divided into two different groups (odd and even hits), from both of which the track time is estimated obtaining $t_{odd}$ and $t_{even}$. The time resolution for a track composed of $N$ hits is estimated as the standard deviation of the distribution of $t_{odd}-t_{even}$ for a track of $2N$ hits, divided by $\sqrt{2}$, since under the assumption that $t_{odd}$ and $t_{even}$ are independent, we have:
$\frac{1}{2}\sigma^2[t_{odd}-t_{even}]=\sigma^2[t_{track}(N)]$. The result of the analysis on time-calibrated cosmic samples is shown in Fig.~\ref{fig:time-resolution-t0}. A time resolution better than 400~ps is achieved for long tracks, with a maximum time resolution of 1~ns for single-cube clusters. As expected, the track time resolution follows an inverse square root dependence on the number of cubes.

\subsection{Neutrino interactions in the SuperFGD and particle identification}
\label{sec:neutrino_interactions}

The SuperFGD allows the identification of the interaction topology by tracking particles in three dimensions and measuring their energy loss with sub-nanosecond timing.
 
A detailed simulation of the detector has been developed and validated on data. The full reconstruction of neutrino interactions is used for the physics analyses: first a 3D image is obtained by matching the three 2D views; then a ``charge sharing'' algorithm assigns the energy loss by each ionising particle to a different cube; the grouping of the 3D hits into different clusters is done with DBSCAN~\cite{dbscan}; after ordering the clusters into different branches with a minimum spanning tree, and selecting those that are track-like (i.e. longer than three cubes), kinks and vertices are identified and the particle position along its track is fitted with a Sequential Importance Resampling particle filter. We defer to a future article for more details about both the detector simulation and the reconstruction algorithms adopted for the analysis of the SuperFGD data. Monte Carlo studies based on Geant4~\cite{GEANT4:2002zbu} show that, thanks to the combined analysis of the range and the Bragg peak, protons produced by muon neutrino interactions can be reconstructed down to about 300~MeV/$c$ momentum and with a resolution better than 5\% at 400 MeV/$c$, and identified with a purity up to 99\%.

In this section, a review of typical neutrino interaction topologies as well as the features provided by the SuperFGD, including particle identification (PID), is provided.

\subsubsection{Neutrino final states with a proton stopping in the SuperFGD}
\label{sec:neutrino_interactions_protons}

\begin{figure}[h!]
\centering
\includegraphics[width=0.65\columnwidth]{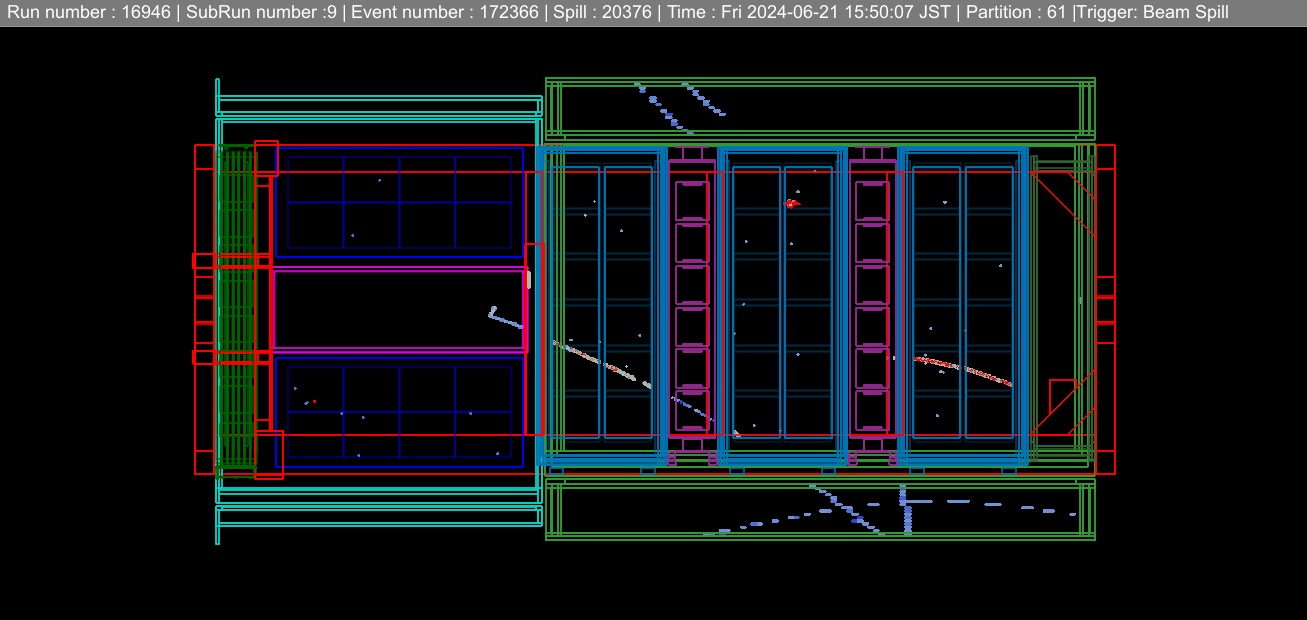}
\includegraphics[width=0.90\columnwidth]{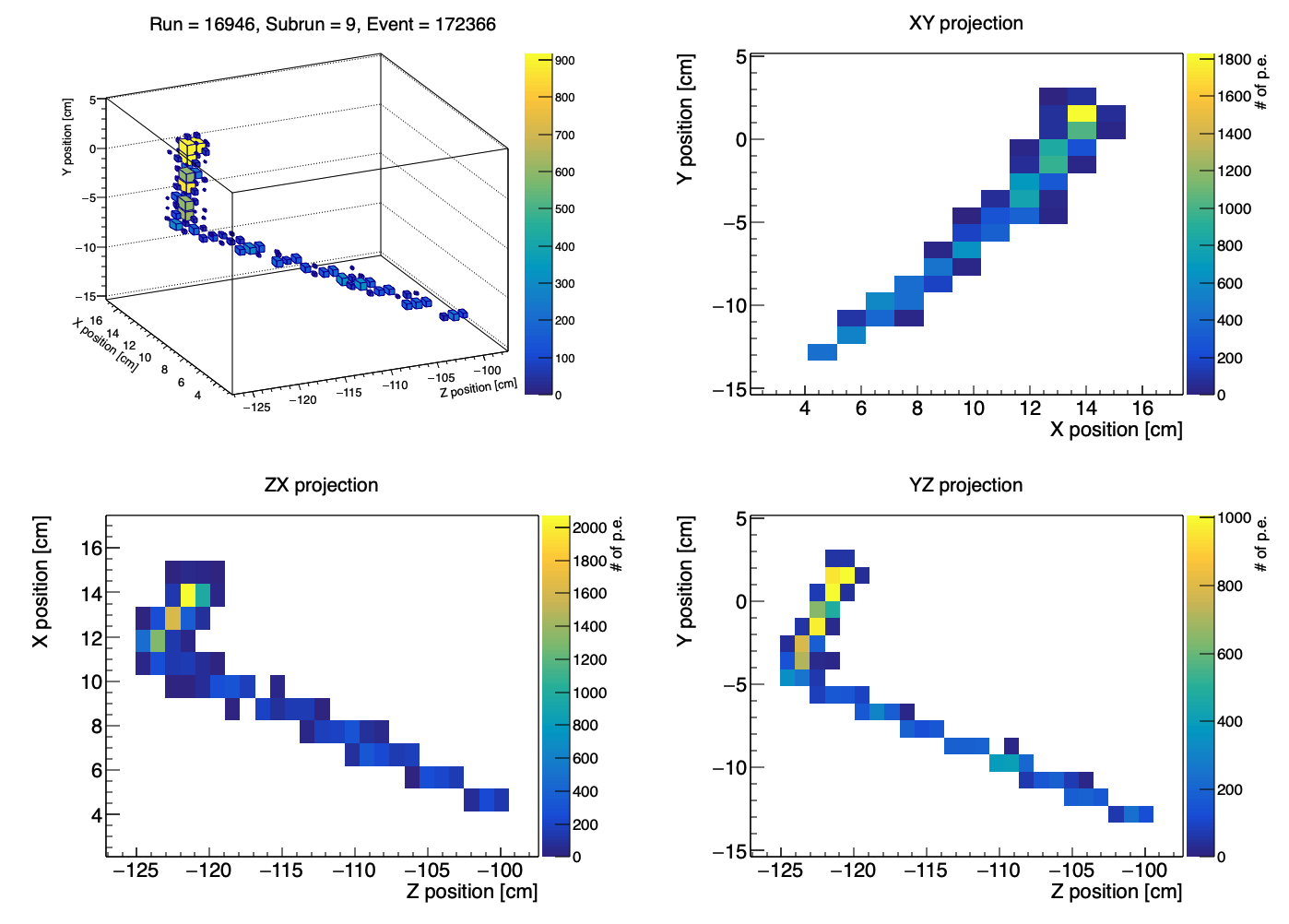}
\caption{
Candidate charged-current (CC) neutrino interaction in the SuperFGD collected during the June 2024 run. The final state comprises a reconstructed muon also detected in the downstream TPC and one reconstructed stopping proton.
Top: event display of the whole ND280.
Bottom: neutrino interaction reconstructed in X, Y, Z coordinates (top left) from the three 2D views of the SuperFGD, shown in the other figures on the top right (XY view), bottom left (XZ view) and bottom right (YZ view).
The colours in the 3D event show the reconstructed energy loss by the particle in arbitrary units (AU).
The size of each voxel is proportional to the energy loss in a specific cube.
The colours in the 2D views are in units of p.e. and the size of each bin is the size of the project cube.
}
\label{fig:neutrino_event_1mu1prot}
\end{figure}

\begin{figure}[h!]
\centering
\includegraphics[width=0.65\columnwidth]{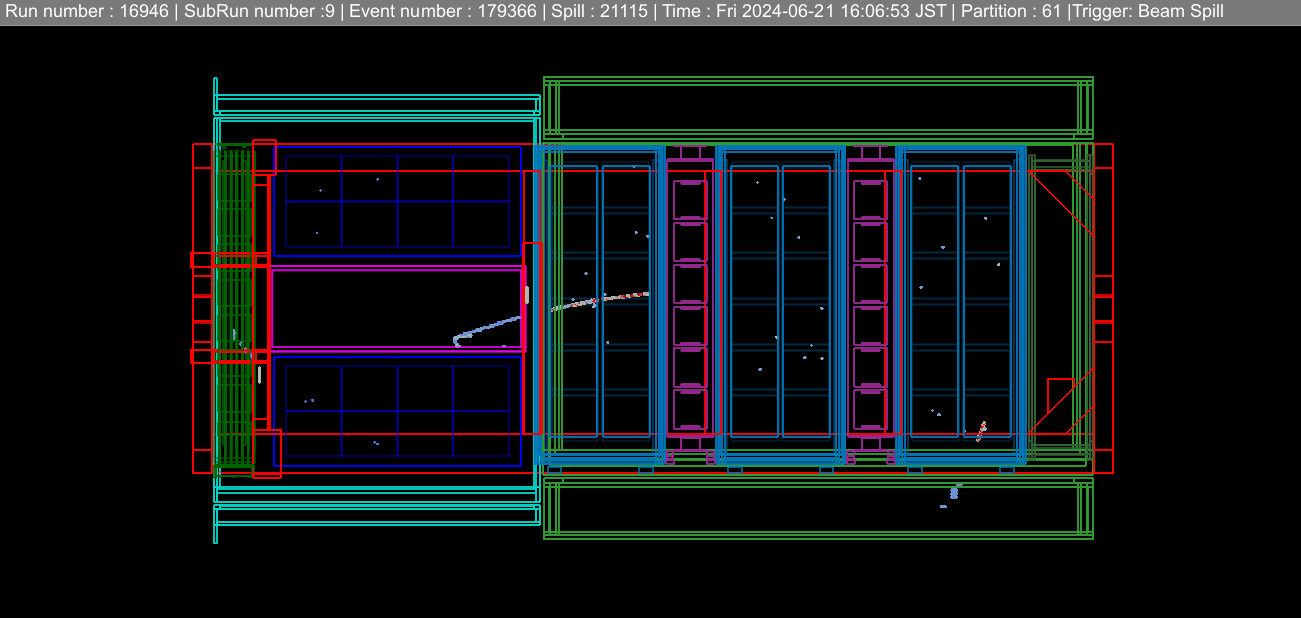}
\includegraphics[width=0.75\columnwidth]{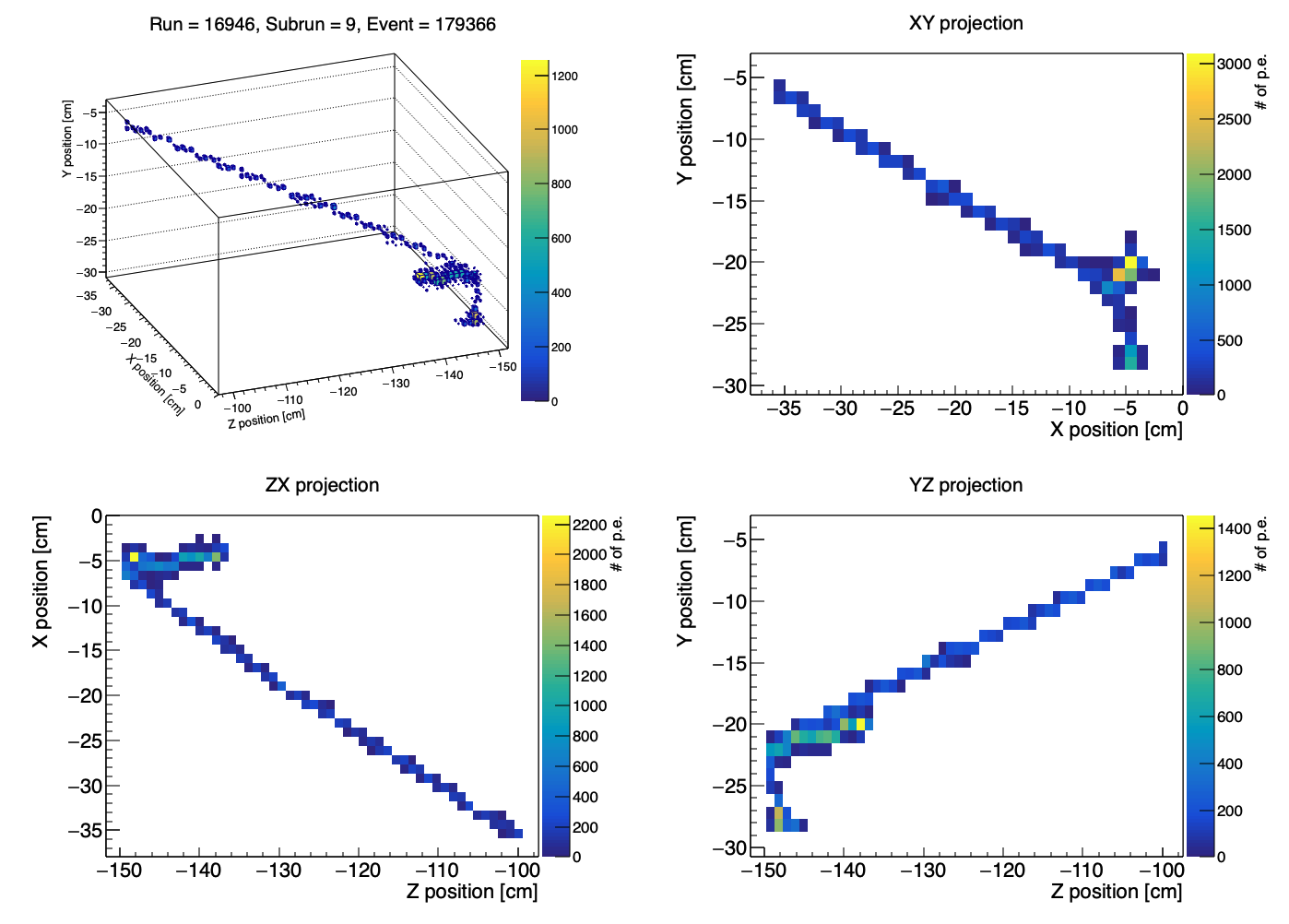}
\caption{
Candidate charged-current (CC) neutrino interaction in the SuperFGD collected during the June 2024 run. The final state comprises a reconstructed muon also detected in the downstream TPC and two reconstructed stopping protons. 
Top: event display of the whole ND280.
Bottom: neutrino interaction reconstructed in X, Y, Z coordinates (top left) from the three 2D views of SuperFGD, shown in the other figures on top right (XY view), bottom left (XZ view) and bottom right (YZ view).
The colours in the 3D event show the reconstructed energy loss by the particle in arbitrary units (AU).
The size of each voxel is proportional to the energy loss in a specific cube.
The colours in the 2D views are in units of p.e. and the size of each bin is the size of the project cube.
}
\label{fig:neutrino_event_1mu2prot}
\end{figure}

Owing to its fine 3D segmentation, the SuperFGD identifies the interaction vertex and distinguishes and tracks charged particles produced at any angle, including protons down to 300 MeV/$c$. In Fig.~\ref{fig:neutrino_event_1mu1prot} and Fig.~\ref{fig:neutrino_event_1mu2prot}, two neutrino interaction candidates with one reconstructed muon and, respectively, one and two reconstructed protons stopping in the SuperFGD are shown.

\subsubsection{Identification of stopping particles}
\label{sec:pid_stopping}

\begin{figure}[h!]
\centering
\includegraphics[width=0.5\columnwidth]{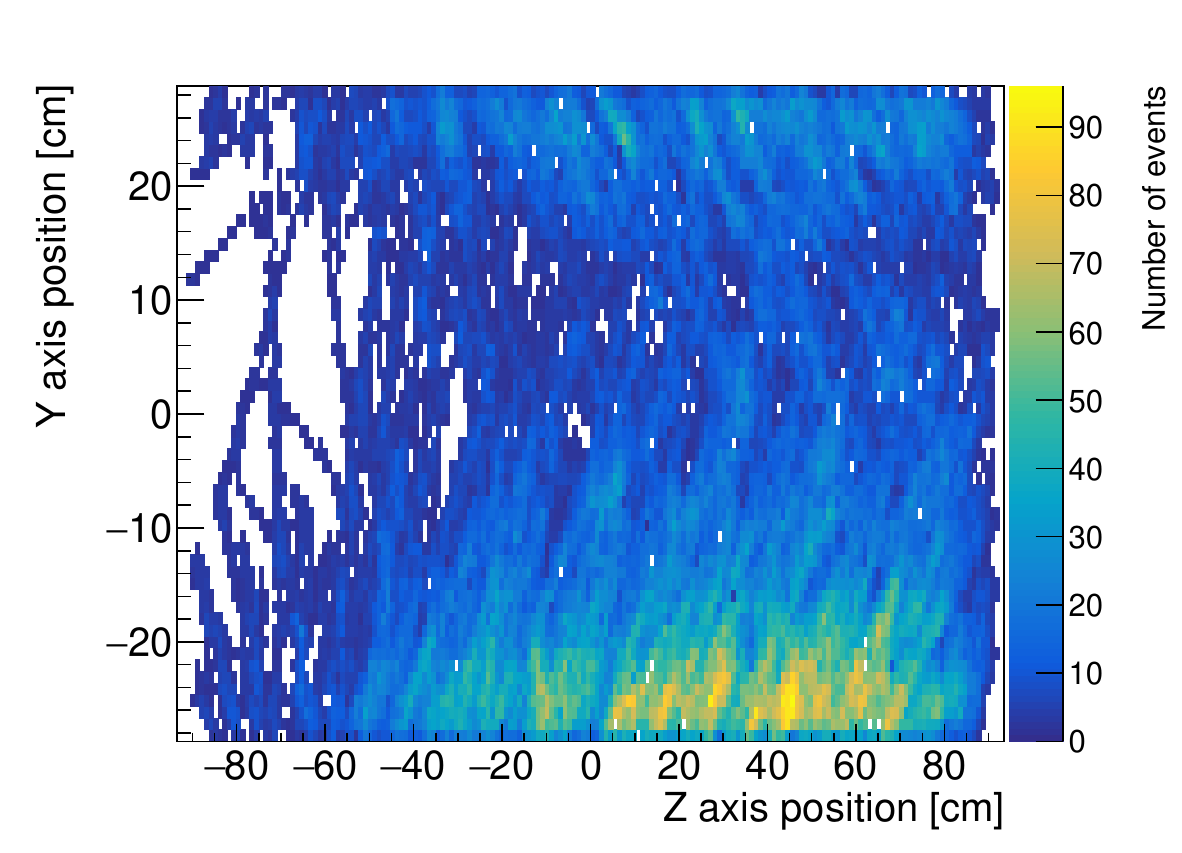}
\includegraphics[width=0.45\columnwidth]{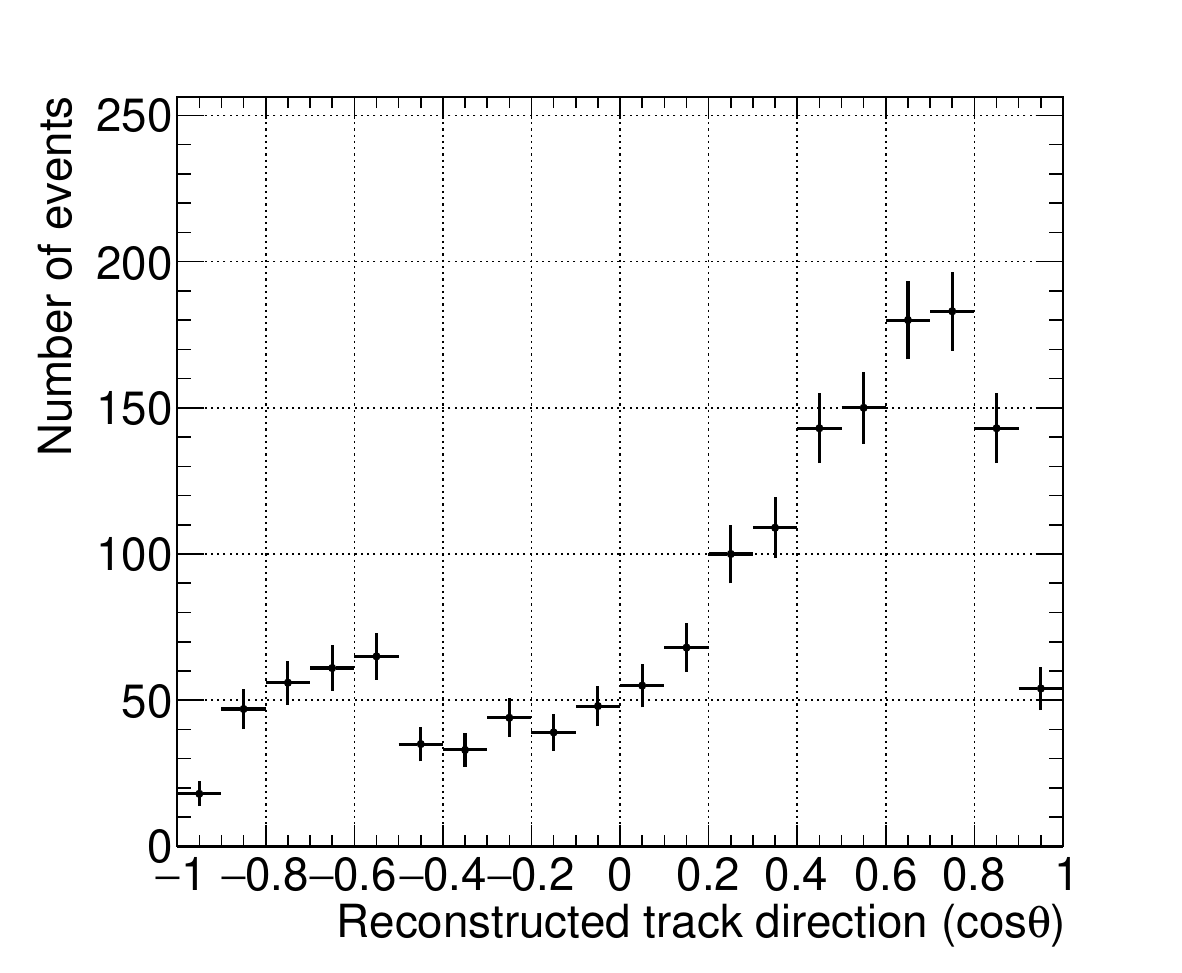}
\includegraphics[width=0.45\columnwidth]{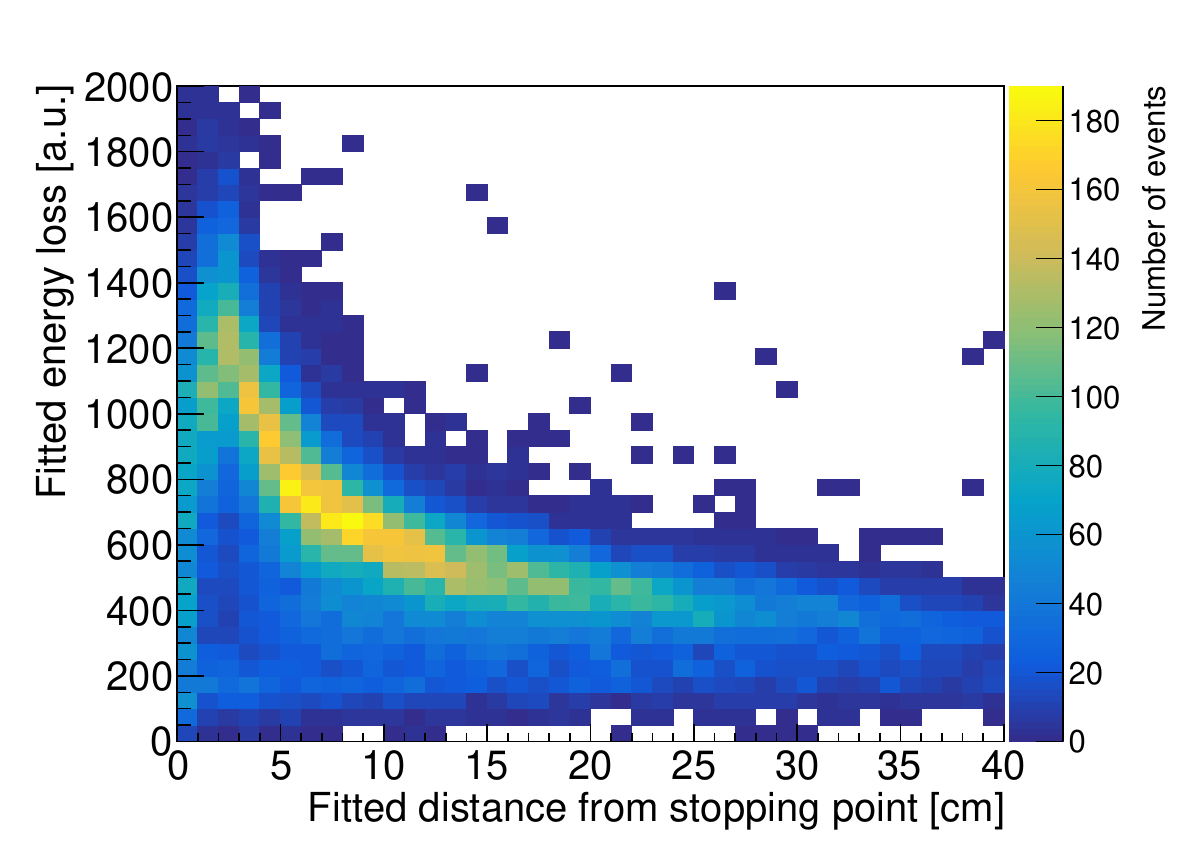}
\includegraphics[width=0.5\columnwidth]{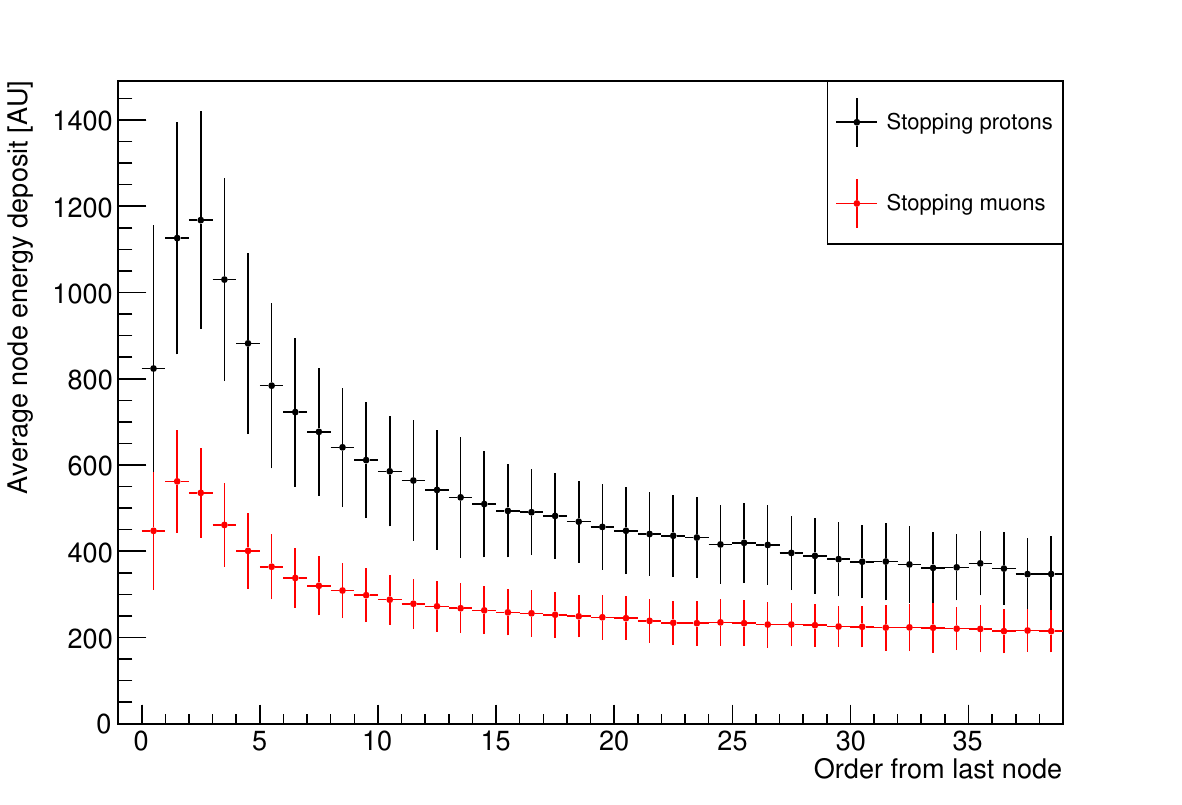}
\caption{
Top left: the accumulated tracks of particles selected in either the top or bottom high-angle TPC, entering and stopping in the SuperFGD are shown in the side (YZ) view.
Top right: distribution of the cosine of the angle of the selected protons with respect to the Z axis. 
Bottom left: distribution of the fitted energy loss by the selected proton stopping in the SuperFGD fiducial volume (arbitrary units) along its trajectory as a function of the fitted distance from the stopping position.
Bottom right: average fitted energy loss by the selected proton (black) or muon (red) stopping in the SuperFGD fiducial volume (arbitrary units) along its trajectory as a function of the fitted distance from the stopping position. The bars show the root mean square of the energy loss distribution in the corresponding distance bin. The graph of the stopping protons is extracted directly from the bottom left plot.
}
\label{fig:pid-particles-all}
\label{fig:pid_stopping_particles}
\end{figure}

The capability of the SuperFGD to correctly identify charged particles (muons, pions, protons and electrons) stopping within its fiducial volume has been quantified. Particles selected and identified in either the top or bottom high-angle TPC of ND280 and stopping in the SuperFGD were selected. The energy loss in the scintillator cubes along the fitted particle trajectory is shown in Fig.~\ref{fig:pid_stopping_particles}. The Bragg peak structure produced by protons stopping in the SuperFGD is clearly visible in the selected data sample showing the excellent PID performance. The relatively small smearing (two cubes at most) on the left side of the Bragg peak is due to a combination of the variable proton path in last cube and optical crosstalk. The few entries peaked around 200~AU are due to minimum ionising particles selected as protons. Fig.~\ref{fig:pid_stopping_particles} also shows how precisely the SuperFGD can distinguish protons from muons at the typical T2K energies by measuring the stopping power along their trajectory. Such separation is especially enhanced when the effect of the Bragg peak becomes more pronounced.

\subsubsection{Neutrino final states with pions stopping in SuperFGD}
\label{sec:neutrino_interactions_pions}

\begin{figure}[h!]
\centering
\includegraphics[width=0.45\columnwidth]{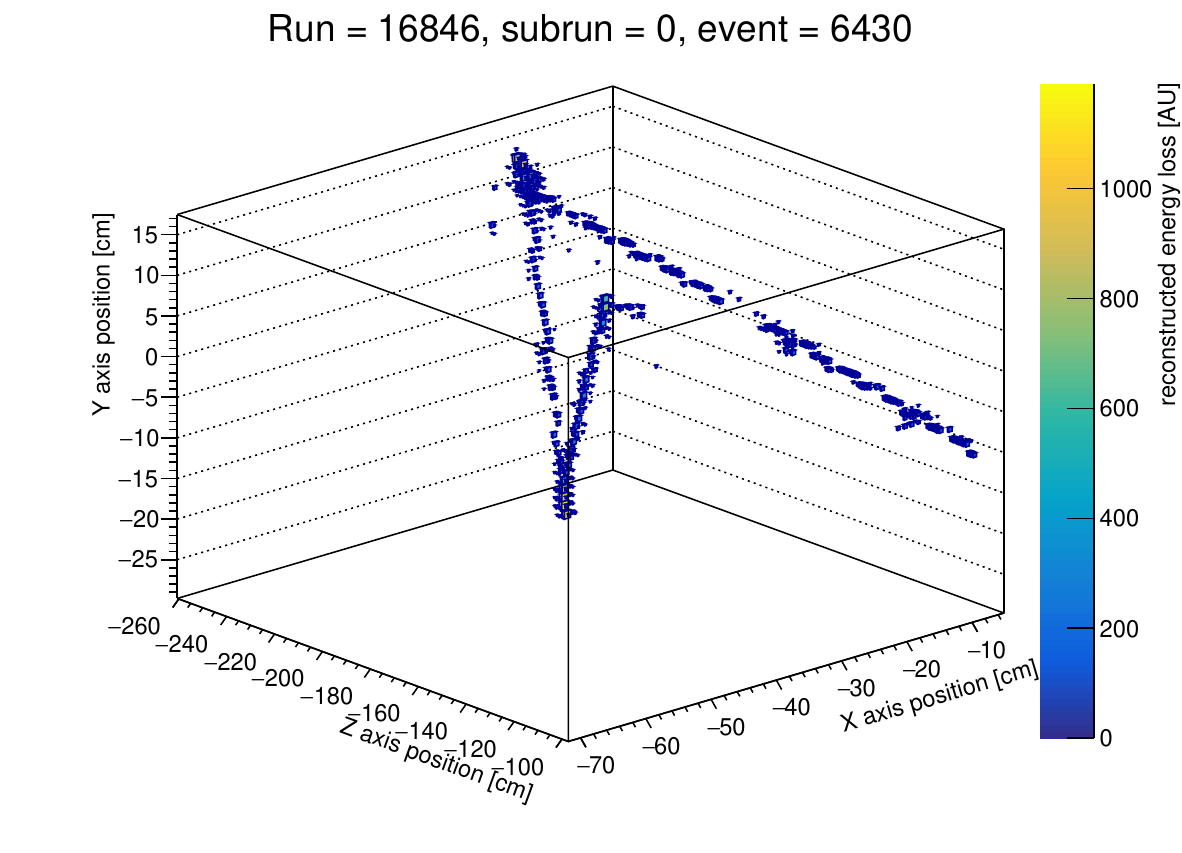}
\includegraphics[width=0.45\columnwidth]{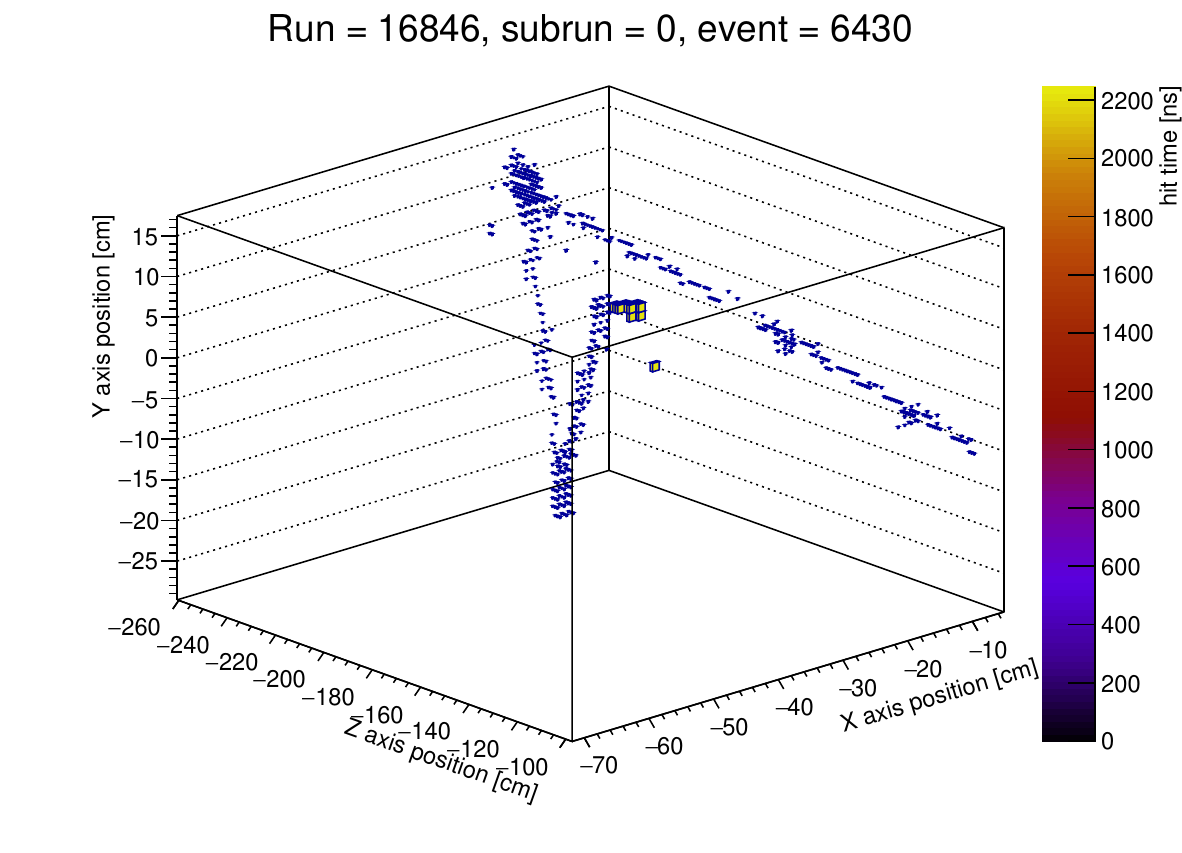}
\includegraphics[width=0.45\columnwidth]{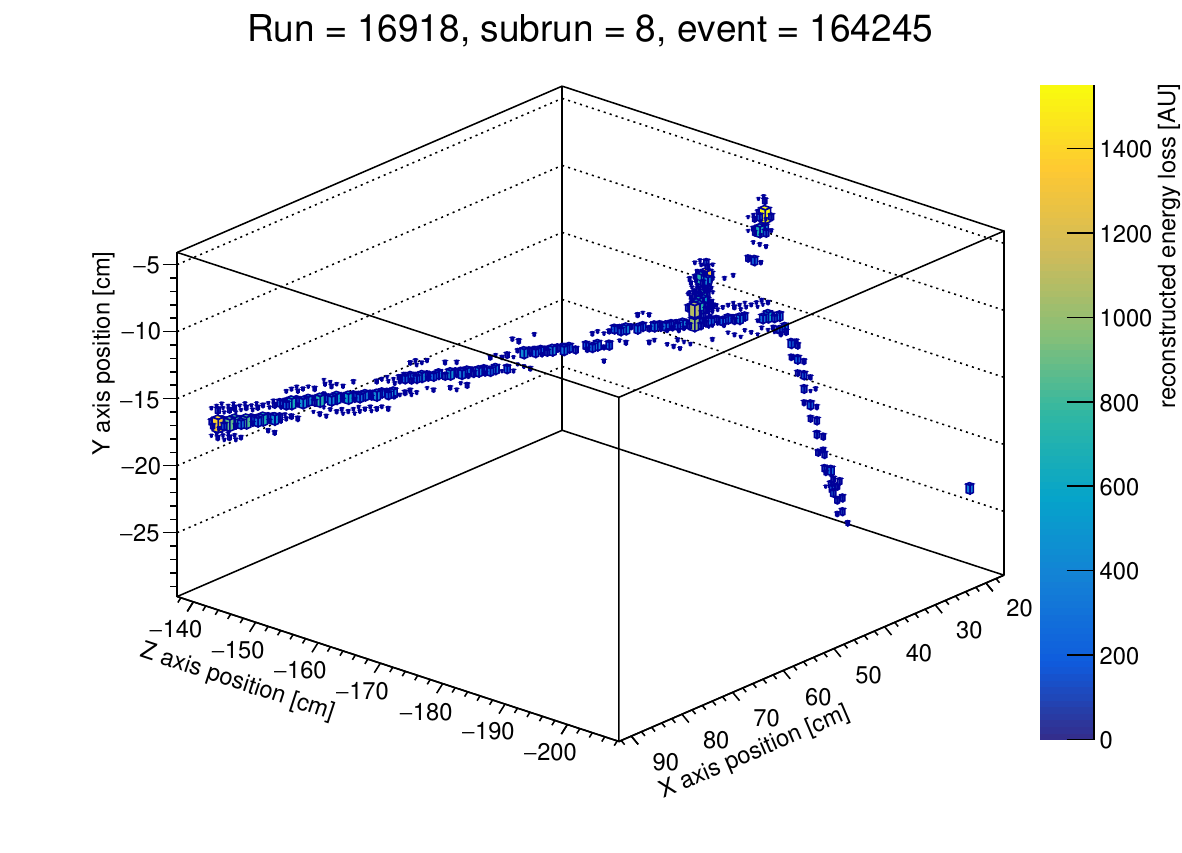}
\includegraphics[width=0.45\columnwidth]{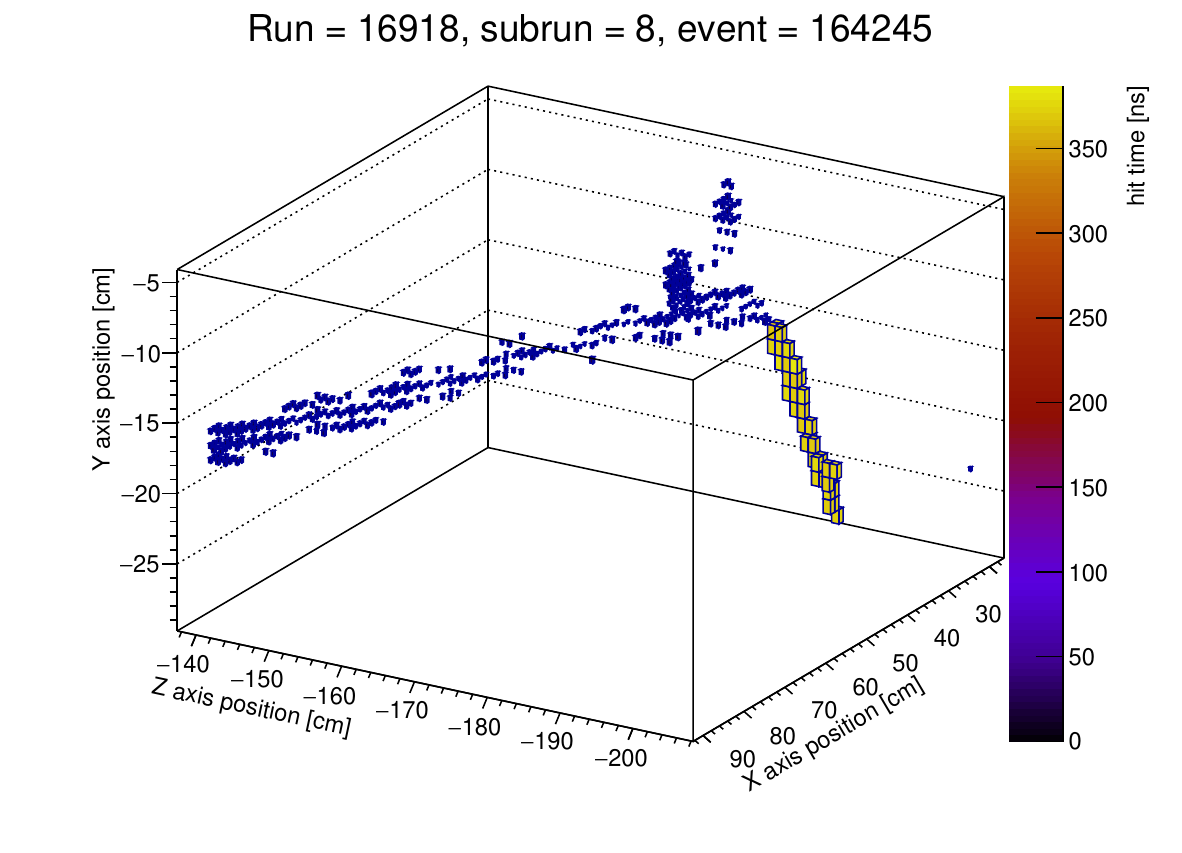}
\caption{
Candidate charged-current (CC) neutrino interactions in the SuperFGD with charged pion candidates collected during the June 2024 run. 
The reconstructed energy loss in AU (left) and the time in nanoseconds (right) are shown for the 3D reconstructed event in SuperFGD. 
Top: the candidate final state particles are a long-track muon and a charged pion undergoing a secondary interaction before decaying into a Michel electron after about 2.2~{\textmu}s.  
Bottom: the candidate final state particles are a long-track muon, a delta ray, and a Michel electron produced with a delay of about 370~ns by the decay of a low-energy pion that does not induce any visible MIP-like track.
}
\label{fig:neutrino_event_1pi}
\end{figure}

\begin{figure}[h!]
\centering
\includegraphics[width=0.7\textwidth]{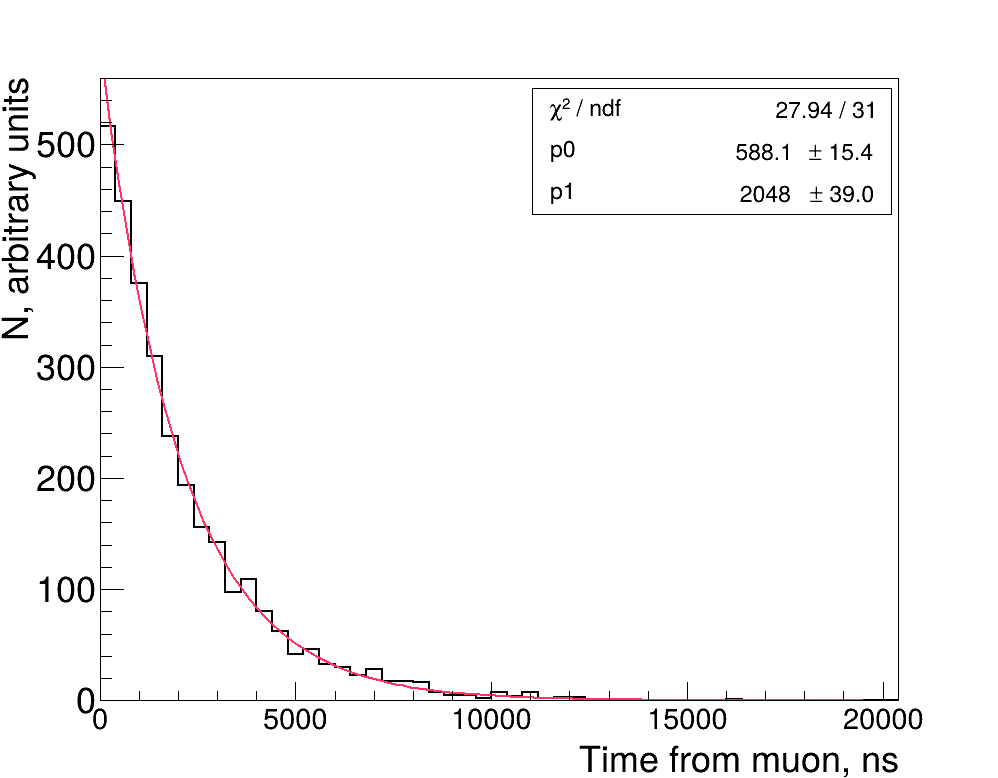}
\caption{
Time distribution of Michel electron events from a sample of $\mu^-$ and $\mu^+$ candidates. Fitting an exponential function shows a decay time of 2.05~{\textmu}s, slightly lower than 2.2~{\textmu}s, due to the capture of a fraction of $\mu^-$ before the decay.
}
\label{fig:michel_electron_decay_time}
\end{figure}

An example of a neutrino interaction producing a pion candidate is shown in Fig.~\ref{fig:neutrino_event_1pi}. The primary tool to identify charged pions stopping in the SuperFGD is the measurement of a time delay with respect to that of the neutrino interaction. The time distribution of particles produced by the decay of a muon stopping within the SuperFGD fiducial volume is shown in Fig.~\ref{fig:michel_electron_decay_time}. As expected, it is compatible with the muon lifetime.

\subsubsection{Neutrino final states with neutrons}
\label{sec:neutrino_interactions_neutrons}

The SuperFGD is the first neutrino detector that can detect high-energy neutrons and reconstruct their energy by measuring the time of flight. Their detection occurs by the identification of isolated clusters or tracks that show a Bragg peak feature with a delay of a few nanoseconds with respect to the neutrino interaction. The measurement of the 3D distance (lever arm) and the time difference between the neutrino vertex and the isolated signature allows the computation of the neutron time of flight and, consequently, its kinetic energy can be inferred.

\begin{figure}[h!]
\centering
\includegraphics[width=0.47\columnwidth]{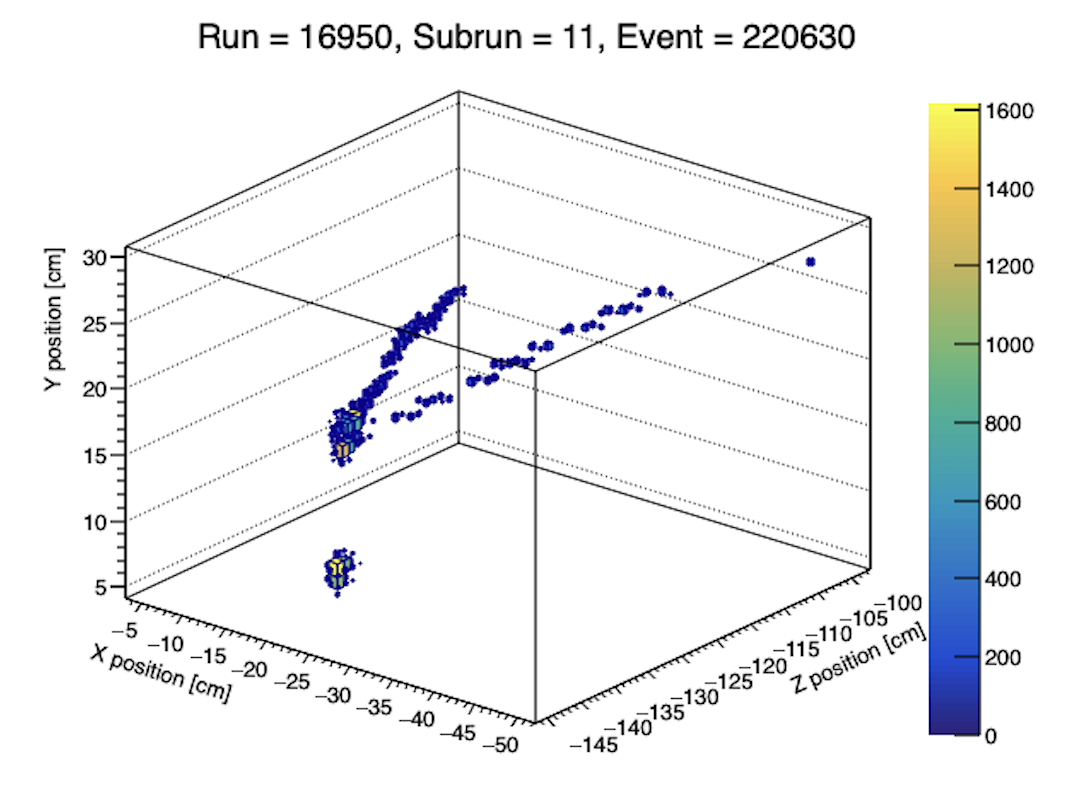}
\includegraphics[width=0.47\columnwidth]{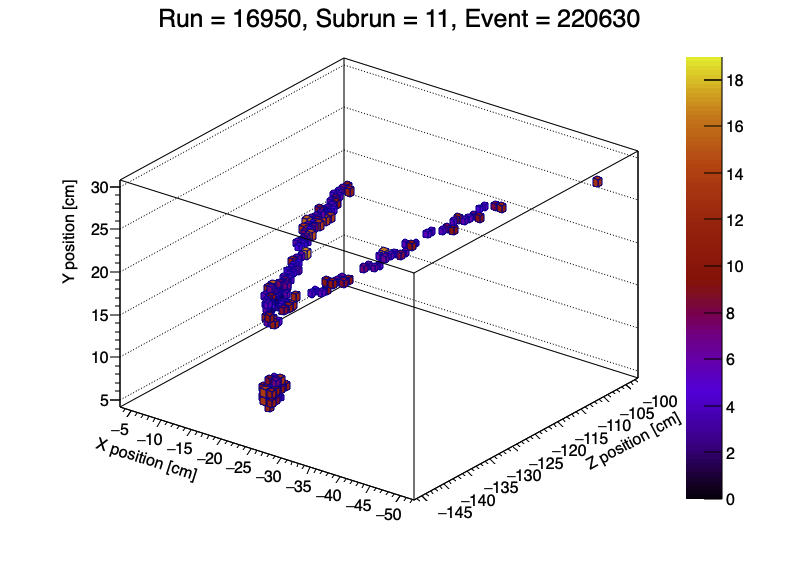}
\caption{
Candidate charged-current (CC) neutrino interaction in the SuperFGD collected during the June 2024 run with a neutron in the final state.
The energy loss in AU (left) and the time in nanoseconds (right) are shown for the 3D reconstructed event in the SuperFGD.
}
\label{fig:neutrino_event_neutron}
\end{figure}

A candidate event is shown in Fig.~\ref{fig:neutrino_event_neutron};
with a measured lever arm of about 122~mm and a time of flight of 1.7~ns, a neutron with a kinetic energy of about 27~MeV is reconstructed.

%% file: conclusions.tex
The novel design of the SuperFGD detector will allow the T2K experiment to detect neutrinos with unprecedented precision. The detector construction was completed in March 2023. Later, the SuperFGD was installed in ND280, then tested, calibrated with cosmic data, and commissioned in October 2023. After detecting the first neutrino interactions during the November-December 2023 technical run, the SuperFGD successfully began collecting neutrino beam data for physics analysis in May 2024. Owing to its innovative design, consisting of almost 2,000,000 scintillating cubes read out by around 56,000 orthogonal WLS fibres, a large sample of neutrino interactions can be reconstructed in three dimensions, independent of the direction of the final-state charged particles.

The fine granularity and high scintillation light yield will allow protons down to about 330~MeV/$c$ to be tracked and identified accurately. The Bragg peak will allow separation of protons from muons and pions, which can also be characterised by the delayed detection of Michel electrons. The 3D granularity plastic scintillator will ensure an  efficient detection of high-energy neutrons and a single-cube time resolution of 0.9~ns will make the measurement of the neutron kinetic energy possible.

The unique features of the SuperFGD make it a state-of-the-art neutrino detector.